\documentclass[letterpaper,usenatbib]{mn2e}

\pdfoutput=1
\usepackage{subfig}
\usepackage{amsmath}
\usepackage{graphicx}
\usepackage{color}
\usepackage{aas_macros}

\newcommand{\width}{0.5\textwidth}
\newcommand{\tserieslong}{Combining spectroscopic and photometric surveys
using angular cross-correlations}
\newcommand{\tseriesshort}{Combining spectroscopic and photometric surveys}

\newcommand{\be}{\begin{equation}}
\newcommand{\ee}{\end{equation}}
\newcommand{\xbf}{\begin{figure}}
\newcommand{\xef}{\end{figure}}
\newcommand{\xbfd}{\begin{figure*}}
\newcommand{\xefd}{\end{figure*}}
\newcommand{\xbt}{\begin{table}}
\newcommand{\xet}{\end{table}}

\newcommand{\mycite}{\citep}
\newcommand{\citedir}{\citet}

\newcommand{\tmpcitetwo}[1]{}
{}
\newcommand{\tmprm}[1]{}
\newcommand{\xfigure}[1]{
\begin{center}
\includegraphics[width=\width]{links/#1}
\end{center}
}

\newcommand{\micecosmo}{$\Omega_m = 0.25$, $\Omega_b = 0.044$, $\Omega_{DE} =
0.75$, $h = 0.7$, $w_0 = -1$, $w_a = 0$, $n_s = 0.95$ and $\sigma_8 = 0.8$ }

\newcommand{\dzbin}{\Delta z}

\newcommand{\sn}{S/N }

\newcommand{\wght}{\mathcal W}

\newcommand{\citeoverlap}{\mycite{cai2011,gazta,cai2012,kirk,mcdonald,deputter}}

\begin{document}

\newcommand{\papertitle}[1]{\title{#1}}

\newcommand{\tmod}{Algorithm and modelling}
\title[\tseriesshort\ I: \tmod]{\tserieslong\ I: \tmod}
\author[Martin Eriksen, Enrique Gazta\~naga]
{\parbox{\textwidth}{Martin Eriksen$^{1,2}$, Enrique Gazta\~naga$^1$}
\vspace{0.4cm}\\
$^1$Institut de Ci\`encies de l'Espai (IEEC-CSIC),  E-08193 Bellaterra (Barcelona), Spain \\
$^2$Leiden Observatory, Leiden University, PO Box 9513, NL-2300 RA Leiden, Netherlands \\
}

\newcommand{\A}[2]{#1}
\maketitle
\begin{abstract}
Weak lensing (WL) clustering is studied using 2D (angular) coordinates, while
redshift space distortions (RSD) and baryon acoustic oscillations (BAO) use 3D
coordinates,  which requires a model dependent conversion of angles and
redshifts into comoving distances. This is the first paper of a series, which
explore modelling multi-tracer galaxy clustering (of  WL, BAO and RSD),
using only angular (2D) cross-correlations  in thin redshift bins. This
involves evaluating many thousands cross-correlations, each a multidimensional
integral, which is computationally demanding. We present a new algorithm that
performs these calculations as matrix operations.

Nearby narrow redshift bins are intrinsically correlated, which can be used to
recover the full (radial) 3D information. We  show that the Limber
approximation does not work well for this task. In the exact calculation, both
the clustering amplitude and the RSD effect increase  when decreasing the
redshift bin width. For narrow bins, the cross-correlations has a larger BAO
peak than the auto-correlation because smaller scales are filtered out by the
radial redshift separation. Moreover, the BAO peak shows a second (ghost) peak,
shifted to smaller angles. We explore how WL, RSD and BAO contribute to the
cross-correlations as a function of the redshift bin width and present a first
exploration of non-linear effects and signal-to-noise ratio on these
quantities. This illustrates that the new approach to clustering analysis
provides new insights and is potentially viable in practice.

\end{abstract}

\section{Introduction}
Galaxy surveys provide data for constraining cosmological models. In the next
years and decades, the constrains will improve from current and upcoming
surveys. The completed CFHTls survey (stage-II) measured shear in the 155 sq.
deg. wide fields to $i < 24.5$ \mycite{cfhtls,cfhtlstomo,cfhtparconstr}. The
dark energy survey has completed the first year of data and plan to observe
5000 sq deg to $i<24.1$ over the next four years. Another examples of ongoing
WL survey are KIDS \& HSC, which aim to map about 1500 sq. deg each to
different depths. In the next decade EUCLID
\mycite{euclid,euclidscience,euclidmission} and LSST \mycite{lsst1,lsst2} will
provide the next generation (stage-IV) of deep lensing surveys, both covering
around 15000 sq. deg.

For spectroscopic surveys, the Wiggle-Z  measured almost 240000 galaxies over
1000 sq. degrees in the redshift range $0.2 < z < 1.0$ \mycite{wigglezsurvey},
while the BOSS survey mapped the redshift of 1.5 million galaxy to $z \approx
0.7$ in 10000 sq. deg \mycite{andersonboss}. A stage IV spectroscopic survey is
DESI \mycite{msdesi}, which is a merge of the previous BigBoss
\mycite{bigboss1,bigboss2} and DESpec collaborations \mycite{despec}. Expected
starting in 2018 at the Mayall telescope, DESI aims at measuring RSD and BAO
through targeting 20 million galaxies and cover between 14000 to 18000 sq.deg.
Also, a new generation of narrow band cosmological surveys will start in the
next years. Through 40 narrow band filter, e.g. the PAU survey will achieve a
high accuracy (0.3 \%) photo-z for $i_{AB} \sim 23$ \mycite{polpau}. The PAUcam in
addition contains u,g,r,i,z filters, so the survey provides a deep photometric
($i_{AB} < 24.1$) over the same area.

How do overlapping photometric and spectroscopic surveys change the
constraints on dark energy and modified gravity? A photometric survey 
with imaging is ideal for WL, while RSD
and BAO benefit from the accurate spectroscopic redshifts. Combining
the spectroscopic and photometric surveys bring additional benefit.
Two overlapping surveys allow cross-correlation of data, e.g. the
foreground spectroscopic galaxies with the background shear. Further,
the overdensities in both surveys trace the same underlying matter
which allows for sample variance cancellations. 

Several groups \citeoverlap, including the authors, have investigated the
effect of overlapping galaxy surveys and find different results for the
benefits. \A{This paper}{This thesis} follows up our previous paper
\mycite{gazta}, where we studied overlapping galaxy surveys by combining a 3D
$P(k)$  for spectroscopic surveys and 2D Cl estimators for photometric surveys.
When doing this, there is the risk of overcounting overlapping modes and not
including the full covariance between them. Here, to simplify the combination
and to avoid assuming a cosmology, both surveys are analyzed using the same
angular cross-correlations. In this respect, our approach is similar to that in
\citedir{asorey,kirk}, but including all the elements in \citedir{gazta}.  The
redshift bin projection in angular correlation removes some radial information
within the bin. However \citedir{asorey} showed that angular cross-correlations
in narrow bins recover the bulk of the available information.

Section \ref{sec_impl}\: discusses the numerical implementation of the
equations for evaluating the angular correlations. The computational time is
especially important for parameter constraints, which often require $10^5$ to
$10^6$ sample points in the parameter space. Including the RSD and lensing for
many thin redshift bins are computationally challenging, especially when also
including multiple galaxy populations, different measurements and finally the
cross correlations between all of them. Further subsection
\ref{subsec_investigating} discusses partial calculations as a method for
evaluating the results.

In section \ref{mod_sec_effects} we study the effect of Limber approximation,
BAO, RSD and the redshift bin width on the auto and cross-correlations.
Analyzing the spectroscopic sample require narrow redshift bins to capture the
radial information. The redshift bin thickness has a large impact on RSD and
BAO for both the auto and cross-correlations. Understanding these are essential
to interpret the forecast in the following papers in this series.  Especially
we note the BAO signal is stronger in the cross-correlations between redshift
bins than in the auto-correlations. The last subsection focus on the expected
error bars.

This paper is the first in a three part series. In this article we study the
modelling of the correlations function. The second article forecast the dark
energy and growth constraints for galaxy clustering, RSD and weak lensing.  In
the third article we investigate the dependence on bias assumptions. A separate
paper compare the constraints for overlapping and non-overlapping photometric
and spectroscopic surveys.

\label{chmod}
\section{Angular correlation function}
\label{mod_sec_corrtheory}
\label{corrtheory}
\subsection{Angular correlation in Fourier space.}
The observables considered here are fluctuations in galaxy number
counts $\delta_g$ and galaxy shapes (or ellipticity) $\delta_\gamma$. 
These fluctuations are subject to intrinsic large scale structure (or
intrinsic alignments), RSD and WL

\begin{equation}
\delta = \delta^{I} + \delta^{RSD} + \delta^{WL}
\label{deltad}
\end{equation}

\noindent
  Calculation of the intrinsic correlations and
the contribution of redshift space distortions are described for
example in \citedir{rsd2dformula}. Following the notation of \citedir{croccemod}
the angular correlation in Fourier space can be calculated by

\begin{equation}
C_l = \frac{1}{2 \pi^2} \int 4 \pi k^2 dk ~ P(k) ~\psi_l^2(k)
\label{cl_main}
\end{equation}

\noindent
For the intrinsic component of galaxy number counts, the kernel $\psi_l(k)$ is

\begin{equation}
\psi_l(k) = \int dz ~\phi(z) ~D(z)~ b(z,k) ~j_l(k r(z))
\label{psi}
\end{equation}

\noindent
Here $P(k)$ is the power spectrum of the underlying dark matter distribution,
$\phi(z)$ is the galaxy selection function (normalized galaxy redshift
distribution in our sample), $D(z)$ is the linear growth of structure 
and $j_l$ the spherical Bessel function. Galaxy overdensities are related to
the matter overdensities through the relation

\begin{equation}
\delta_g(z,k) = D(z) ~b(z,k) ~\delta_m(z=0, k)
\end{equation}
\noindent
where $\delta_g$ and $\delta_m$ are the galaxy and matter overdensities.
Therefore the power spectrum $P(k)$ can be expressed as $P(k)= D^2 b^2 P_m(k)$.
The bias ($b$) relates the galaxy and matter overdensities, with details being
discussed in paper-III.

Including redshift  space distortions adds an additional contribution 

\be
\psi_l(k)= \psi_l(k)^{Real} + \psi_l(k)^{RSD}
\ee

\noindent
to the real space contribution \mycite{rsdcl1,rsdcl2,rsdcl3}. The RSD term in
linear theory is given by \citedir{kaiser,linrsdrev}

\begin{eqnarray}
\nonumber
\psi_l^{\text{RSD}}& = &\int dz ~f(z) ~\phi(z)~ D(z) \\\nonumber
       &   &[L_0(l)~ j_l(kr) + L_1(l)~ j_{l-2}(kr) + L_2(l)~ j_{l-2}(kr)] \\\nonumber
\label{psi_r}
L_0(l) & \equiv & \frac{(2 l^2 + 2l -1)} {(2l+3)(2l-1)} \\
\nonumber
L_1(l)  & \equiv  &- \frac{l(l-1)}{(2l-1)(2l+1)} \\
L_2(l)  &  \equiv &- \frac{(l+1)(l+2)}{(2l+1)(2l+3)}
\label{psi_rsd}
\end{eqnarray}

\noindent
where $f(z)$ in the growth rate, which we write as $f(z) \equiv
\Omega_m(z)^\gamma$. Note that we assumed that velocities are the same for the
galaxies as for matter, so there is no bias term in the RSD contribution.

Weak gravitational lensing changed the galaxy ellipticities and the number
densities through magnification effects. Both of these can be described by the
convergence field $\kappa$. The convergence in a redshift bin $j$ caused by
dark matter lenses at z is \mycite{bartscheider}

\begin{equation}
p_{\kappa_j}(z) \equiv \frac{3 \Omega_{m0} H_0 r(z)}{2 H(z) a(z) r_0}
 \int_z^\infty dz' \frac{r(z'; z)}{r(z')} \phi(z')
\label{lensing_contribution}
\end{equation}

\noindent
where $\Omega_{m0}$ and $H_0$ are the matter density and Hubble distance
at $z=0$. The quantity $r(z'; z)$ is the angular diameter distance between
$z'$ and $z$. In order to estimate the lensing power spectrum, we can
use Eq. \ref{cl_main} and to evaluate the lensing kernel we need
to replace $\phi(z)$ by $p_{k_j}(z)$ in Eq. \ref{psi} (with $b= 1$), i.e.:

\begin{equation}
\psi_l(k) = \int dz ~p_{\kappa_j}(z)   ~D(z) ~j_l(k r(z))
\label{psi_lens}
\end{equation}

Gravitational lensing changes the observed number counts through two effects.
A galaxy observed close to a foreground matter overdensity will appear
brighter, which change the number of galaxies entering a magnitude limited
sample. This change depends on the slope of the number counts ($s_n$). Weak
lensing magnification also affects the area. The observed area in the
background of a matter overdensity will appear larger and would reduce the
galaxy density for a fixed number of galaxies. Combining these two effects the
change in $\delta_g$ from weak lensing magnification is

\begin{equation}
\delta^{WL}_g \propto 5 s_n(z_i) - 2
\end{equation}

\noindent
where the slope of the number counts $s_n$ comes from 

\begin{equation}
s_n(z_i) \equiv \frac{dlog_{10} N_n(<m, z_i)}{dm}
\end{equation}

\noindent
and $N_n(<m, z_i)$ is the number of galaxies at redshift $z_i$ with apparent
magnitudes less than $m$. Therefore the lensing component of galaxy number
count fluctuations in Eq. \ref{deltad} is

\begin{equation}
\delta^{WL}_g \approx (5s -2) \delta_{\kappa} \equiv \alpha \delta_{\kappa}
\end{equation}

\noindent
where the last equivalence defines $\alpha$. For simplicity, the galaxy
ellipticity $\gamma$, is assumed here to be directly proportional to $\kappa$
($\delta^{WL}_{\gamma} =  2 \delta_{\kappa}$), but we could in principle
easily include additive and multiplicative observational biases in the
calculation. 

\subsection{The Limber approximation}
Two approximations greatly simplify the evaluation of the analytic correlation
functions.  The first one is the narrow bin approximation, assuming no redshift
evolution within a redshift bin. For narrow bins it can be a good
approximation. Second is the Limber approximation, using the relation 
\mycite{limber,extlimber,limberjeong}

\begin{equation}
\frac{2}{\pi} \int k^2 dk ~j_l(kr) ~j(kr') = \frac{\delta^D (r-r')}{r^2}
\end{equation}

\noindent
can remove one additional integration. The symbols $r$ and $r'$ are the distances
to the two redshifts to correlate. In the case of $r \neq r'$, which is the case
for cross-correlations between redshift bins, the contribution is zero for the Limber
approximation. Later we will compare the exact calculations and the Limber approximation
in detail. 

In the notation $C_{A_i B_j}$ then $A$ and $B$ is the observable, i.e. galaxies
($g$) or shear ($\gamma$). An additional letter behind $g$ (e.g. gF) indicates a specific
galaxy population. The indices $i$ and $j$ denote the redshift bin and $i = j$
is the auto-correlation, while $i \neq j$ is a cross correlations. Below follow
a short summary of the formula given in \mycite{gazta}. To simplify the
notation we define

\newcommand{\calp}{\mathcal P}
\begin{equation}
\calp(k, z) \equiv \frac{P(k,z)}{r_H(z) r^2(z)}
\end{equation}

\noindent
where $\calp(k, z)$ is the power spectrum and $r_H(z) \equiv \partial r(z) /
\partial z$.  The galaxy clustering can then be written

\begin{equation}
C_{gn_i gm_j} \approx [b_{n_i} b_{m_j} \frac{\delta^K_{ij}}{\Delta_i} + \alpha_{m_j}
  b_{n_i} p_{ij}] \calp_i + \alpha_{n_i} \alpha_{m_j} C_{\kappa_i \kappa_j}(l)
\label{c_gg}
\end{equation}

\noindent
where $b$ are the galaxy biases, $\delta^K_{ij}$ is the Kronecker delta. The
second term is the correlation between the intrinsic galaxy lenses and the
magnified galaxy counts. This magnification term include the lensing potential

\begin{equation}
p_{ij} = \frac{3 \Omega_{m0} H_0}{2 H(z_i) a_i} \frac{r_i r(z_j; z_i)}{r_0 r_j}
\label{lensing_eff}
\end{equation}

\noindent
which is evaluated in the narrow bin approximation. As before, the $r(z; z')$
notation indicates an angular diameter distance between $z$ and $z'$. The term
$\Omega_{m0}$ denote the matter density at $z=0$ and $r_0 = c/H_0$. Last term
in \eqref{c_gg} correlates magnified lenses with magnifies sources. In practice
the two first terms dominate.

The galaxy-shear correlation is

\begin{equation}
C_{gn_i \gamma_j} \approx b_{n_i} p_{ij} P_i + 2 \alpha_{n_i} C_{\kappa_i \kappa_j}
\label{c_gk}
\end{equation}

\noindent
when $z_i < z_j$, otherwise zero. Finally, there is the $C_{\kappa \kappa}$ term,

\begin{equation}
C_{\kappa_i \kappa_j} \approx \int_0^{z_i} \frac{dz}{r_H}
 {(\frac{3 \Omega_m H_0}{2 H a r_0})}^2 \frac{r(z_i; z) r(z_j; z)}{r_i r_j} \calp(k,z)
\end{equation}

\noindent
which is proportional to the shear-shear ($C_{\gamma \gamma} = C_{\kappa
\kappa} / 4$) signal and is also part of the calculations for $C_{g g}$ and $C_{g \kappa}$
in Eq. \ref{c_gg} and \ref{c_gk}. One integration remains, since the lensing is
affected by all the matter in front of the redshift bin. Using a thin bin and
only integrating over the lens or source bin would lead to wrong results
\mycite{gazta}. When showing correlation with the Limber approximation, we use
the expressions above.

\section{Algorithm for 2D correlations}
\label{sec_impl}

\subsection{Motivation}
Estimating the angular correlation function involves integrating equation
\eqref{cl_main} using \eqref{psi} and \eqref{psi_rsd} for the intrinsic and RSD
contributions, and then \eqref{lensing_contribution} to add lensing. For the
intrinsic and redshift space distortion the calculations are three dimensional
integrals, two for each of the redshift bins and one over scale. When adding
lensing one should, to be correct, use two more integration,
corresponding to the dark matter lensing the source and the lens redshift. This
section looks at how to collapse the multi dimensional integrals into matrix
multiplications. It results in a both efficient and understandable algorithm.

Next level of complication includes using multiple observations, like galaxy
counts and shear, splitting tracers into multiple populations and doing the
analysis with a large number of redshift bins. One could approach this problem
by constructing a function or equivalent returning the correlation for a given
observation, tracer and pair of redshift bins. But this approach is not very
efficient.  In general organizing a code introducing additional layers help the
organization, while removing layers improve the speed. Part of this section
discusses how to simultaneously calculate the correlations for different
tracers, observations and pairs of redshift bins. The idea is to save time by
reusing parts of the calculations.

Multiple dimensional integration over
spherical Bessel functions are a potential source for
numerical errors. Two common approaches for testing the
accuracy  is to compare against other codes and to increase resolution settings
within the code. We have of course tried both.
A third approach is to inspecting if partial results of the calculations
make sense. In subsection \ref{subsec_investigating} we discuss how this can
be done in practice and explore potential problems in the integration.

One alternative to do these calculations is to use publicly available software
like CAMB Sources \mycite{cambsources1,cambsources2} or CLASS
\mycite{xclass1,xclass2,xclass3,xclass4,xclass5,xclass7}. But these only became
available in 2011 after the project had already started.  Moreover, integrating
the codes to be able to use arbitrary $n(z)$, bias parameterization, different
galaxy populations and magnification slopes would itself be a significant
addition.  We hope the formalism provided here gives another view on how to
evaluate the correlations in Fourier space, which is also quite efficient and
produce very fast results for a give accuracy.

\subsection{Implementation}
\label{subsec_impl}
As detailed in section \ref{corrtheory}, the starting point is that the fluctuations, in both
 galaxy counts $\delta_g$ and galaxy shear $\delta_\gamma$,
are made up of three contributions: intrinsic, 
redshift space and lensing:

\begin{equation}
\delta_A(k, z) = \delta_A^I(k,z) + \delta_A^{RSD}(k,z) + \delta_A^\text{WL}(k,z).
\end{equation}
where $A$ can be one of the two probes: $A=g$ for galaxy counts or
$A=\gamma$ for shear ellipticity in galaxies.
\noindent
When correlating these overdensities in two redshift bins labeled $i$ and $j$:

\begin{equation}
C_{ij}(l) = \int dk \int_{\text{Bin}\: i} dz_i \int_{\text{Bin}\: j} dz_j \left< \delta(k, z_i), \delta(k, z_j)
  \right>
\end{equation}

\noindent
Thus, the final correlation includes nine different terms for each probe or
cross-combination (i.e. $\delta_g \delta_g$, $\delta_\gamma\delta_\gamma$ and
$\delta_g \delta_\gamma$), which are not all equally important. For the time
being all the effects will be included in the calculation without further
approximation. Following this general approach leads to a simple implementation
with a good performance. 

\subsection{Tomographic integration}
Numerical deterministic integration of a function $f$ over a finite interval
can be expressed as \footnote{Integration algorithms can also be stochastic.
For one example in astronomy, MPTBreeze use the Vegas algorithm to efficiently
evaluate the two-loop propagator \mycite{mptbreeze}. Further, some integration
algorithms use knowledge of the function derivatives.}

\begin{equation}
\int dy ~f(y) = \sum_x w_x f(y_x)
\end{equation}

\noindent
where $w_i$ is a set of weights and $y_i$ is a set of sample points, which
differs between algorithms. Adaptive algorithms are often on the form above and
then subdividing the integral domain where the required accuracy has not been
achieved. 

Ignoring multiple tracers and different probes by now, the integration to evaluate
the Cls can be written as

\begin{equation}
C_{ij}(l) = \int dk ~G_i(k)~ G_j(k) \equiv \sum_x ~w_x ~G_i(k_x)~ G_j(k_x)
\label{cij_tomo_int}
\end{equation}

\noindent
where the form of $G(k)$ follows from Eq.\ref{cl_main} and $i$ and $j$ denote two redshift
bins. One could evaluate the integral \eqref{cij_tomo_int} for each pair
$(i,j)$ of redshift bins. Alternatively by defining

\begin{align}
H_{sx} \equiv \sqrt{w_x} ~G_s(k_x)
\label{H_def}
\end{align}

\noindent
the integration \eqref{cij_tomo_int} can be rewritten in terms of \eqref{H_def}
as 

\begin{equation}
C_{ij}(l) = \sum_x H_{ix} H_{jx}.
\label{cl_as_H}
\end{equation}

\noindent
In this form the matrix $H$ can be constructed once and then used to compute
the correlations between all bins. More importantly, the form \eqref{cl_as_H}
is closely related to the matrix product. If we consider $C$ to be a matrix
where $C_{ij}$ is the correlation between bin $i$ and $j$, the whole $C$
can be calculated as

\begin{equation}
C = H H^T
\label{cl_Hdot}
\end{equation}

\noindent
where $T$ denotes the transpose. The calculations are normally expressed as
loops over $i$, $j$ and $k$. Expressing the operations as matrix multiplication
makes it possible to evaluate the expression using DGEMM from level-3 
BLAS\footnote{http://www.netlib.org/blas/}. This is particularly important in
higher level languages, like Python, where looping is very slow. Also FORTRAN,
c and c++ should benefit since DGEMM has highly efficient implementations like
MKL from Intel and the open source OpenBlas. In addition the expressions looks
readable and require less lines of code.

One suitable algorithm for evaluating oscillating integrands is the Clenshaw-Curtis
(CC) quadrature. The appendix \ref{cc_int} includes a brief introduction and how
to handle changes of integral domain for the tomographic integration and here
we include the explicit integration formulas.

Using the CC-algorithm one needs to split \eqref{cl_Hdot} into two parts 

\begin{equation}
C = H^{+} {(H^{+})}^T + H^{-} {(H^{-})}^T 
\end{equation}

\noindent
where

\begin{align}
H^{+}_{sx} &= \sqrt{k_w \wght_x} G_s(\bar{k} + k_w \cos{\frac{n \pi}{n}}) \\
H^{-}_{sx} &= \sqrt{k_w \wght_x} G_s(\bar{k} - k_w \cos{\frac{n \pi}{n}}) \\
\bar{k} &= \frac{1}{2} (k_{min} + k_{max}) \\
k_w &= \frac{1}{2} (k_{max} - k_{min}) \\
\end{align}

\noindent
and the weights $\wght$ are given in the appendix \ref{cc_int}.

\subsection{Intrinsic correlations and RSD.}
This subsection focus on the expression for $G$ in equation \ref{cij_tomo_int},
taking into account the intrinsic correlation and RSD contribution, while next
subsection explains the lensing contribution.

The integration over the redshift binning can be done through the relations

\begin{align}
G^{I}_i &= \tilde{G} \int_{\text{Bin i}} dz \psi^I(z,k) \\
\label{g_exp1}
G^{RSD}_i &= \tilde{G} \int_{\text{Bin i}} dz \psi^{RSD}(z,k) \\
\label{g_exp2}
\tilde{G} &= \frac{2}{\pi} k \sqrt{P(k)}
\end{align}

\noindent
using equations \eqref{cl_main}, \eqref{psi} and \eqref{psi_r}. As stated earlier, the goal
is to express the integration through matrix multiplication. First the redshift
range where some bin has support is divided into a grid. For narrow top-hat bins
one can simply use the bins themselves. The function $\phi_i(z)$ in Eq. \ref{psi}
denote the probability of a galaxy in bin $i$ having true redshift $z$. In
photometric surveys the bins are not top-hat, but are for each bin
given by a probability distribution. The probability is found by binning in
photometric redshift and the comparing with the spec-z in the calibration
sample.

The probability distributions are then combined into one matrix

\begin{equation}
\phi \equiv
\begin{pmatrix}
\phi_{00} & \phi_{01} & \dots &  \phi_{0n} \\
\phi_{10} & \phi_{11} & \dots & \phi_{1n} \\
\dots & \dots & \dots & \dots \\
\phi_{n0} & \phi_{n1} & \dots & \phi_{nn} \\
\end{pmatrix}
\label{prob_combined}
\end{equation}

\noindent
where $\phi_{ij}$ is the part of $\phi_i$ overlapping with the underlying grid
bin $j$. In the case of narrow non-overlapping redshift bins using the bins
itself as a grid, then $\phi_{ij} = \delta_{ij} \phi_{ij}$. Integration in
redshift is also done using the Clenshaw-Curtis algorithm inside each of the
redshift grid bins. The evaluation points in redshift, using $N_z$ integration
points inside grid bin $j$, are

\begin{align}
z^{+}_{jx} &\equiv \bar{z}_j + z^w_j \cos{\frac{n \pi}{N_z}} \\
z^{-}_{jx} &\equiv \bar{z}_j - z^w_j \cos{\frac{n \pi}{N_z}} \\
\bar{z}_j &\equiv \frac{1}{2} (z_j^{Min} + z_j^{Max}) \\
z^w_j &\equiv \frac{1}{2} (z_j^{Max} + z_j^{Min}) \\
\end{align}

\noindent
with $z^{+}_{j}$ and $z^{-}_{jx}$ denoting two contribution to the integral
over bin $j$. In practice one concatenates the two 1D arrays $z^{+}_{j}$ and
$z^{-}_{jx}$ into a larger array before evaluating the probability functions.

One also needs weights for integrating over the redshift bins. The weight 
arrays $w_i ... w_{ngrid}$ for each of the $ngrid$ redshift grid bins are
then concatenated into the array 

\begin{equation}
\wght_z \equiv [w_0, w_1,\dots,  w_{ngrid}]. 
\label{w_z}
\end{equation}

\noindent
If one uses the same number of integration points in each bin, as we do, then
the operation reduces to repeating the same weight matrix $ngrid$ times. The
probability functions \eqref{prob_combined} and \eqref{w_z} can then be
combined into 

\begin{equation}
(W^{Gal})_{ij} = \phi ~\wght_z
\end{equation}

\noindent
where the multiplication is with the second index in $\phi$. Integration over
redshift bins in \eqref{g_exp1} and \eqref{g_exp2} can, dropping the superscript,
be written as 

\begin{equation}
G = \phi ~y(z,k)
\label{G}
\end{equation}

\noindent
where the $z$ binning is the one used for the redshift grid when evaluating
$\phi$. The function $y(z,k)$ is defined through $\psi(z,k) = \phi(z,k) y(z,y)$,
and can explicitly be written as

\begin{equation}
y^I(z,k) = \tilde{G}(k) ~D(z) ~b(z,k) ~j_l(k r(z)).
\end{equation}

\noindent
In the case of redshift space distortions one should add a similar term

\begin{align}
\label{yrsd}
y^{RSD}(z,k) =& \enskip  \tilde{G}(k) ~f(z) ~D(z) ~ [L_0(l) j_l(kr) 
\\
\nonumber
& + L_1(l) ~j_{l-2}(kr) + L_2(l) ~j_{l-2}(kr)]
\end{align}

\noindent
where $L_0(l)$, $L_1(l)$ and $L_2(l)$ are defined in \eqref{psi_r}. To
implement this, one can construct splines of the spherical Bessel functions.
Instead of evaluating $j_l$, $j_{l-2}$ and $j_{l+2}$, the linear combination
used in \eqref{yrsd} is calculated once and then stored in splines.

\subsection{Weak gravitational lensing}
Weak gravitational lensing affects the galaxy shapes and counts at the source
redshift from the foreground matter, while the intrinsic correlations and RSD
contributions are caused by the matter overdensities at the same redshift. In
addition to integration over the scale and the two redshift bins, evaluating
the lensing contribution require integrating over the foreground dark matter.
While five dimensional integrals sounds tricky, they can be evaluated
efficiently by reusing terms and considering the correlations between all
redshift bins at once.

To include the lensing effect described in Eq.
\eqref{lensing_contribution}, we use

\begin{equation}
\eta(z_j, z_i) \equiv  \frac{3 \Omega_{m0} H_0 r(z)}{2 H(z) a(z) r_0} \frac{r(z'; z)}{r(z')}
\end{equation}

\noindent
where $z_i$ is the lens and $z_j$ the source redshift. Defining the second
index for the lens redshift allows to later add lensing using left
multiplication. The lensing contribution is then

\begin{align}
G^{WL} &\equiv (\phi \wght_z) ~(\tilde{\eta} \wght_z) ~y^{Mat} \\
y^{Mat}(z,k) &\equiv \tilde{G} ~D(z) ~ j_l(k r(z))
\end{align}

\noindent 
where $\tilde{\eta}(z_j, z_i) \equiv \alpha(z_j) \eta(z_j, z_i)$ either include
the magnification factor for galaxy counts or is set to unity for cosmic shear.
When evaluating $\eta$ one use the same redshift binning as $\phi$. In this
notation the same $\phi$ is also used for the intrinsic correlations and
redshift space distortions. The disadvantage is that we need to use the highest
redshift resolution required. However this allows us to reuse e.g. the
evaluated spherical Bessel functions $j_l$, for all contributions to the
overdensities.

\subsection{Combining multiple terms}
In the previous subsections the focus was an efficiently evaluation the 
cross-correlations including the intrinsic correlation, RSD and weak lensing.
These contributions were included in the terms $G^I$, $G^{RSD}$ and $G^{WL}$ and
to account for all effects we just have to sum them

\begin{equation}
G = G^I + G^{RSD} + G^{WL} \\
\end{equation}

\noindent
and calculate the $C_l$ (Eq. \ref{cl_Hdot}). These calculations alone could
require 7 nested loops, if implemented in a straight forward and naive
approach. In addition a forecast or MCMC run require the following layers

\begin{itemize}
\item Cosmological parameters
\item l value
\item Galaxy population in Bin 1
\item Galaxy population in Bin 2
\item Observable in Bin 1
\item Observable in Bin 2
\end{itemize}

\noindent
The efficiency of the integration depends on the matrix order. Matrix
multiplication is an associative operation, i.e.

\begin{equation}
A (B C) = (A B) C
\end{equation}

\noindent
where $A$, $B$ and $C$ are matrices. In term of implementation the order affects
the number of the operations. Assume the matrix dimensions are 

\begin{align}
A &= k \times m \\
B &= m \times n \\
C &= n \times s
\end{align}

\noindent
Evaluating $A (B C)$ require $m n s + k m s$ operations, while (AB)C require $k
m n + k n s$ operations. Depending on the k,m,n values one ordering is the most
efficient. For the accuracy used in our implementation, we calculated the
number of operations needed for different orders of integrating using the
default resolution in redshift, scale and around 100 redshift bins.
The most efficient choice was first integrating out lensing, then binning in
redshift and last integrating over the scale. This order was in previous
subsections reflected both in the formulas and in the presentation. 

\subsection{Investigating partial calculations.}
\label{subsec_investigating}

The previous subsections described an algorithm for calculating the 2D
correlations in Fourier space including the intrinsic correlation, redshift
space distortions and lensing. In addition to comparing the resulting Cls with
public software, one can directly check steps of the calculations. The
algorithm first integrates over all redshift variables and last over the scale.
In the notation of Eq.\ref{cl_Hdot}, one can construct a cumulative sum

\newcommand{\kcum}{k_{\text{c}}}

\begin{equation}
C_{ij}^{\text{Partial}}(\kcum) = \sum_x^{k(x) < \kcum} H_{ix} H_{xj}
\label{cpartial}
\end{equation}

\newcommand{\cpart}{C^{\text{Partial}}}
\noindent
of the correlations, where $k(x) < \kcum$ means the sum of 
indices $x$ until the one corresponding to $\kcum$. Note: the $\kcum$ used in 
Eq. \ref{cpartial} is only defined to see which scales contribute to a
correlation and is not the maximum $k$ considered in the forecast.

In the case of insufficient integral precision in redshift, the numerical
artifacts often enters at high $k$ due to the product $k r(z)$ in
$\psi_l(k)$ (Eq. \ref{psi}). These errors can create serious problems
for Fisher matrices, where high precision is needed are easily detectable looking
at $C^{\text{Partial}}$, but difficult to spot looking at the final
correlations.

In the remainder of this subsection we present figures of $C^{\text{partial}}$
for counts-counts auto and cross-correlations, and counts-shear, shear-shear
auto-correlations.  These figures are not only useful for detecting errors, but
also helps to understand which scales contributes to the correlations. The
fiducial cosmological model used is $\Lambda CDM$ with \micecosmo corresponding
to the values used in the MICE simulations \mycite{mice,mice2}. Galaxies are
bias through the relation $b(z) = 2 + 2 (z - 0.5)$, except for the thick
redshift bins in Fig. \ref{delta4} which has a galaxy bias of $b(z) = 1.2 + 0.4
(z - 0.5)$. These bias values are chosen to exactly match the Bright and Faint
population introduced in paper-II. The power spectrum used is Eisenstein-HU
(EH) \mycite{eishu}.

\newcommand{\sumk}[1]{Cumulative contribution to #1 at $l = 256$ from different
scales ($\cpart$). The correlations includes k-values until the limit displayed
on the x-axis.}

\begin{figure}
\xfigure{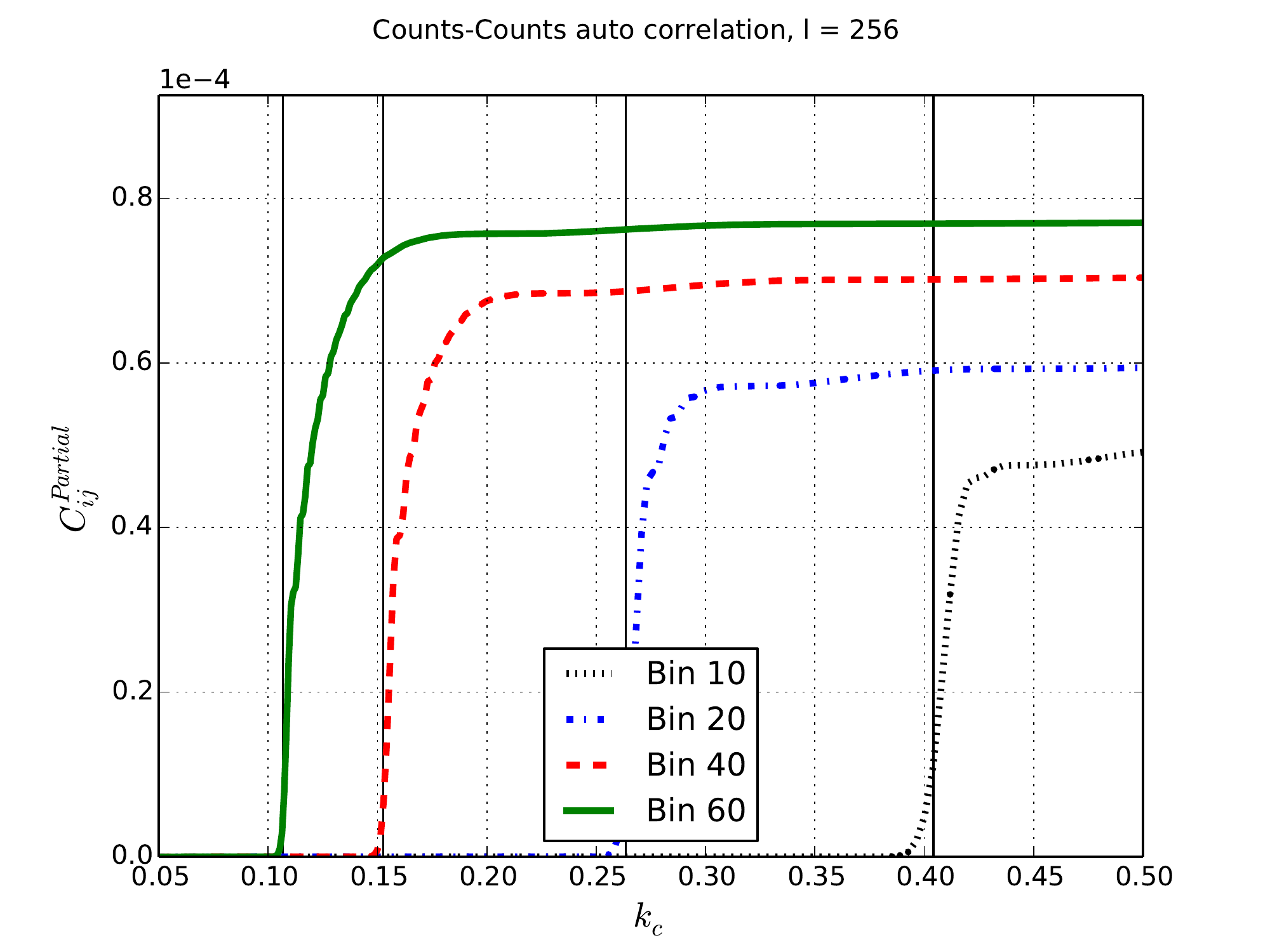}
\caption{\sumk{Counts-Counts auto-correlations} Lines in figure corresponds to
the redshift bins $0.22 < z_{10} < 0.23$, $0.34 < z_{20} < 0.36$, $0.64 <
z_{40} < 0.65$ and $1.00 < z_{60} < 1.02$}
\label{delta_auto}
\xef

Fig. \ref{delta_auto} includes four $C^{\text{Partial}}$ lines for the auto
correlations at $l = 256$ of galaxy counts in four narrow redshift bins. These
auto-correlations include redshift space distortions and a sub-dominant
magnification term. From the Limber approximation, one expects the largest
contribution from $k = (l + 1/2)/\chi(z_m)$, where $\chi(z_m)$ is the mean
comoving distance of the redshift bin. For the fiducial cosmology one expect
the main contribution to the correlations around 
$k = 0.40, 0.26, 0.15, 0.11$ Mpc $h^{-1}$ for the redshift bins $z = 0.23,
0.35,0.65,1.01$ as shown by the vertical lines in Fig. \ref{delta_auto}.  The
estimated scale for the main contribution to the correlation agrees well with
the figure.

\xbf
\xfigure{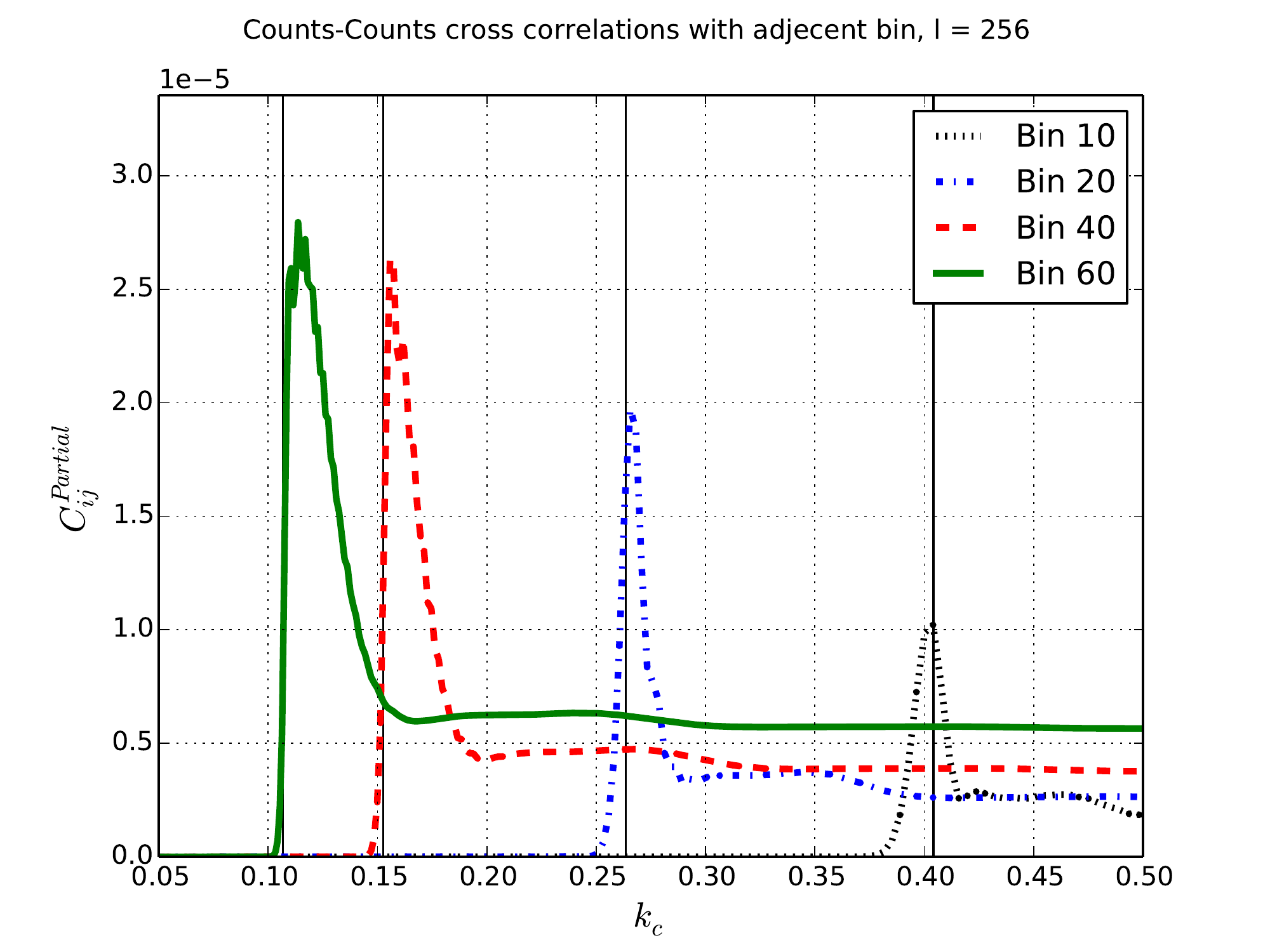}
\caption{\sumk{Counts-Counts cross-correlations} The first correlation is between 
$0.22 < z_{10} < 0.23$ and $0.23 < z_{11} < 0.24$,
the second between $0.34 < z_{20} < 0.36$ and $0.36 < z_{21} < 0.37$, 
the third between $0.64 < z_{40} < 0.65$ and $0.65 < z_{41} < 0.67$, 
and the fourth between $1.00 < z_{60} < 1.02$,  and $1.02 < z_{61} < 1.04$.}
\label{delta_cross}
\xef

In figure \ref{delta_cross} four thin redshift bins are cross-correlated with
the adjacent redshift bin. Each cross-correlation in the figure corresponds to
one auto correlation in Fig. \ref{delta_auto}.  Comparing the two figures,
similar scales contribute to both the auto and cross-correlations. A
characteristic feature of the cross-correlation is the sharp peak. An
auto-correlation has only positive contribution as a function of scale, while
for a cross-correlation the spherical Bessel functions are slightly out of
phase. This results in $k$ values with negative contributions and result in
a filtering of small scales (see \ref{baosubsec}).

\xbf
\xfigure{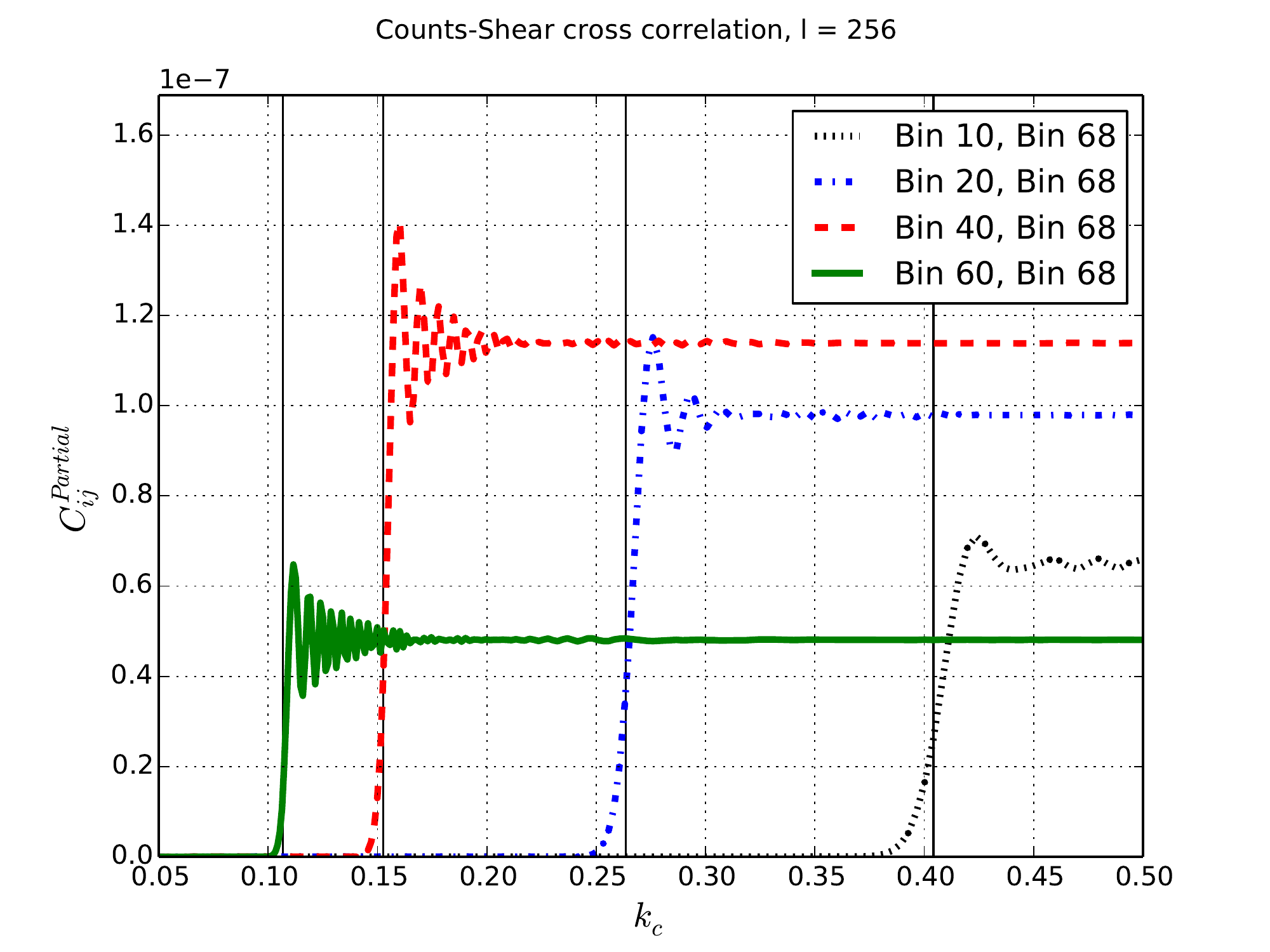}
\caption{\sumk{Counts-Shear cross-correlations} For all the correlations the
background is $1.16 < z_{68} < 1.19$, while the foreground redshift bins are $0.22
< z_{10} < 0.23$, $0.34 < z_{20} < 0.36$, $0.64 < z_{40} < 0.65$ and $1.00 <
z_{60} < 1.02$}
\label{delta_gk}
\xef

Fig. \ref{delta_gk} cross-correlate overdensities of foreground galaxies with
background shear. All lines use the source redshift bin $1.16 < z <
1.19$, while the foreground redshift bins equals the foreground bins in the 
previous two figures (Fig. \ref{delta_auto} and \ref{delta_cross}).
The scales contributing to the counts-shear cross correlation
is precisely the ones contributing to the auto correlation. That is expected
from looking at the Limber equations for counts-counts Eq. \ref{c_gg} and 
counts-shear Eq. \ref{c_gk}, both including the power spectrum evaluated at
$k = (l + 0.5) / \chi(z_m)$. Further, in the auto-correlation the amplitude 
increases with redshift because galaxy bias increase with redshift.
For counts-shear the correlation peaks with redshift and is lower for the highest
redshift bin. This effect comes from the lensing efficiency (Eq. \ref{lensing_eff}),
which has a similar peak in redshift. One can also see oscillations around the
peak in $k$. These oscillations comes from the galaxy counts being negatively
correlated with a redshift range of nearby matter which again lenses the
background shear.

\xbf
\xfigure{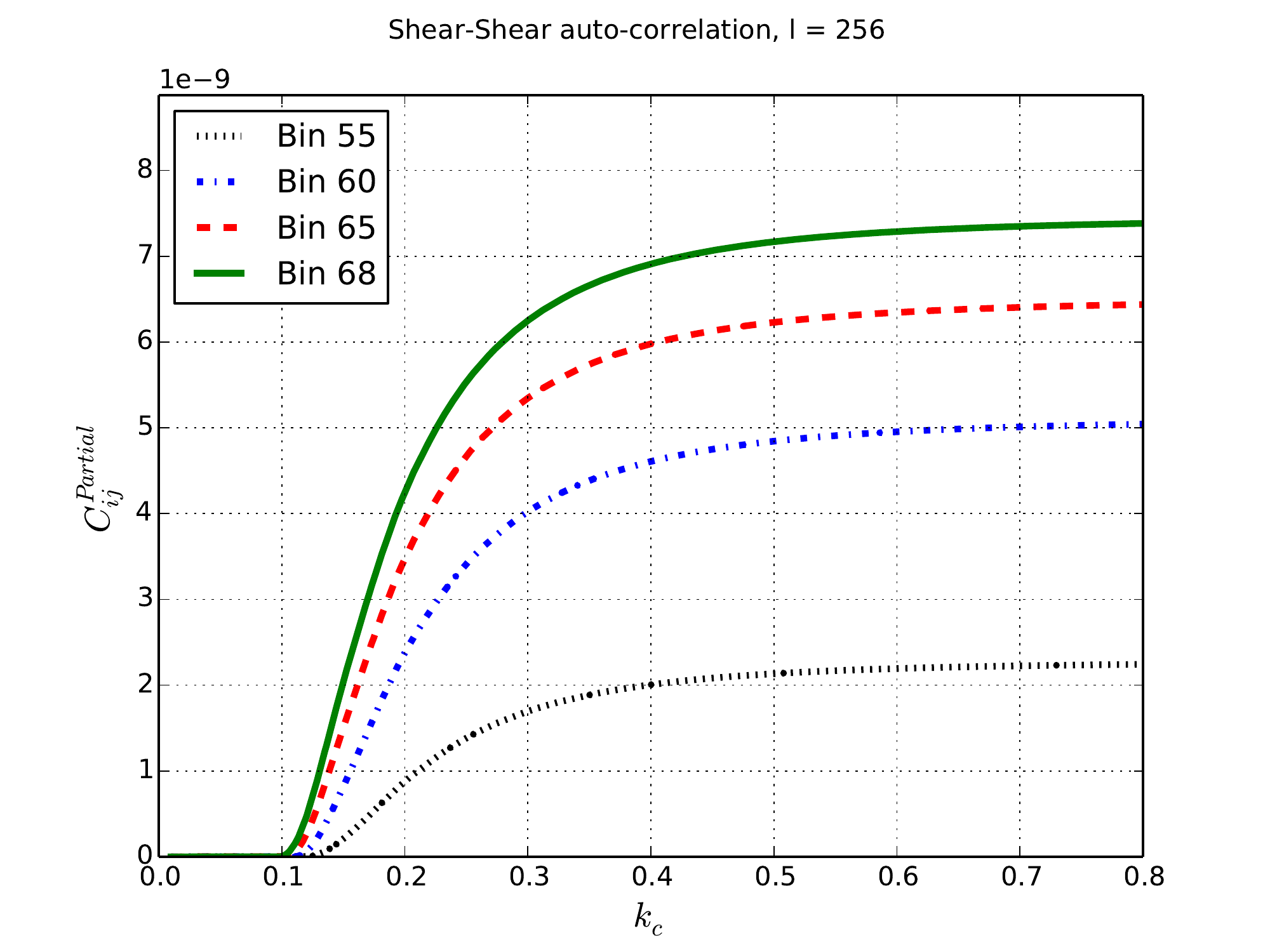}
\caption{\sumk{Shear-Shear auto-correlations} Lines correspond to the following
thin redshift bins: $0.90 < z_{55} < 0.92$, $1.00 < z_{60} < 1.02$, $1.10 <
z_{65} < 1.12$, $1.16 < z_{68} < 1.19$}
\label{delta_kk}
\xef

Fig. \ref{delta_kk} is the shear-shear auto-correlation for four redshift bins.
These redshift bins differ from previous figures (Fig. \ref{delta_auto} and
\ref{delta_cross}) since the lensing signal is stronger for higher redshift.
The signal results from a range of scales because the lensing kernel is broad
and the shear-shear correlations convolve two lensing kernels. This is in
contrast to the counts-counts and counts-shear which peaks around a specific
scale. Even if the lensing kernel is broad, correlating with a narrow
foreground redshift bin results in a contribution from a narrow range of
scales. Therefore one often describe the shear-shear as a 2D signal, while
counts-counts and counts-shear being 3D.

\xbf
\xfigure{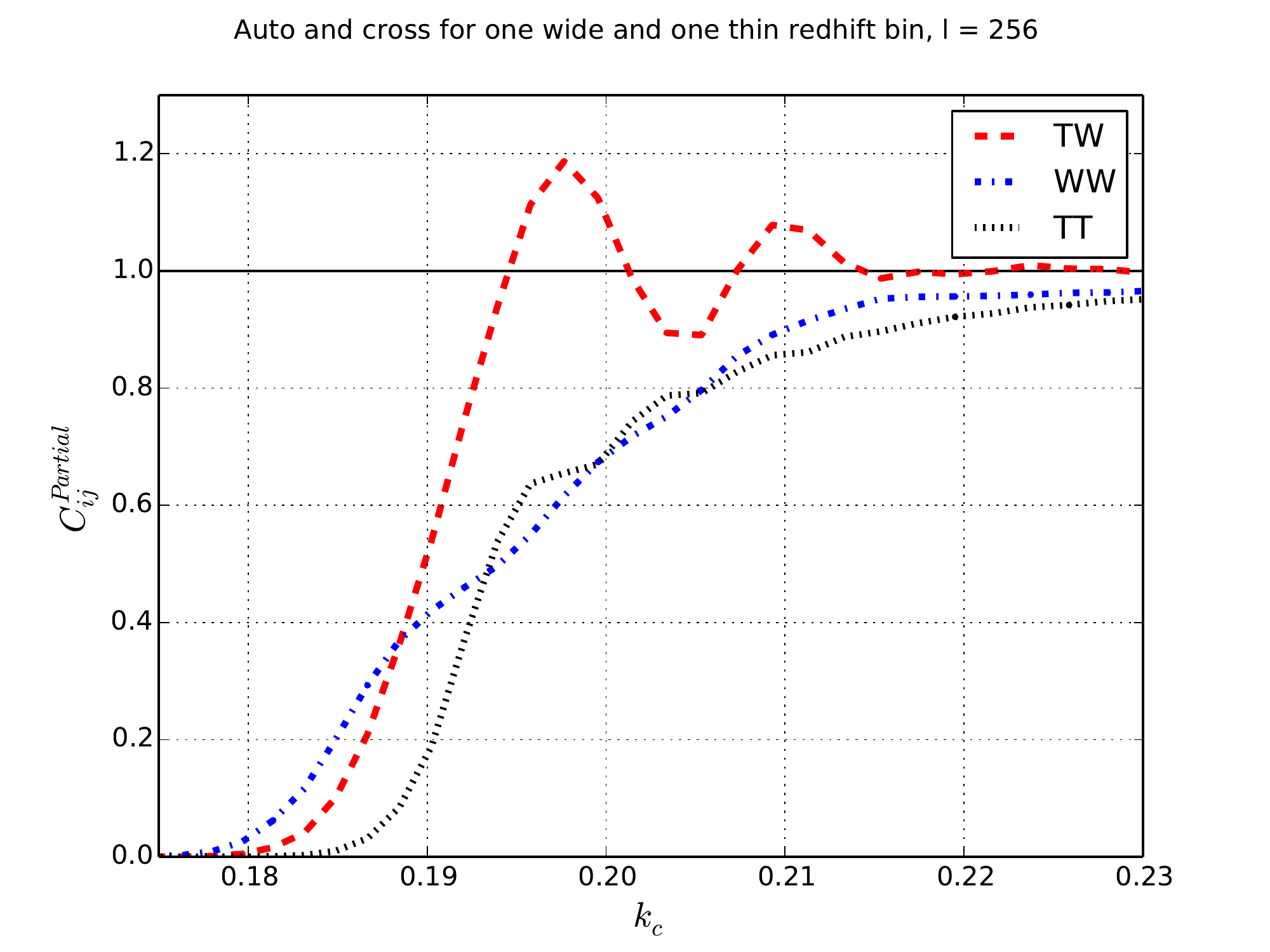}
\caption{Cumulative contribution for different scales. Auto correlations and
the cross correlations for two overlapping thin and thick redshift bins.  The
thin redshift bin (label T) $0.497 < z < 0.512$ and the thick one (label W)
$0.44 < z < 0.54$. These bins are selected to include $z = 0.5$ and the thick
bin is 6.7x wider than the thin redshift bin. To show all three lines together,
they are normalized to 1 at the asymptotic value and the scale range is
limited.}
\label{delta4}
\xef

Fig. \ref{delta4} includes the count-count auto-correlation for a thin and a
thick overlapping redshift bin, together with their cross-correlation. As
expected, the thick redshift bin has contributions from a wider range of
scales. For a decreasing redshift bin width the correlations does not approach
a delta function in scale.  Further, as seen in the decline of $\cpart$,
cross-correlations of overlapping redshift bins has both positive and negative
contributions. One can understand the effect by decomposing the
cross-correlation in the overlapping and non-overlapping regions in redshift.
For the overlapping part the cross-correlation behaves similar to the
auto-correlation in figure \ref{delta_auto}, while the non-overlapping parts
are like the cross-correlations between adjacent bins in figure
\ref{delta_cross}. The cross-correlation of overlapping bins combines these
contributions, but is closer to the auto-correlation which has the strongest
signal.

\subsection{Converting Cls to $w(\theta)$}
So far, this paper has expressed the correlations in Fourier space. Equivalently they
could be defined and calculated in $w(\theta)$, which is the 2D correlation
function in configuration space. Converting from Cls to $w(\theta)$ is a
linear combination \mycite{dodelson}

\begin{equation}
w(\theta) = \sum_l \frac{2l+1}{4 \pi} C(l) L_l(\cos{\theta})
\end{equation}

\noindent
where $L_l$ is the Legendre polynomial of order $l$. The sum is theoretically
infinite, while in practice one sum until the result converges. In this
subsection we explicitly show how to convert from Cls to $w(\theta)$ using
matrix multiplication, to convert multiple correlations in one multiplication.
This is afterwards extended to also efficiently integrate
of the angular bins in one multiplication.

Let $C_{xl}$ be the 2D correlations in Fourier space stored with 
the first (x) and second (l) index respectively being the observable and
the l-value. Converting from Cl to $w(\theta)$ is done through the matrix
multiplication

\begin{equation}
w_{xi} = \sum_l C_{xl} S_{li}
\label{theta_conv}
\end{equation}

\noindent
where $S$ is defined by

\begin{equation}
S_{li} \equiv \frac{2l+1}{4 \pi} L_l(\cos{\theta_i}).
\label{S_intangle}
\end{equation}

\noindent
and $\theta_i$ denote the mean of angular bin $i$. The angular bins has a thickness,
therefore the correct solution is

\begin{equation}
w[\theta_A, \theta_B] = \frac{1}{\theta_B - \theta_A} 
\int_{\theta_A}^{\theta_B} d\theta w(\theta)
\end{equation}

\noindent
when considering $w(\theta)$ in an angular bin $[\theta_A, \theta_B]$. When
using a linear binning in angle, this effect can often be neglected, but
results in problems at large angles if using logarithmic spacing in angle and too
few bins.  The formulas for the integration below uses the Clenshaw-Curtis
algorithm. 

\newcommand{\npointw}{\text{N}_{\text{wbin}}}
\newcommand{\nth}{\text{n}}

To integrating over angular bins we first define
\begin{align}
k_m &\equiv \cos{\left( \frac{m \pi}{\npointw} \right) } \\
x^\pm [\theta_A, \theta_B]_n &\equiv \theta_B + \frac{1}{2} (\pm k_m - 1) (\theta_B - \theta_A)
\end{align}

\noindent
where $\npointw$ is the number of integration points inside each angle bin. The
expressions $x^\pm$ are the integral points for the two contributions to the
integration over an angular bin $[\theta_A, \theta_B]$. If $A$ and $B$ denote
the edges of the angular bins and $n$ the number of angular bins, then 

\newcommand{\phiint}{\phi^{\text{Int}}} 
\newcommand{\thetaint}{\theta^{\text{Int}}}
\begin{equation}
\thetaint \equiv
\left( x^{-}[\theta_0, \theta_1]  \mid x^{+}[\theta_0, \theta_1] \mid \dots \mid
x^{-}[\theta_\nth, \theta_{\nth+1}] \mid x^{+}[\theta_\nth, \theta_{\nth+1}]
\right) \\
\label{thetaint}
\end{equation}

\noindent
gives a vector with all intermediate angles in the integration for all angular bins. The
integration weights is combined in

\begin{equation}
\phiint \equiv 
\left( \begin{array}{ccccccc}
w & w &   &   &   &   &   \\
  &   & w & w &   &   &   \\
\dots &  & \dots &  & \dots &  & \dots  \\
  &   &   &   &   & w & w
\end{array}
\right)
\label{phiint}
\end{equation}

\noindent
where the non-included entries are zero and the matrix is block diagonal.
Converting from Cl to $w(\theta)$, also integrating over angle, is done using

\begin{align}
S' &\equiv S (\phiint)^T \\
w &= C_l^T S'
\end{align}

\noindent
One could instead use a simpler algorithm leading to less complicated formulas.
A key to implementing these formulas is using a mathematical library or
language with user friendly array calculations. For example with Numpy (Python
library), Eq. \ref{phiint} can be reduced to a one-line expression.
Constructing $s'$ is compared to calculating the correlations very fast. In
addition these formulas requires estimating $S'$ only once for one specific
angular bin and maximum summation value in l. When estimating a Fisher matrix
or running a MCMC chain, these matrices only needs to be computed once.

\section{Impact of Limber, RSD and BAO}
\label{mod_sec_effects}
\label{subsec_correffects}
First we quantify the importance of using the exact integrals instead of the
Limber approximation. Then we study in detail the effects of RSD
and BAO on the auto correlations and cross-correlations. In
particular because the importance of these effects depends strongly on the
redshift bin width. Since a 2D analysis is most widely applied to
photometric surveys in broad redshift bins (i.e. $\dzbin \approx 0.1$), it is
important to investigate here the effect for the narrower  redshift bins
($\dzbin = 0.01$) that we are proposing to constrain cosmological models
including the effects of BAO and RSD in paper-II. Galaxies are bias through the
relation $b(z) = 2 + 2 (z - 0.5)$, except for the fixed bin in Fig.
\ref{comp_rsd_mean}.

The first and second subsection focuses respectively on the auto and
cross-correlations. For many of the figures, the same correlations
are presented both in Fourier space (Cl) and configuration space $w(\theta)$.
Cl plots are directly related to the formalism presented in section
\ref{mod_sec_corrtheory}, but the effect of e.g. BAO are easier to understand
using $w(\theta)$. For forecasts the Cls are preferred, since the
Gaussian, unmasked and full-sky covariance is block diagonal in l-values.  For
analyzing data one might prefer $w(\theta)$, since Fourier space correlations
can be harder to interpret. Therefore, we include $w(\theta)$ plots to make
this section more general than only supporting the forecast.

Signal-to-noise and error bars is an essential part of observational physics.
Decades of preparation and billions of dollars are spent, taking a narrow
perspective, to reduce the error bars on the measured correlations to improve
constraints on cosmological parameters. Effects entering in the correlations
are mostly interesting when being comparable large to the error bars. A naively
promising signal, might be uninteresting due to low signal-to-noise. The last
subsection study the signal-to-noise and the errors on the correlations. 

A special case is cross-correlations between partially overlapping bins, which
has traits of both auto and cross correlations. These and the non-linear
effects are studied in appendix \ref{effcompare}.

\subsection{Auto correlations}
\subsubsection{Comparing effects as a function of bin width.}

\xbf
\xfigure{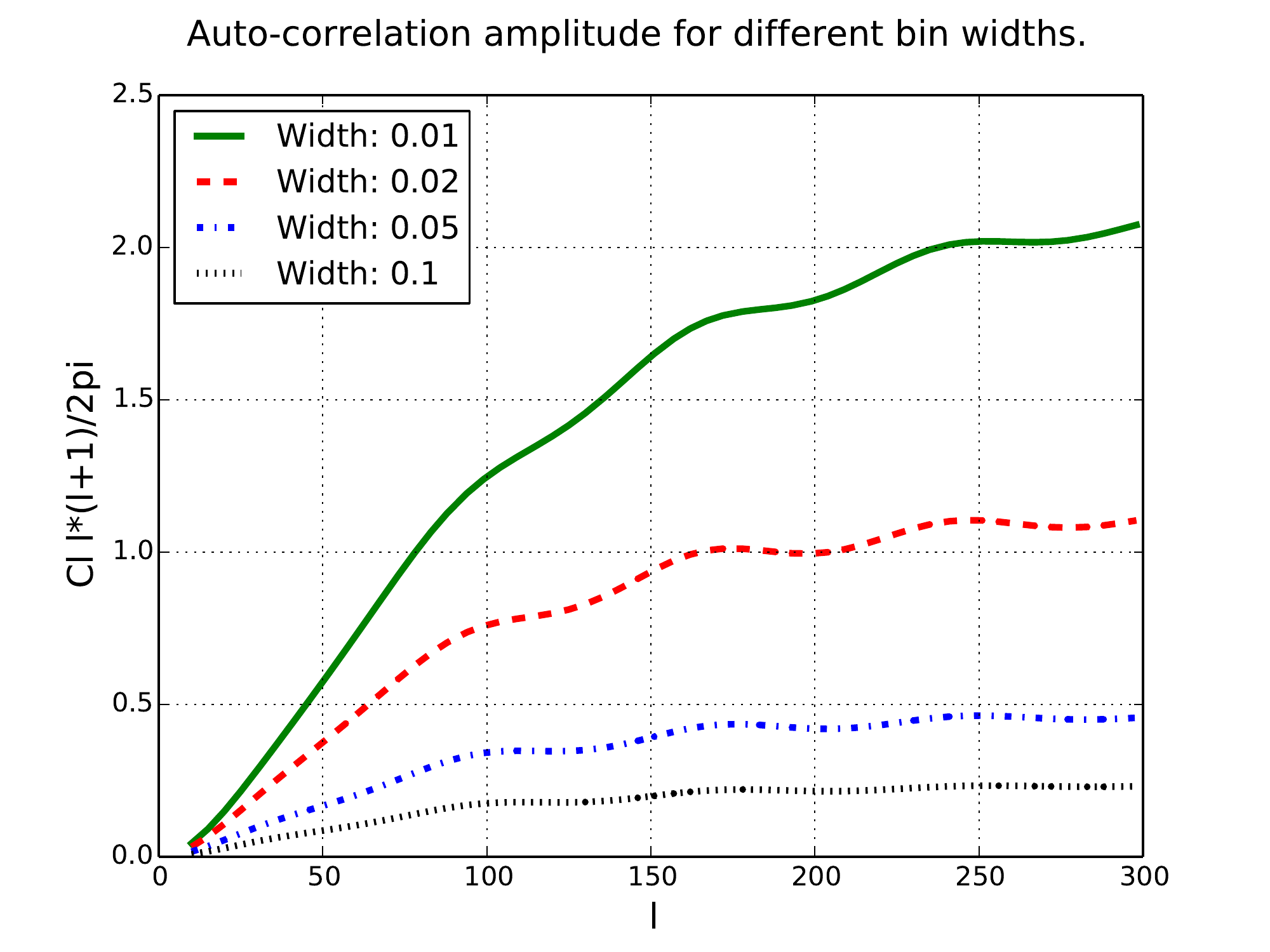}
\xfigure{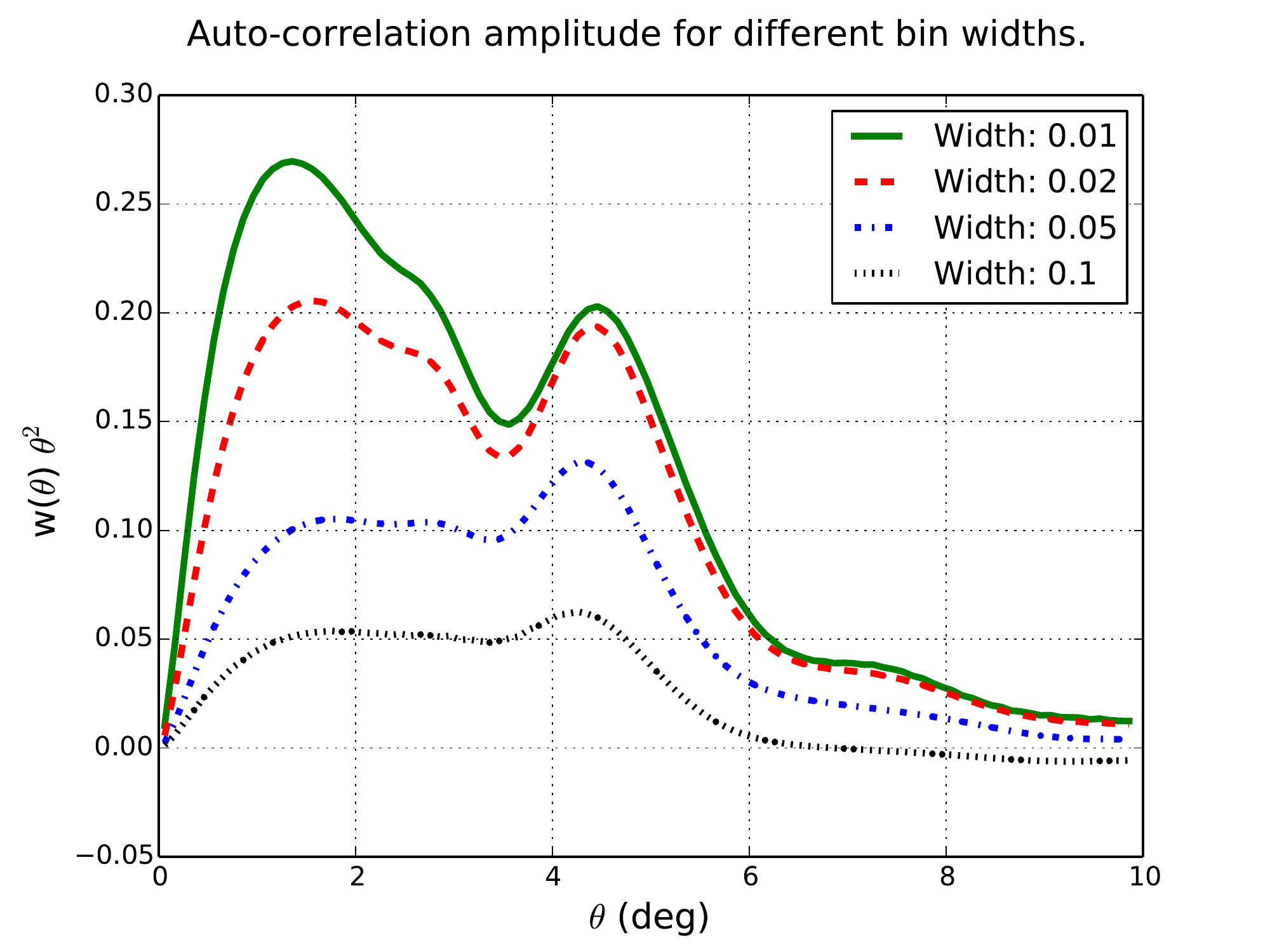}
\caption{The amplitude of auto-correlations for different redshift bin
widths. In both panels the redshift bin centred in $z = 0.5$, with the four lines
corresponding to redshift bin widths $\dzbin = 0.01, 0.02, 0.05, 0.1$. The
top and bottom panels correspond respectively to Cl and $w(\theta)$, with the
y-label indicating prefactors.}
\label{auto_comp_amp}
\xef

Fig. \ref{auto_comp_amp} show the Cl and $w(\theta)$ auto-correlations for
different bin width $\dzbin$ . We can see how the amplitude of the overall
correlations and the contrast of the BAO wiggles in $C_l$ or BAO peak in
$w(\theta)$ decrease as we increase $\dzbin$.

\xbf
\xfigure{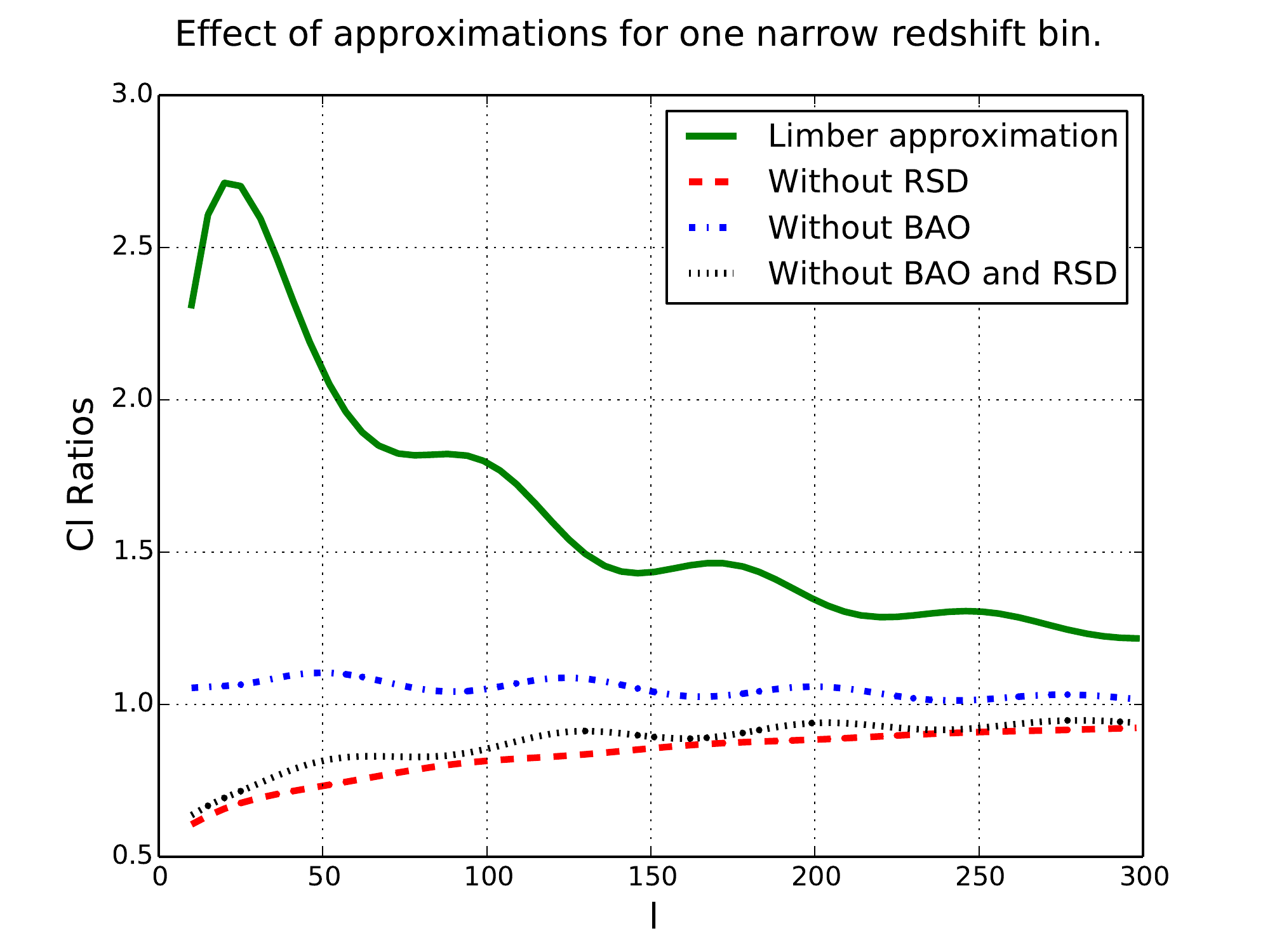}
\xfigure{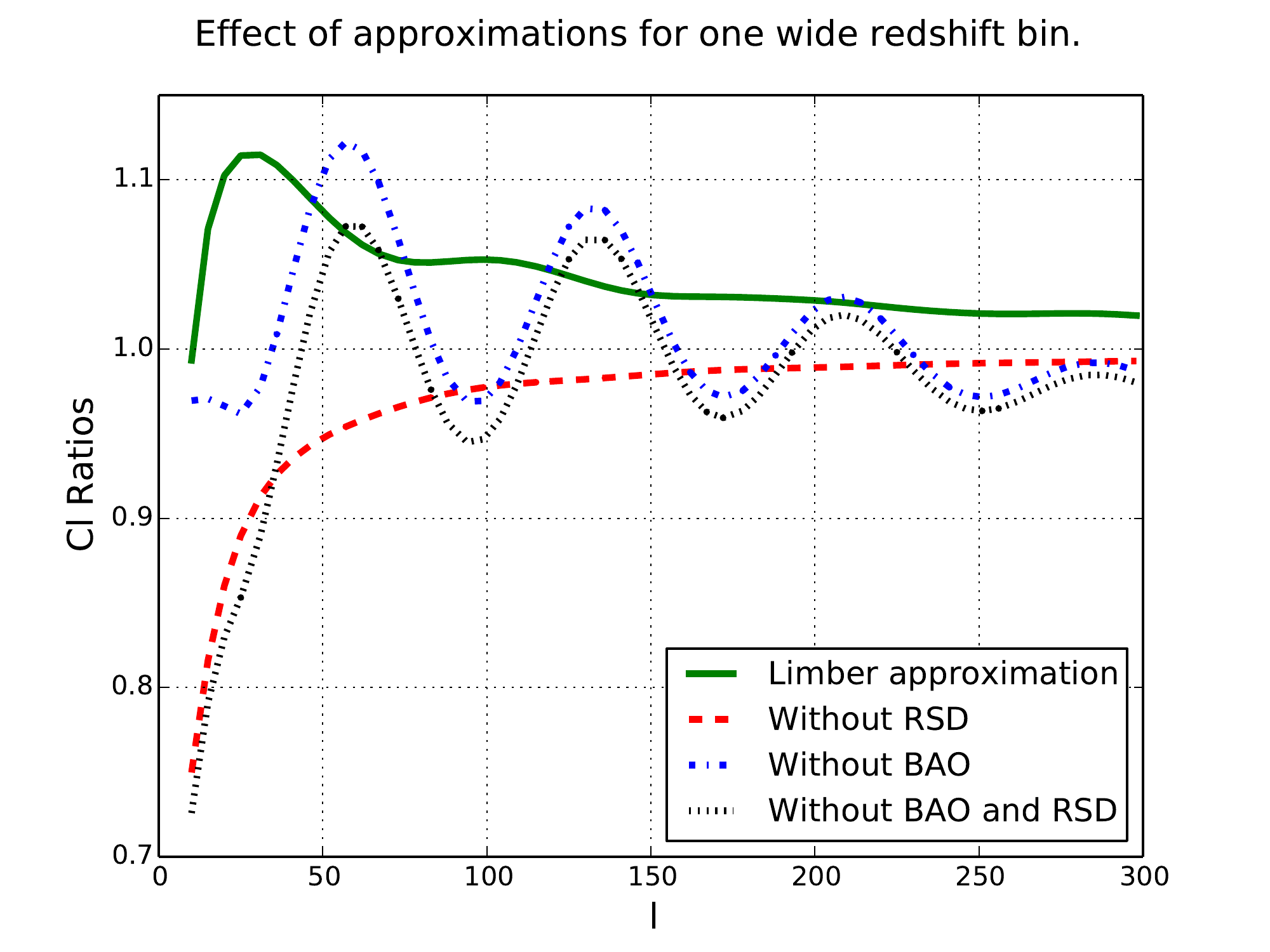}
\caption{Comparison of the
effect of Limber approximations, redshift space distortions and BAO 
The redshift bin is centred around $z = 0.5$ and width 
$\dzbin = 0.01$ (top) or $\dzbin = 0.1$ (bottom). Ratios plotted with respect
to the exact fiducial calculation (including RSD and BAO), except for the
Limber case. Since the Limber approximation is in real space, the ratio is
with respect to the real space correlations (i.e. without RSD).}
\label{all_approx}
\xef

The continues line in Fig. \ref{all_approx} shows the ratio of the Limber
approximation to the exact calculation (both without RSD) for $C_l$, with the
top and bottom panel respectively using a narrow ($\dzbin = 0.01$) and wide
($\dzbin = 0.1$) redshift bin. All the correlations are centred around $z =
0.5$. The other lines show the ratio after removing RSD (dashed), BAO
(dot-dashed) or both (dotted). For RSD the ratio is below one, meaning the
redshift space distortions contribute positively to the correlations. Including
the correlations for two bin widths in Fig.\ref{all_approx}, show how the
relative size of different effects depend on the redshift bin width.

When comparing the effects in a wide and narrow redshift bin, the largest
effect comes from the inaccuracy of the Limber approximation in narrow bins.
The Limber approximation is known to break down for thin bins. This can be seen
in Eq. \ref{c_gg}, where the division on the redshift bin width would result in
infinite correlations for infinitely thin bins. At $l = 150$ the Limber
approximations account for $3\%$ for bin width 0.01, which can be tolerated
depending on the survey accuracy, while for a narrow redshift bin the effect is
close to $50\%$. Which for all purposes is too large.

\subsubsection{Limber approximation}
\xbf
\xfigure{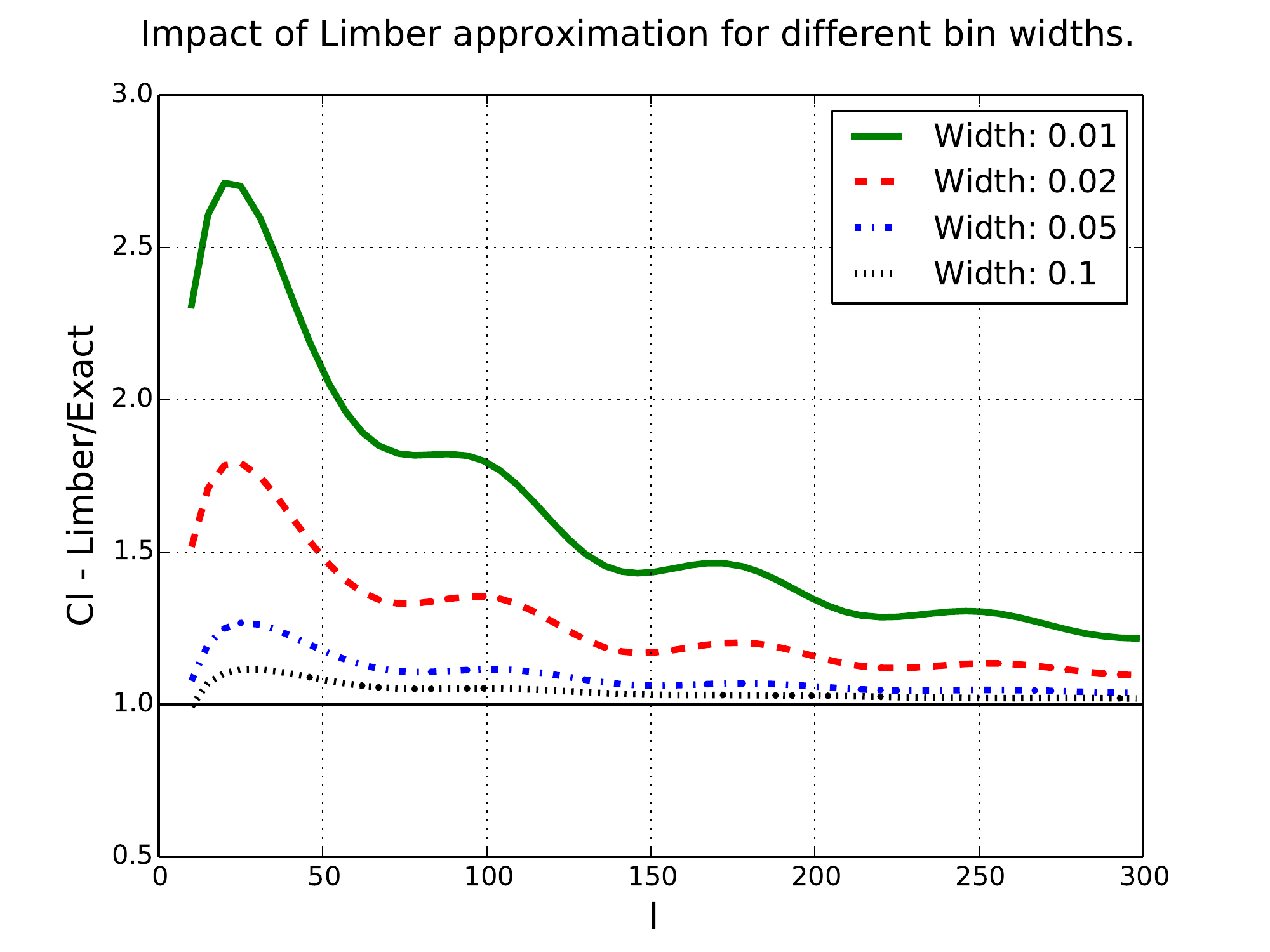}
\xfigure{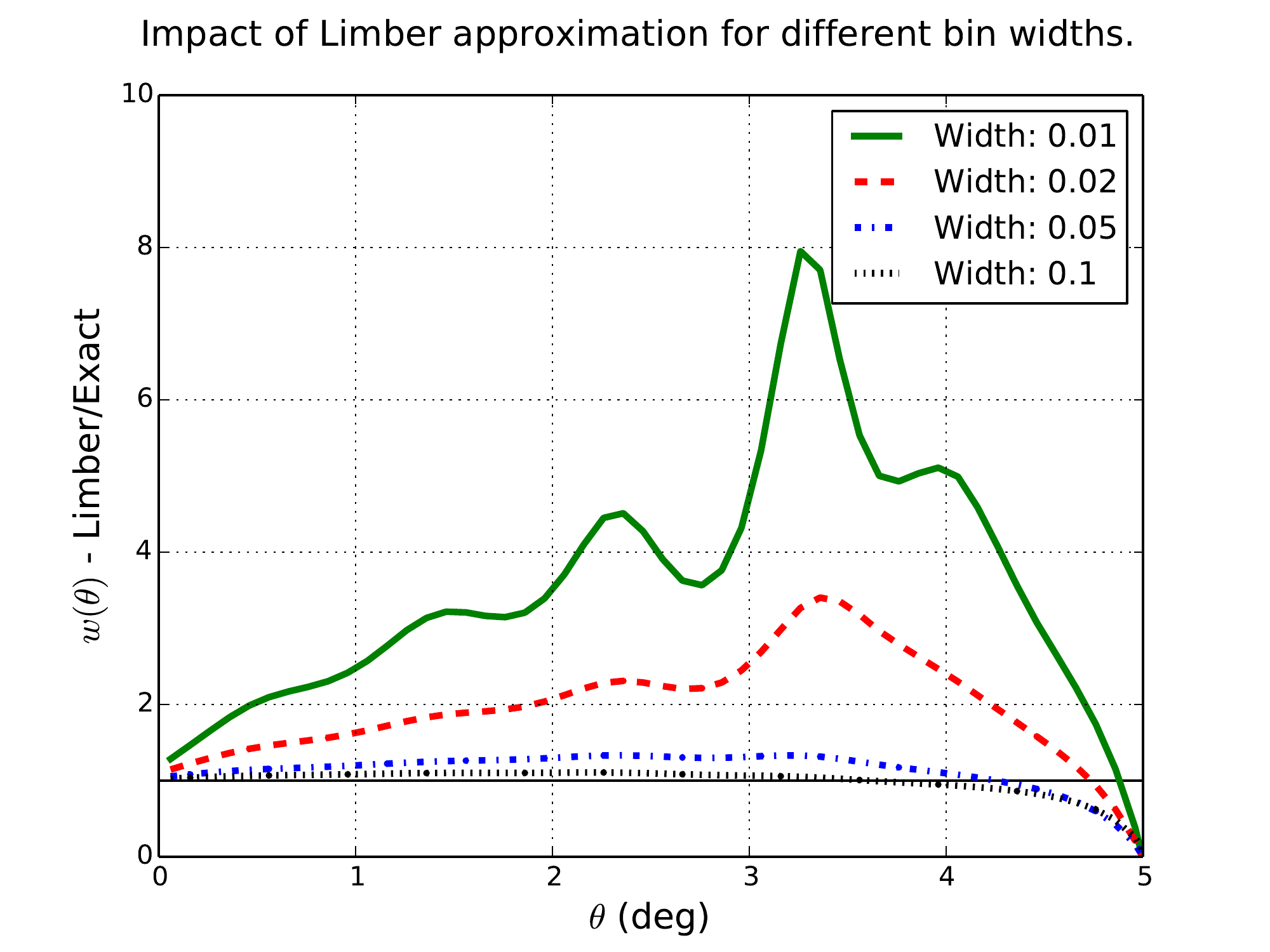}
\caption{The lines corresponds to bin width 0.01, 0.02, 0.05 and 0.1, centred
in $z = 0.5$. Both the exact calculations and the Limber approximation is in
real space.}
\label{limber_impact}
\xef

The Fig. \ref{limber_impact} includes in real space the ratio of the Limber
approximation to the exact calculations for both the
Cls and $w(\theta)$. From the ratios of the correlations in Fourier space,
there is a huge difference is using a wide or a narrow redshift bin. A goal of
the forecast in paper-II is to include radial information in the spectroscopic
sample. Next subsection show how cross-correlation between adjacent redshift
bins which can be used for measuring radial correlation and this requires bins
$\dzbin = 0.01 (1+z)$ for our choice of kmax. From Fig. 8, for our purpose the
Limber approximations is unusable even for the auto-correlations.

\subsubsection{Redshift space distortions}
\xbf
\xfigure{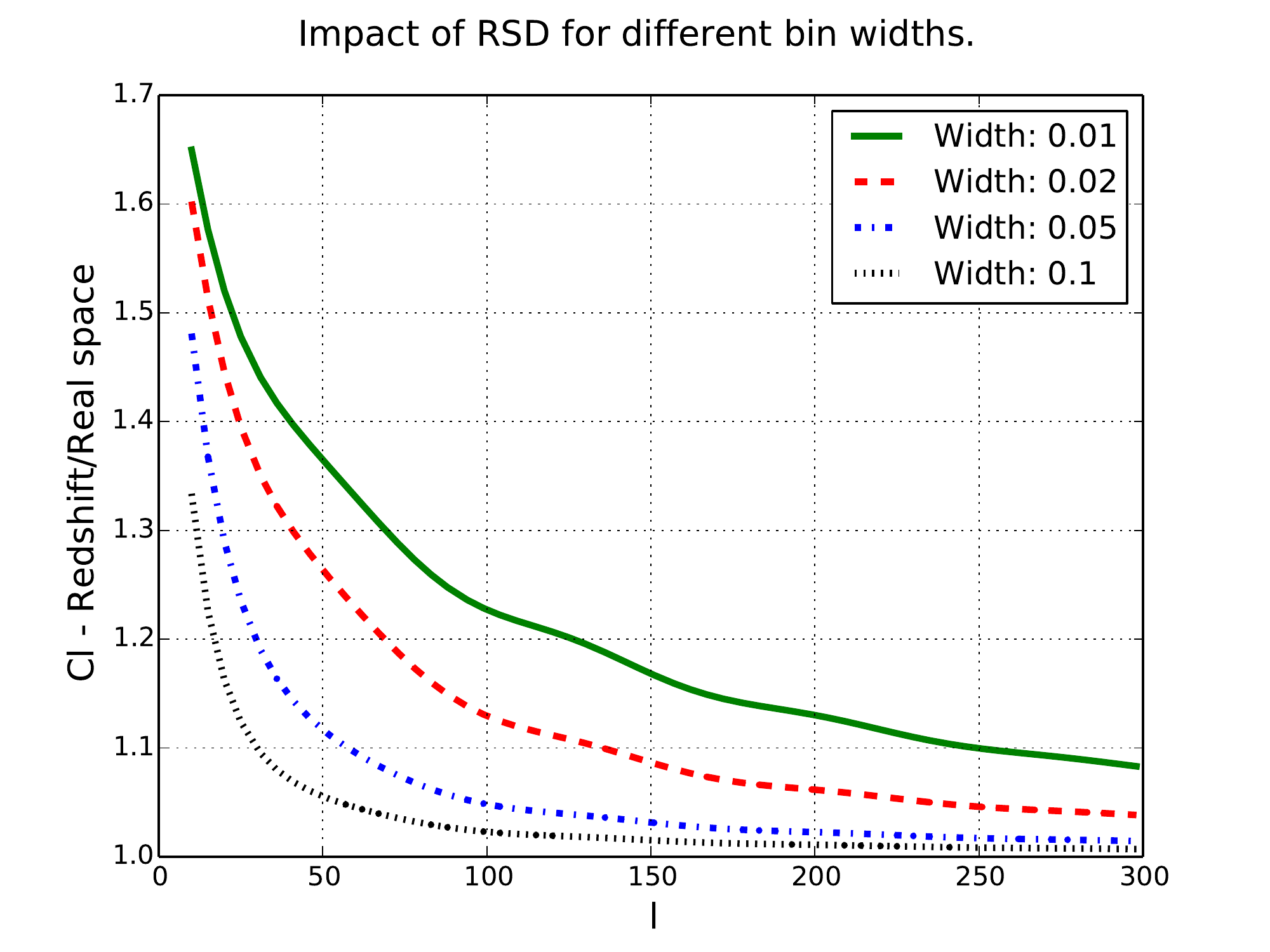}
\xfigure{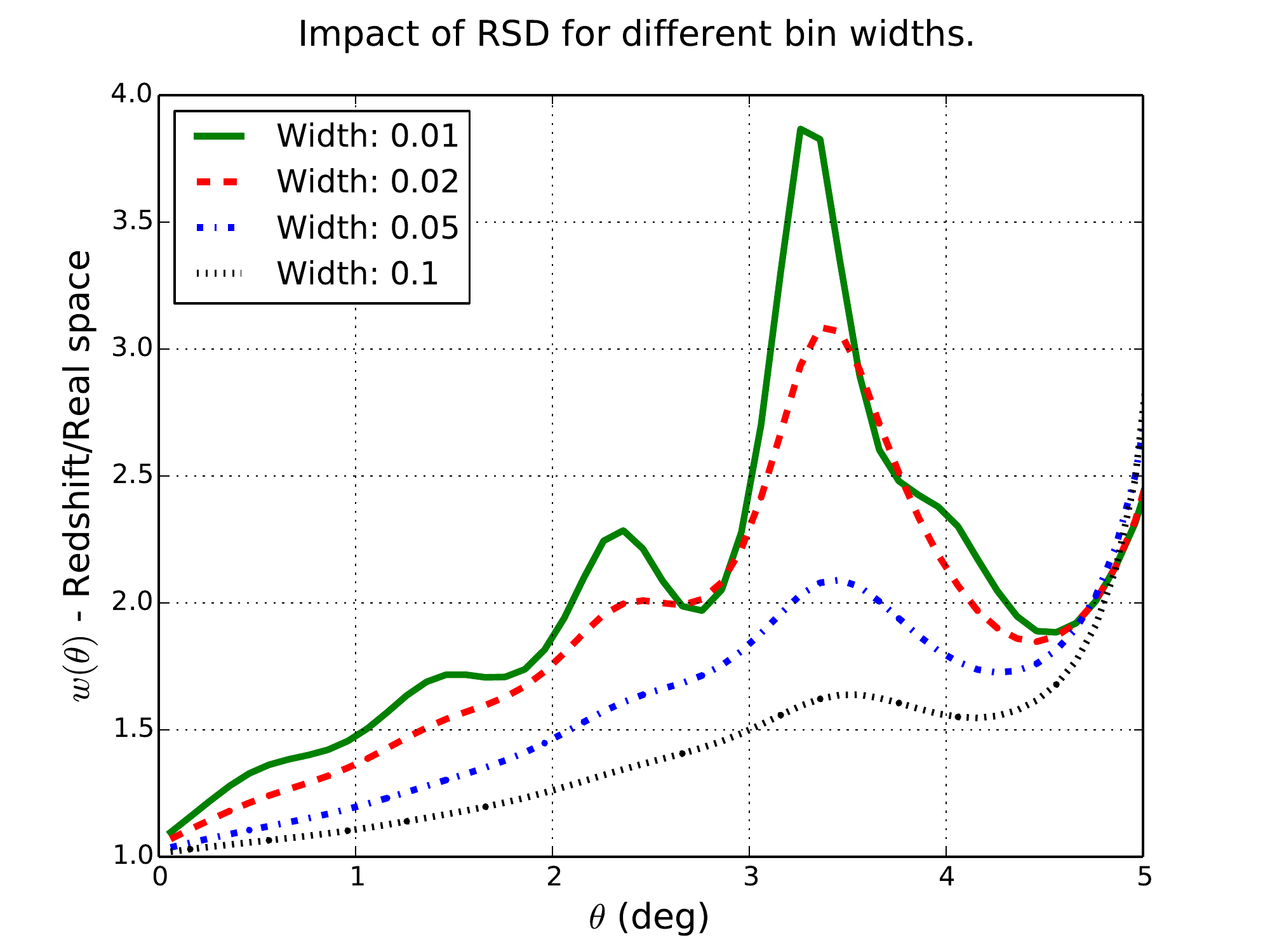}
\caption{Redshift/real-space ratio in Fourier (top) and configuration space
(bottom). The lines corresponds to redshift bin width $\dzbin = 0.01,0.01,
0.02, 0.05, 0.1$, centred around $z = 0.5$. In this figure ratios about unity
means RSD increase the amplitude.}
\label{impact_rsd}
\xef

Redshift space distortions (RSD) do not change the angular positions of
galaxies, but they do change their angular correlation when selected in
observed redshift bins as large-scale motions move structures across the
boundaries in a spatially coherent way (see section \ref{corrtheory} and
references therein). Fig. \ref{impact_rsd} study the redshift to real-space
ratio when varying the redshift bin width. When analyzing galaxy clustering in
photometric surveys, the standard approach use 2D-correlations in thick
redshift bins. A broad band photo-z scatter RMS is around $0.01(1+z)$ to
$0.05(1+z)$ depending e.g. on the magnitude, filters, exposure times and
calibration sample. Because of the photo-z scatter, analyzing the data in
narrower bins would give little improvements. With narrow bins one would need
to model photo-z transitions between redshift bins and their uncertainty
\mycite{gazta}. For a spectroscopic survey one can analyze the data in narrow
bins.

The effect of RSD result in a significant amplitude increase as shown in Fig.
\ref{impact_rsd}. A lower bin width results in a higher amplitude.  For low
values, $10 < l <30$, the effect for the broad bin ($\dzbin = 0.1$) is
$10-30\%$, while the effect is $45-50\%$ for the narrow bin ($\dzbin = 0.01$).
More importantly, the scales affected depend on the bin width. For the thick
redshift (top panel), redshift space distortions only contribute significantly
for $l < 50$, while for $\dzbin = 0.01$ the effect is still ~10\% at $l=300$.
Physically, the redshift space distortions in the 2D-correlations is a boundary
effect. When decreasing the bin width, the bulk decrease and the RSD boundary
effect becomes more important. This is why Fig.\ref{impact_rsd} looks similar
to Fig.\ref{limber_impact}, as the Limber approximation also can be cast as a
boundary effect in real space.

In configuration space (Fig. \ref{impact_rsd}, bottom panel), the redshift
to real space ratio peaks around 3.5 degrees, shifting only slightly depending
on the bin width. For the thinnest bin ($\dzbin = 0.01$), the RSD/real space
ratio nearly doubles compared to 3 degrees. Increasing the bin width cause the
peak to flatten. The higher contribution around 3.5 degrees is caused by BAO
in redshift space. From Eq.\ref{psi_r}, the RSD contribution in Fourier space
consist of three contributions proportional to spherical Bessel functions of
different orders. A small l-value shift give an angular shift in $w(\theta)$.
This result in a BAO contribution in $w(\theta)$ which is shifted in angle. We
label this contribution the Ghost BAO peak and study its cosmology dependence
in subsection \ref{ghostbao}.

\subsubsection{Baryon acoustic oscillations (BAO)}
\xbf
\xfigure{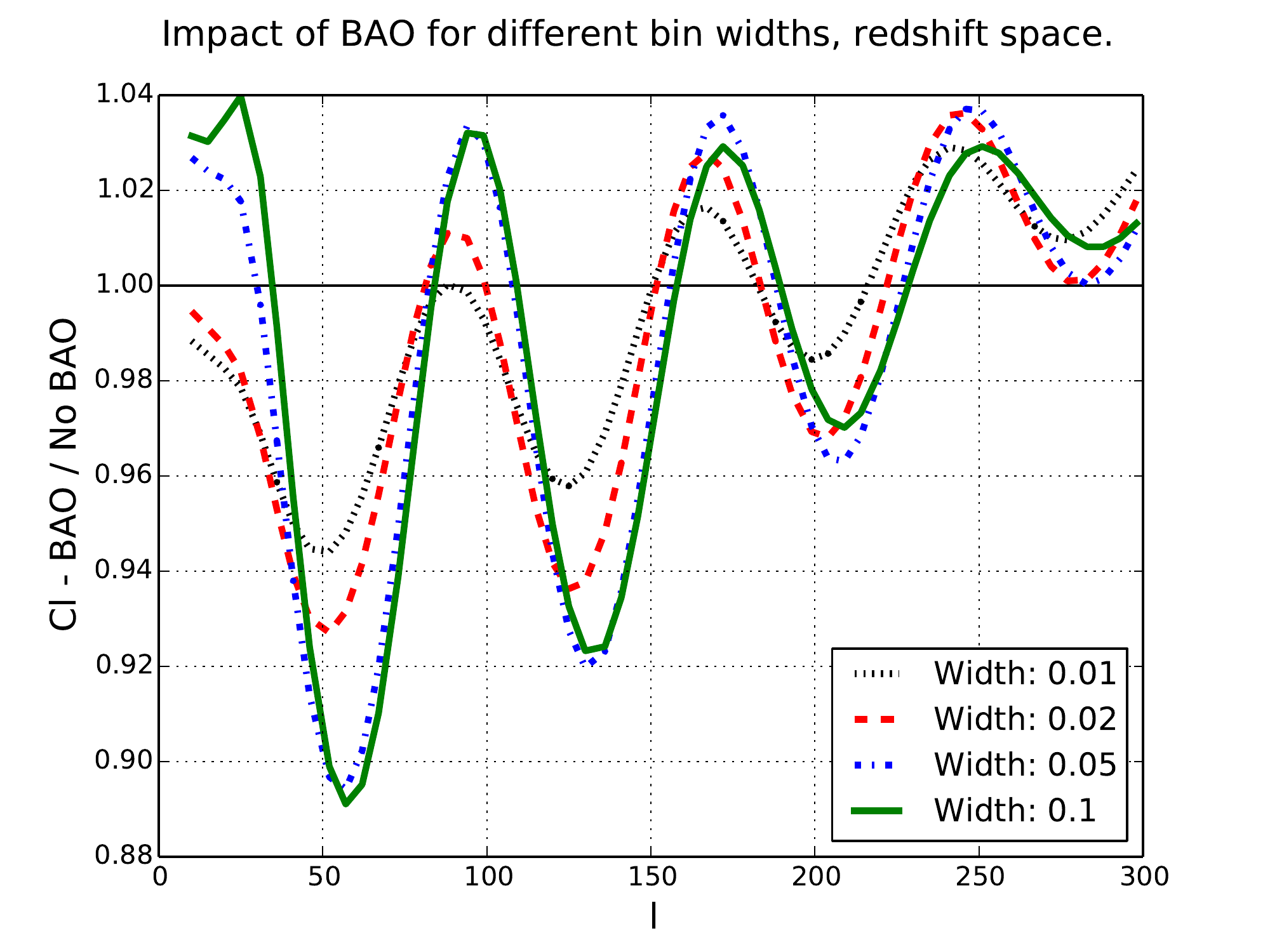}
\xfigure{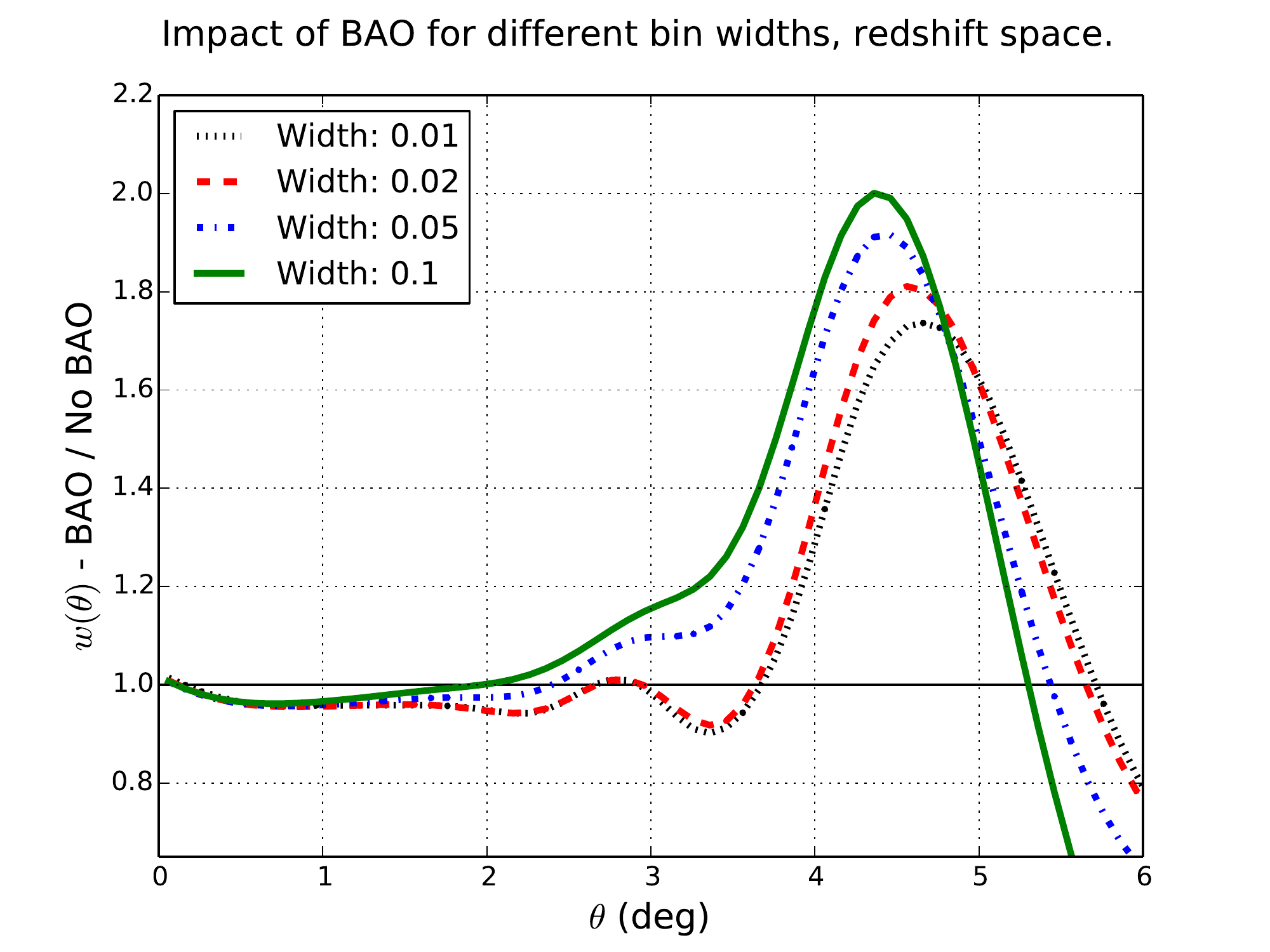}
\xfigure{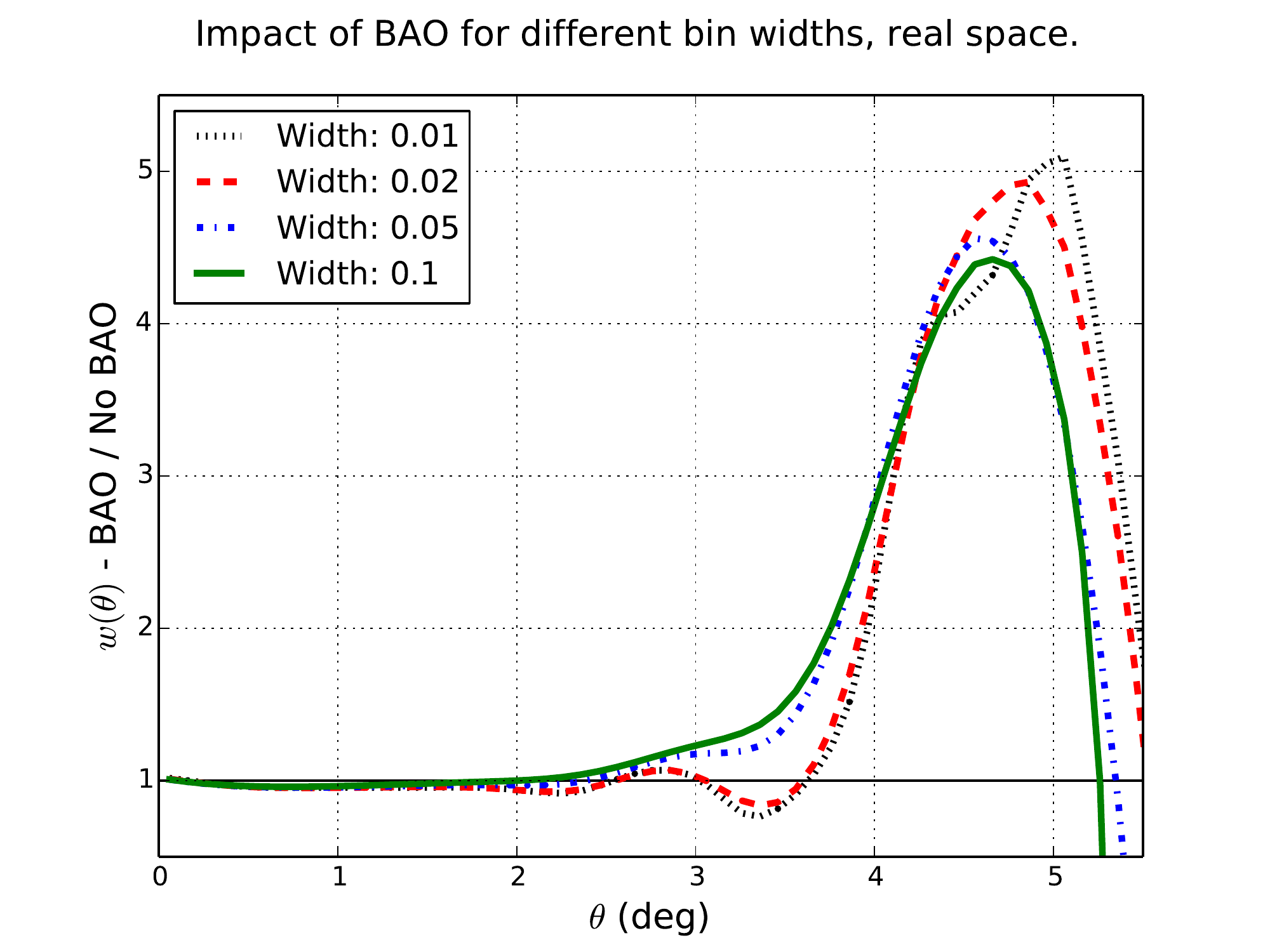}
\caption{The ratio of correlations including BAO wiggles to a model removing
the BAO peak in the EH power spectrum. The first two panels show the ratios in
redshift space for Cl and $w(\theta)$. To discuss the effect of redshift space
distortions on the BAO peak, the third panel shows the angular correlation in
real space. All correlations uses a mean redshift of $z=0.5$ and the bin widths
$\dzbin = 0.01, 0.02, 0.05, 0.1$.}

\label{impact_bao}
\xef
The three panels in Fig. \ref{impact_bao} are included to study the impact
of BAO. The first panel show the correlations in Fourier space.
One can see how including BAO or using a no-wiggles models leads to oscillation
in the ratios. The middle panel shows how the $w(\theta)$ correlation peak
around 4.5 degrees for $z=0.5$, shifting towards lower angles for higher
redshifts. One see the effect of BAO in redshift space increases when using
wider redshift bin. This is counterintuitive, since integrating over the
redshift bin results in a convolution in angle. This effect can be understood
from redshift space distortions. The bottom panel displays the same ratios in
real space, where thicker redshift bins leads to a slight decrease in the BAO
signal and a shift in the angle due to projection scales. Including redshift
space distortions increases the correlation amplitude, but lower the BAO ratio.
The last effect enters since the RSD effect is stronger at lower angles than
where BAO peak. For thin bins contribution of redshift space distortions has a
narrower peak, which explains why thick bins see a higher contribution on BAO
in redshift space.

\subsubsection{The Ghost BAO peak}
\label{ghostbao}
\xbf
\xfigure{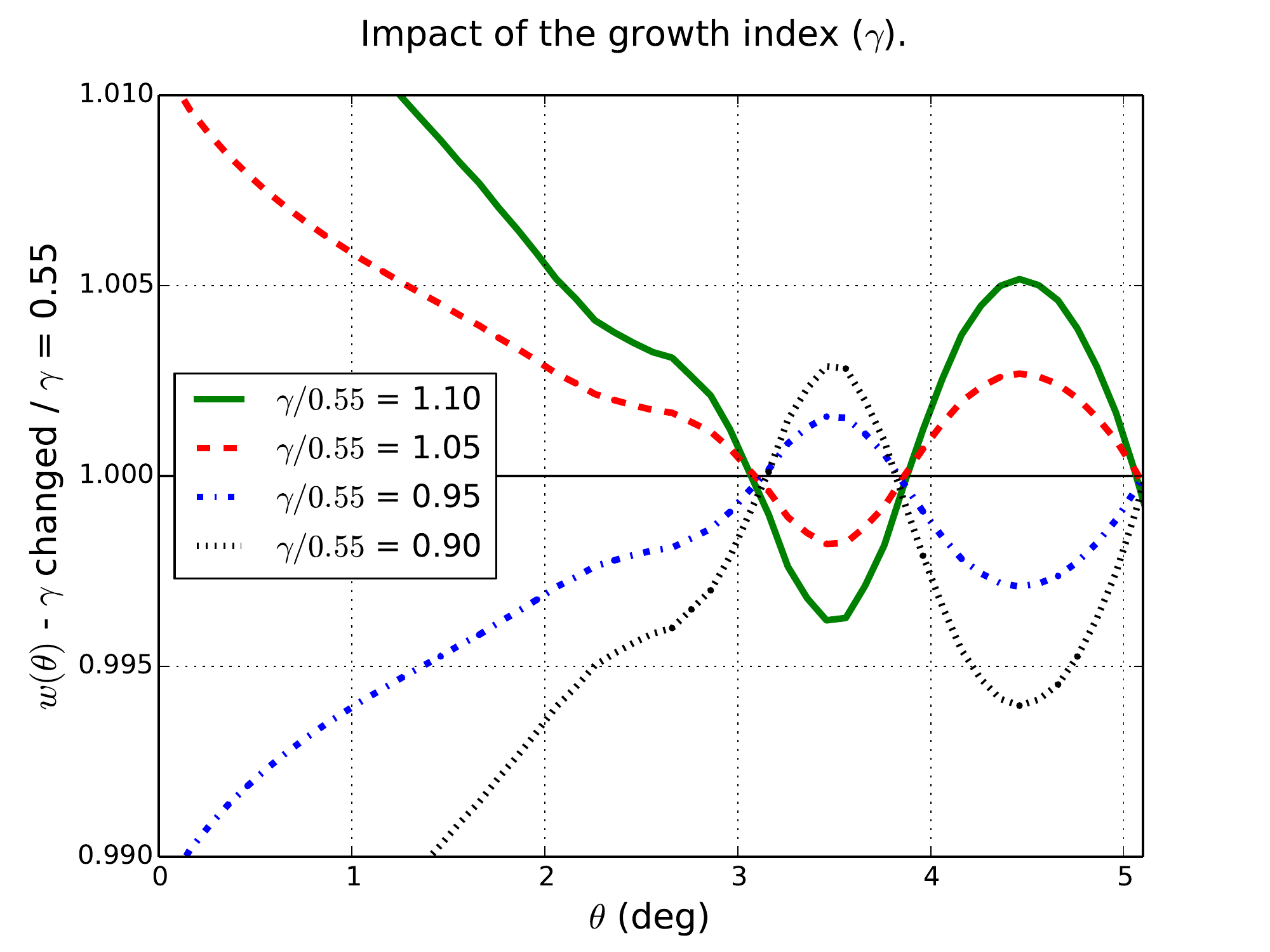}
\caption{The $w(\theta)$ auto-correlation ratio between a changed growth rate parameter
and GR cosmology ($\gamma = 0.55$). The auto-correlation use a thin bin
($\dzbin = 0.01$), centred in $z = 0.5$. Values of $\gamma$ in the ratios
are $\pm 5\%, \pm10\%$ of the fiducial value (see legend).}
\label{impact_gamma}
\xef

Fig. \ref{impact_gamma} illustrates the cosmological dependence of the
Ghost BAO peak. As shown in Fig. \ref{impact_rsd}, the RSD contribution peaks
at different angular values when measured in narrow redshift bins. In previous
figures we assumed the growth rate $\gamma = 0.55$ of general relativity.
Modified gravity models can change the growth rate \mycite{bellido_growth}.
Increasing $\gamma$ leads to a higher amplitude of the clustering, while
lowering the effect of redshift space distortions. This follows from
$\frac{\partial D}{\partial f} < 0$ and $f \equiv \Omega_m(z)^\gamma$. In Fig.
11, for higher $\gamma$ the correlations increase for all angles, except around
3.5 degrees where the RSD contribution peaks. While being interesting, note
that the amplitude difference is low and the effect might be difficulty to
measure.

\subsection{Cross correlations between redshift bins}
\label{subsec_cross}
The auto-correlations are the correlation of an observable with itself.
Examples are the shear-shear or counts-counts correlation of overdensities in
the same redshift bin. A cross-correlation can either come from correlating
different quantities as galaxy populations or using different redshift bins for
the same quantity. Correlating foreground galaxies with the background
shear \mycite{jointgalshear} or correlating two populations of galaxies
\mycite{mcdonseljak,asorey2} are examples of cross-correlations. In the
previous subsection we studied the auto-correlation of galaxy counts in narrow
redshift bins. This subsection focus on the 2D cross-correlations between galaxy
counts in nearby redshift bins.

\xbf
\xfigure{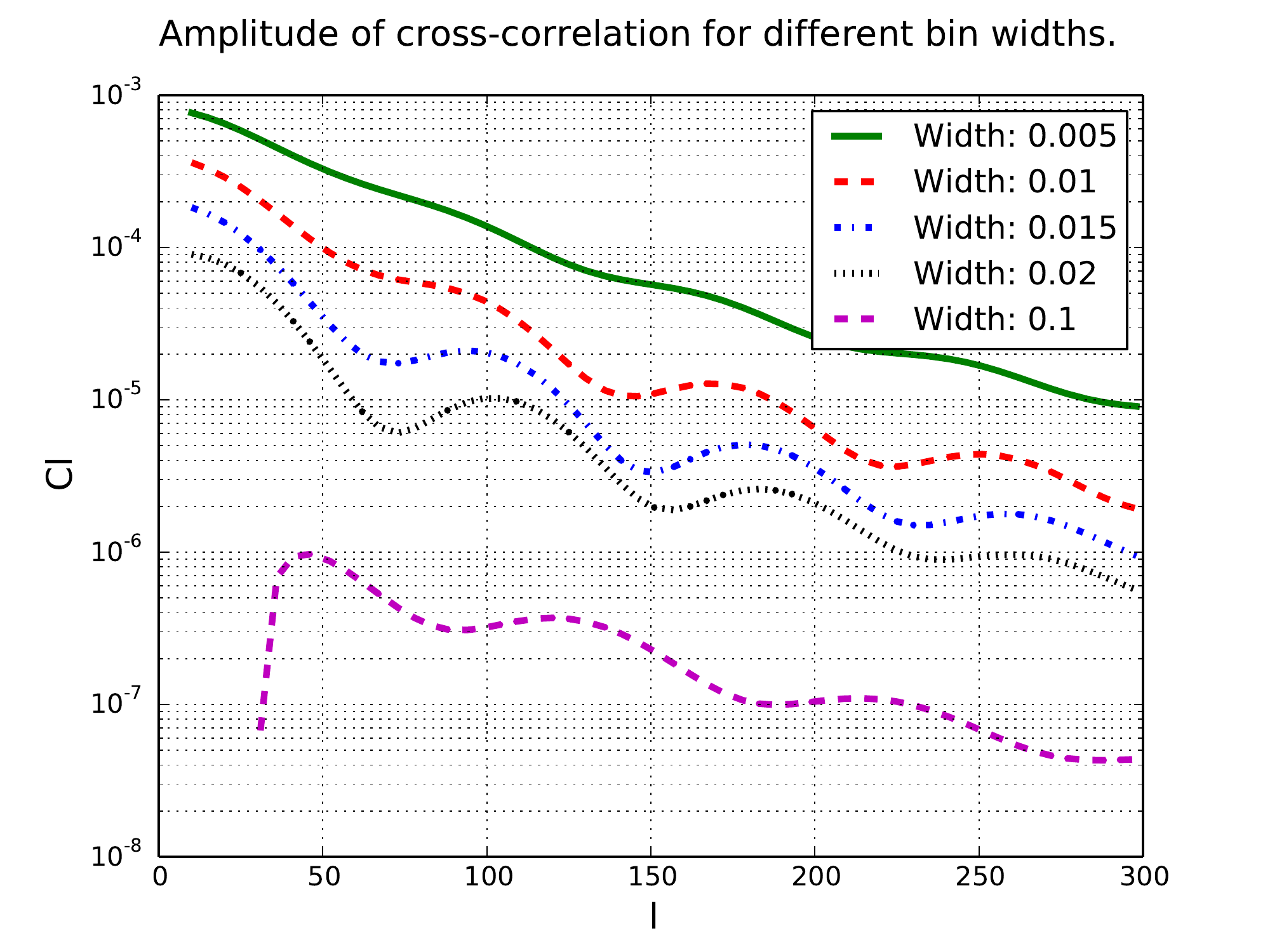}
\xfigure{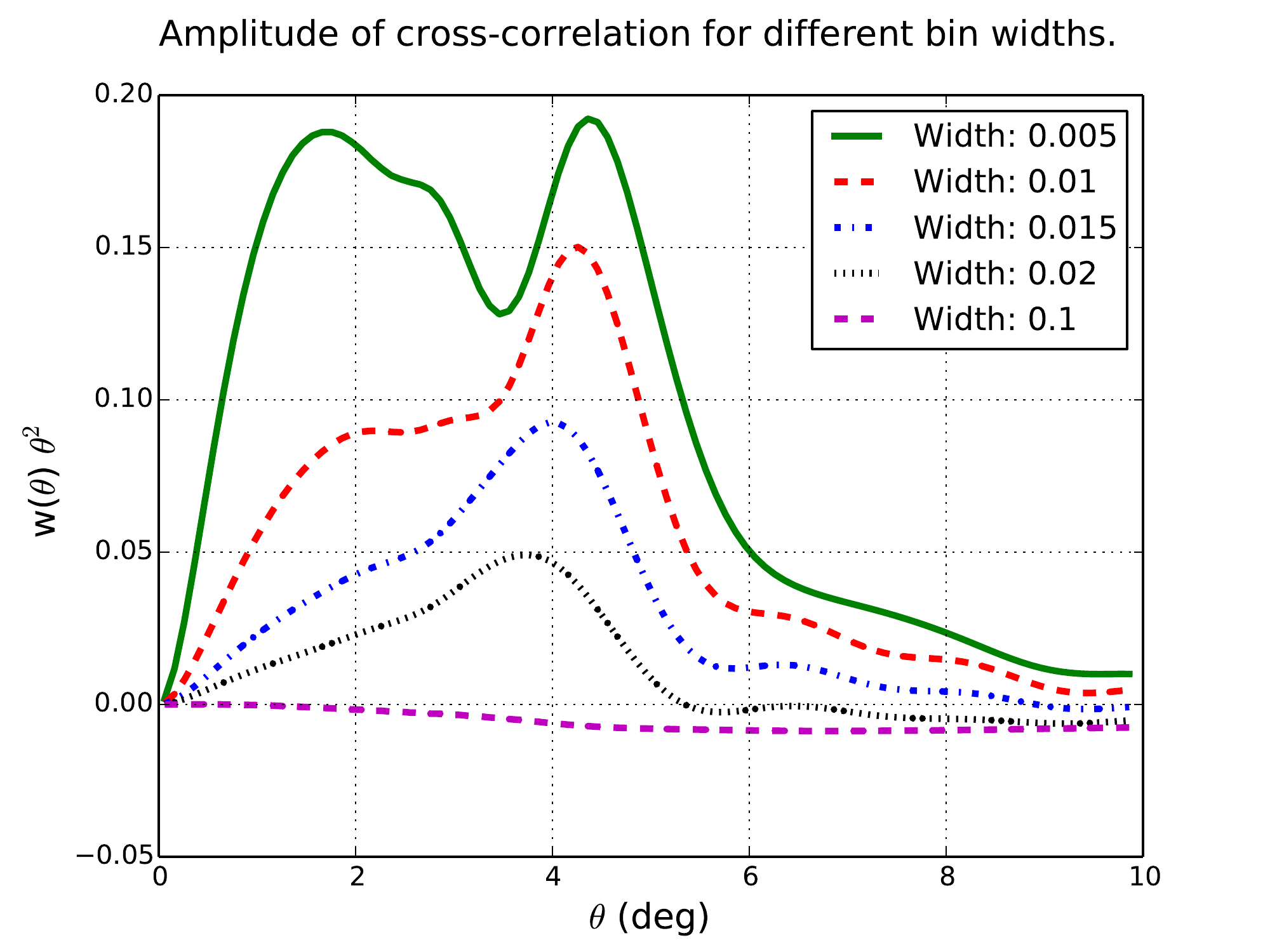}
\caption{Amplitude of cross-correlations between adjacent redshift bins with
equal redshift width. The first redshift bin starts at $z=0.5$ and the five
lines correspond to biw width $\dzbin/(1+z) =0.005, 0.01, 0.015, 0.02, 0.1$.
The top and bottom panels respectively respectively show the Cls and
$w(\theta)$ correlations. Here $\dzbin = 0.01$ corresponds to the default
forecast binning of a spectroscopic sample, while $\dzbin/(1+z) = 0.005$ is
included to motivate the potential gain by using even thinner redshift bins.}
\label{cross_amp}
\xef

\subsubsection{Amplitude of correlations and comparing effects}
The intrinsic correlations between two redshift bins is weaker than the
auto-correlations and depends strongly on the separation between the redshift
bins. Note that the redshift bin cross-correlation presented here are due to
correlation of the matter distribution and not from bins overlapping in photo-z
space. This distinction is important if studying photo-z surveys in wide
redshift bins.  The observed cross-correlations $\tilde{C}_{ij}$ including
photo-z effects are approximately \mycite{gazta}

\begin{equation}
\tilde{C}_{ij} \simeq r_{ij} C_{jj} + r_{ji} C_{ii} + r_{ii} r_{jj} C_{ij}
\end{equation}

\noindent
where $r_{ij}$ is the fraction of galaxies actually in bin $j$, but observed in
bin $i$ due to photo-z inaccuracies. If the two first terms dominate, then the
cross-correlations are dominated by the tail of the redshift distribution and
not the intrinsic cross-correlation.

Fig. \ref{cross_amp} shows the cross-correlations where the first bin starts at
$z=0.5$ and the adjacent of two bins. When increasing the bin
width from $0.005$ to $0.02$, which is a factor of 4, the amplitude change with
an order of magnitude in Fourier space. The rapid decline with increasing bin
widths is also seen in the angular correlations. In addition one see a trend
where the small scales are affected more than the BAO scale. The amplitude
double at 2 degrees using a width of 0.005 instead of 0.01, while the change is
30\% at the BAO peak.

\xbf
\xfigure{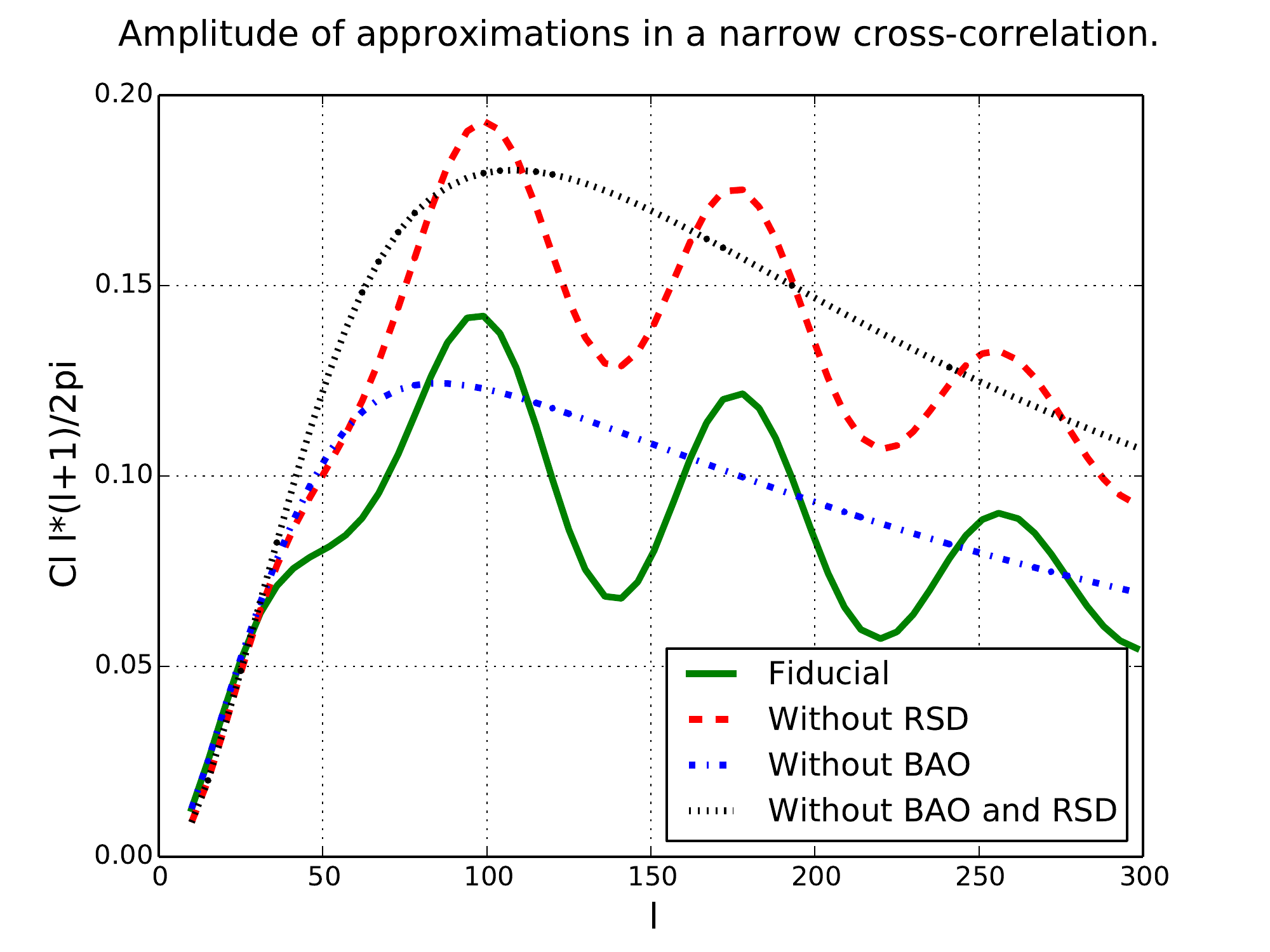}
\xfigure{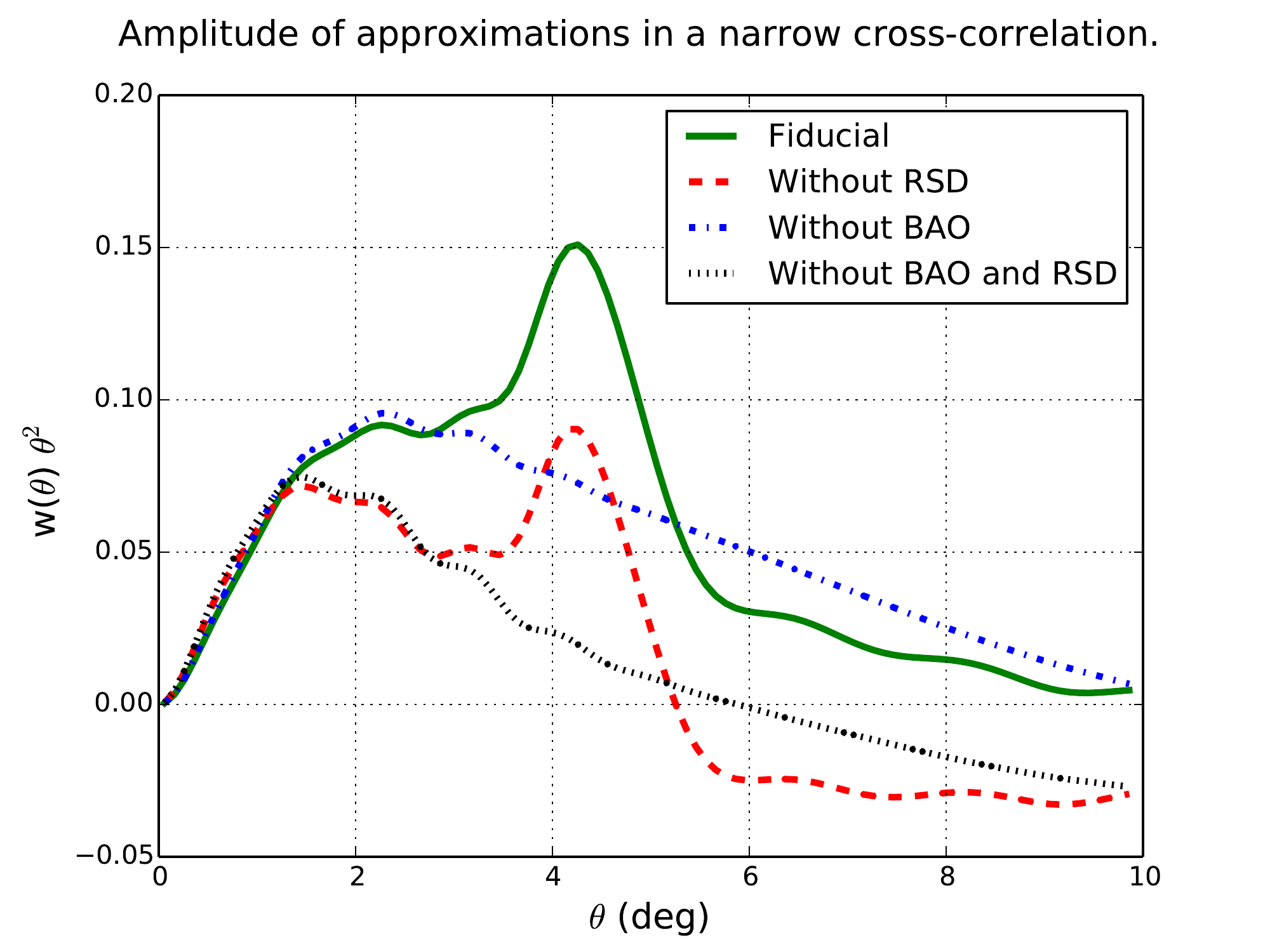}
\caption{Cross-correlations between adjacent redshift bins. Both bins are
$\dzbin = 0.01 (1+z)$ wide and the first bin starts at $z=0.5$. The four lines
include different effects, with the fiducial line including both RSD and BAO.
The top and bottom panel respectively show Cls and $w(\theta)$.}
\label{cross_effects}
\xef

\xbf
\xfigure{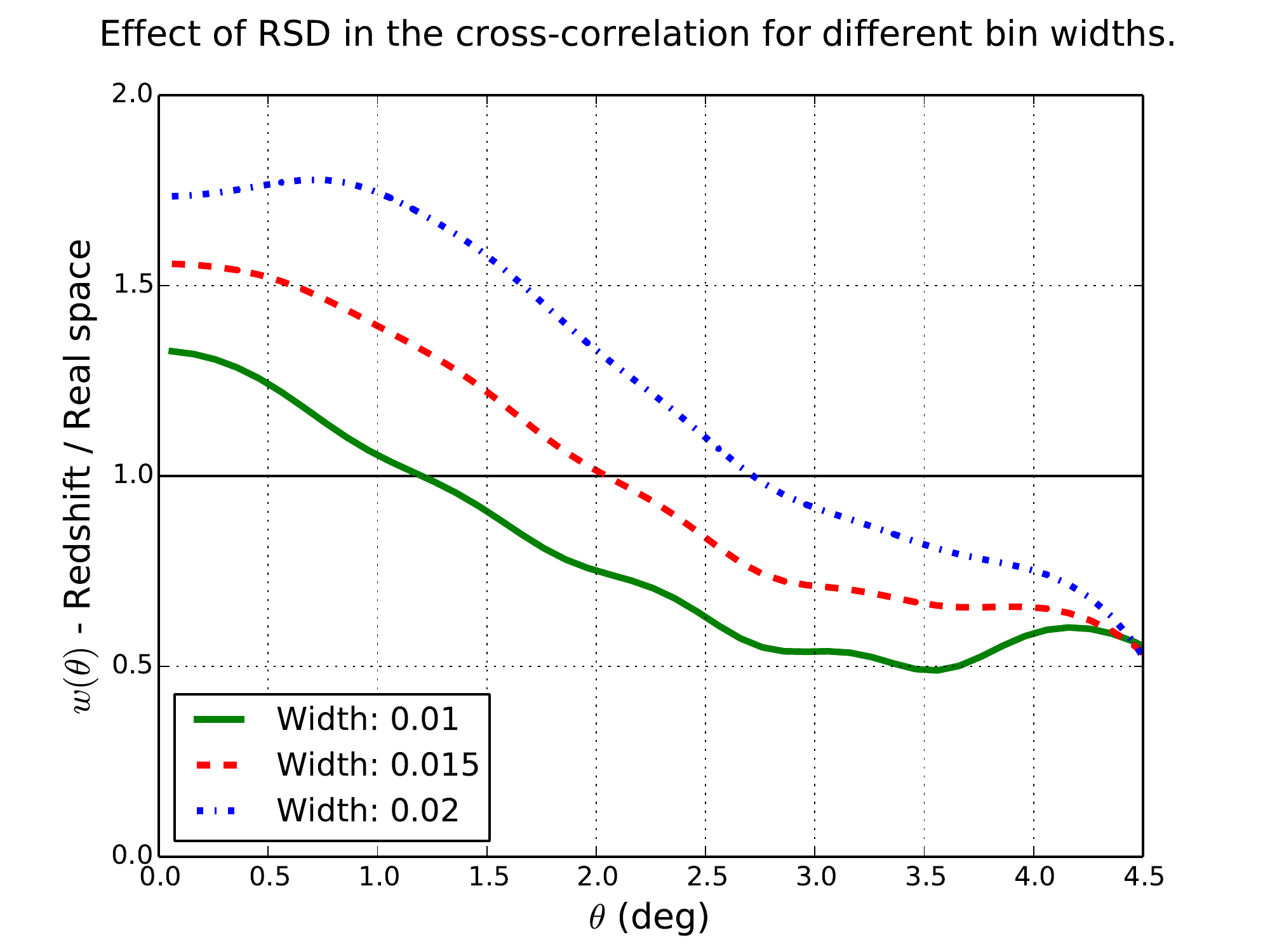}
\caption{Redshift/real space ratios for the cross-correlation of galaxy counts
between adjacent bins. The first bin starts at $z = 0.5$ and the three lines
corresponds to $\dzbin/(1+z) = 0.01, 0.015, 0.02$. Ratios below unity mean
the redshift space distortions suppress the cross-correlations.}
\label{cross_rsd_ratios}
\xef

Fig. \ref{cross_effects} demonstrates how the cross-correlations are affected
by BAO, RSD and both effect together. These effects are also strong in the
cross-correlations. In the auto-correlations the effect of redshift space
distortions increased when decreasing the redshift bin width. Fig.
\ref{cross_rsd_ratios} show the effect of RSD for 3 different bin widths. The
effect of RSD depend strongly on the separation. Also, for thinner bins the RSD
suppresses the signal down to smaller angles.

\xbf
\xfigure{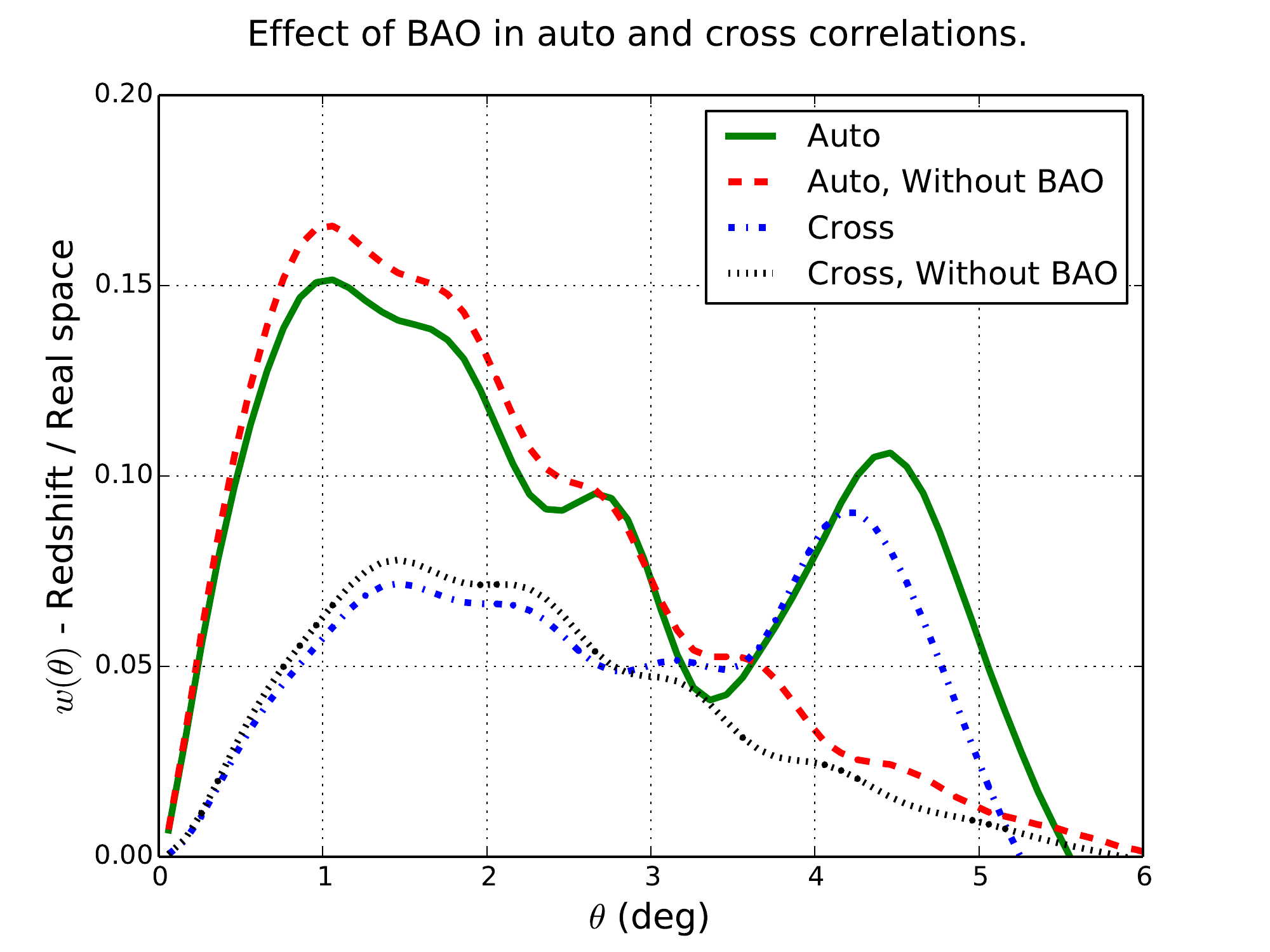}
\caption{Effect of BAO in the auto and cross correlations. The auto
correlations starts at $z=0.5$ with $\dzbin = 0.01(1+z)$. In the cross-correlation
the second bin is the adjacent bins, which also use $\dzbin = 0.01(1+z)$. Both
the auto and cross-correlations are shown with and without BAO.}
\label{auto_cross_bao}
\xef

\subsubsection{Baryon Acoustic Oscillations.}
\label{baosubsec}
A characteristic effect in the cross-correlations is an enhancement of BAO.
While the effect is present in Fourier space, the physical explanation is
simpler in configuration space. Fig. \ref{auto_cross_bao} show together an
auto- and cross-correlation with and without BAO. Around 1 degree the auto and
cross correlations differs by a factor of 2, while they are comparable around
the BAO peak. A geometrical interpretation follows from the galaxy pair
separation. The galaxy pair separation $r$ can be decomposed into

\newcommand{\rpar}{r_{\parallel}}
\newcommand{\rperp}{r_\perp}
\begin{equation}
r^2 = \rpar^2 + \rperp^2
\end{equation}

\noindent
where $\rpar$ and $\rperp$ are the distance parallel and perpendicular to the
line of sight. The $\rperp$ is measured in an angular separation $\theta$ (on
the sky) and converted to a distance through $\rperp \approx \chi(z) \theta$, with
$\chi(z)$ being the comoving distance to the closest galaxy. Looking at one
fixed angle for the BAO scale corresponds roughly to selecting galaxy pairs
with one transverse separation. Galaxy pairs within a thin redshift bins are
mostly radially separated around the BAO scale, therefore measuring the angular
diameter distances. The cross-correlations between close redshift bins is
dominated by the radial BAO, therefore measuring the comoving distance
\mycite{radialbao}.

The auto and cross-correlations differ radically in the $\rpar$ distribution of
galaxy pairs. For the auto-correlation in $\Delta R$ wide top-hat bins, the
probability is highest for zero radial separation and decreases linear towards zero at
the bin edges. In the cross-correlation of adjacent bins, the highest
probability corresponds to $r_{||} = \Delta R$ and decreases linearly towards
the lowest and highest separations. For cross-correlations the redshift bin
separation act as a filter around a characteristic distance. In
Fig.\ref{auto_cross_bao}, the result can be seen from the small scales being
suppressed, while the change is less around the BAO scale.

\xbf
\xfigure{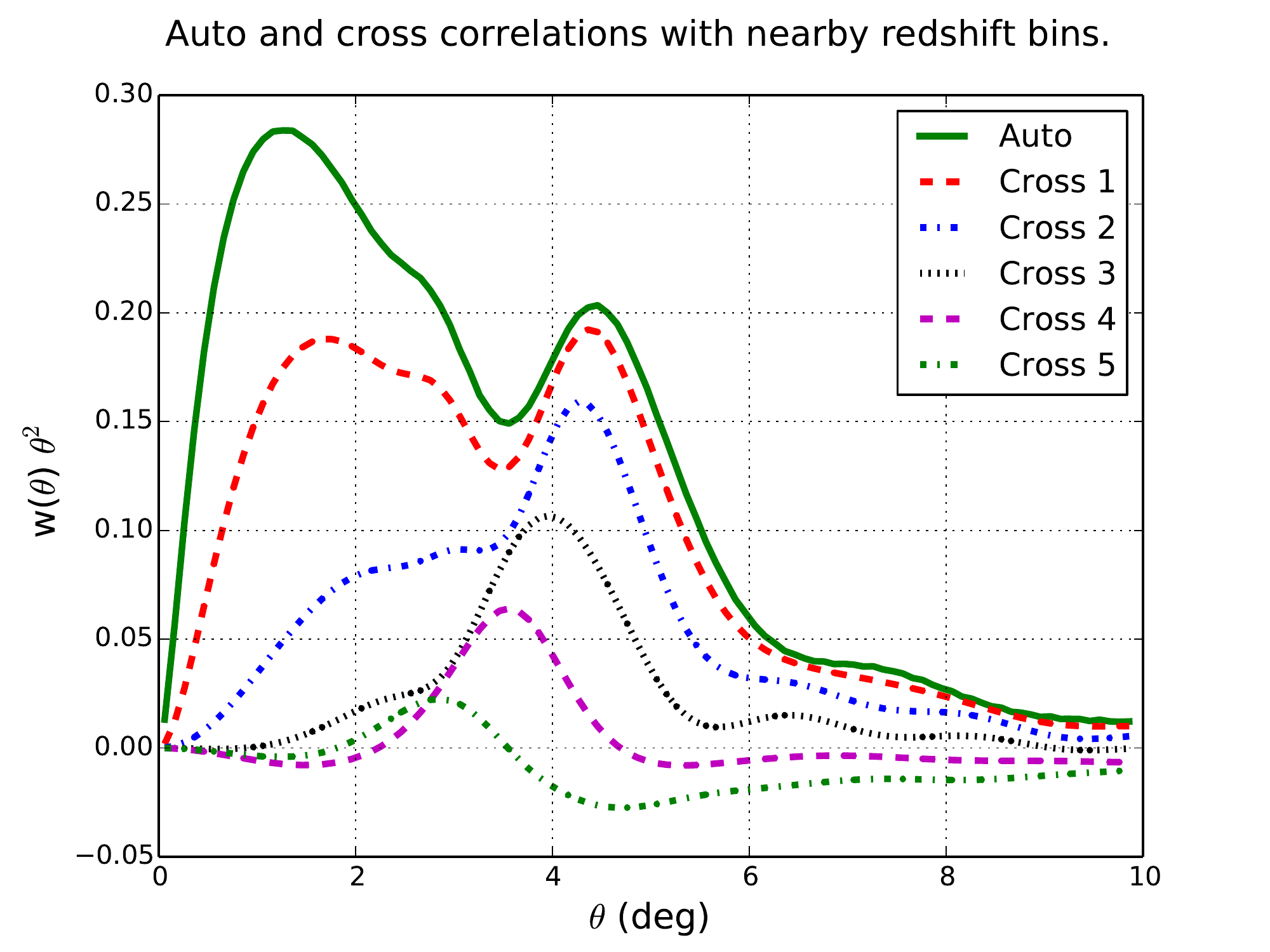}
\vspace{-1.7\baselineskip}
\xfigure{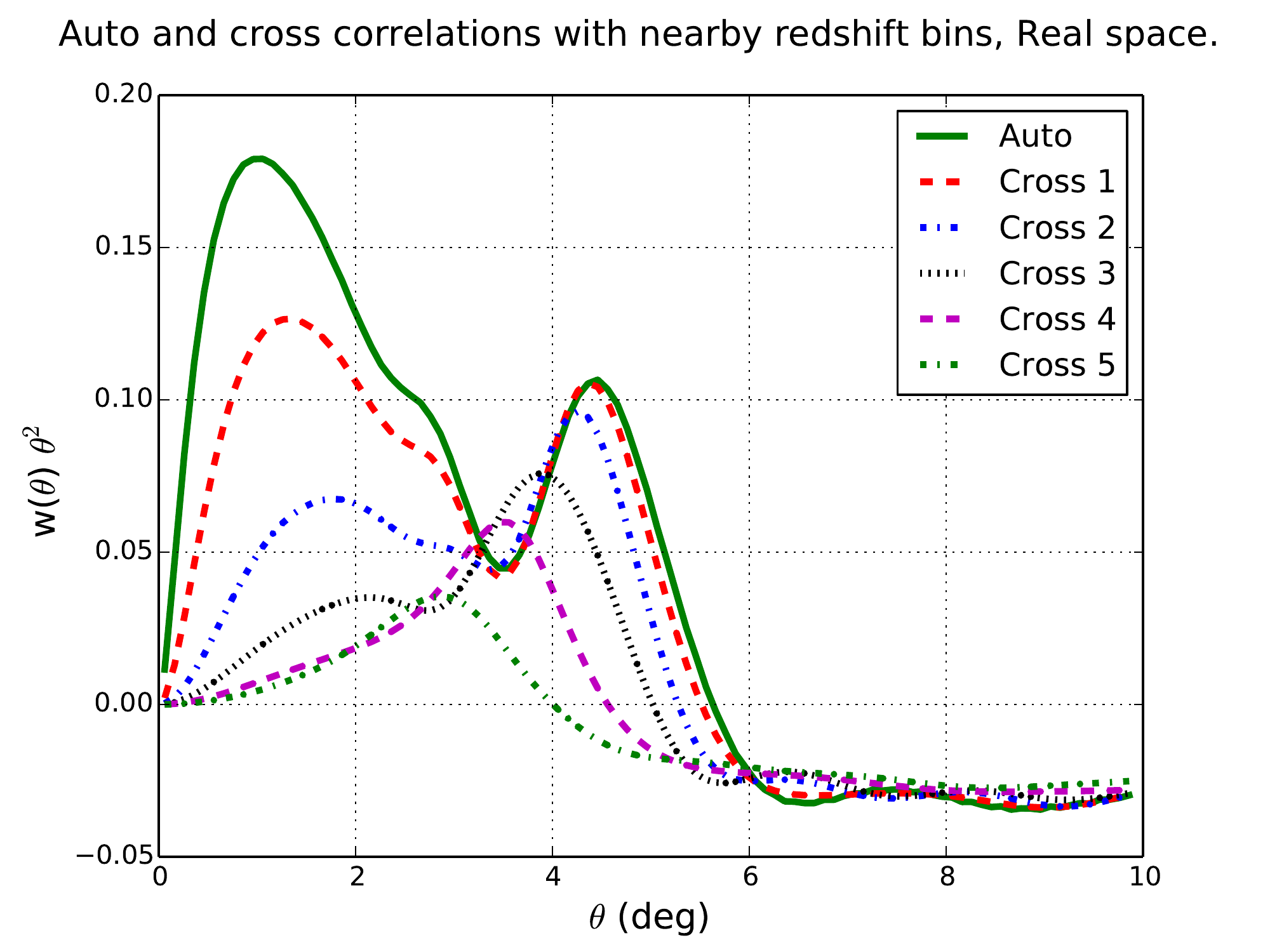}
\vspace{-1.7\baselineskip}
\xfigure{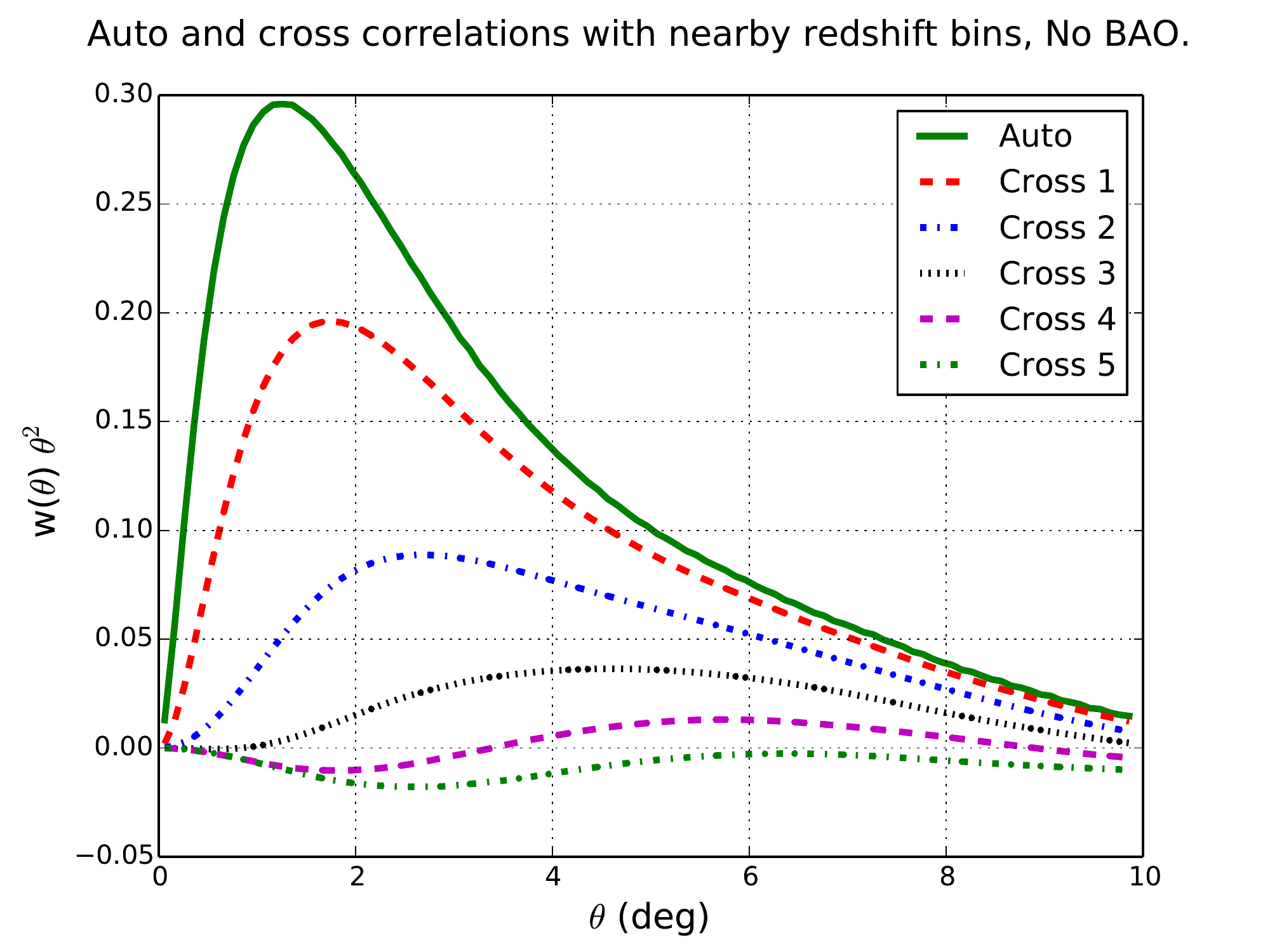}
\caption{The auto and cross-correlations between close redshift bins.  The
first redshift bin always starts at $z=0.5$ and all the redshift bin widths are
$\dzbin = 0.005 (1+z)$. We show the auto correlation (auto) and correlation
with the adjacent bin (cross 1) and the four next closest redshift bins at
higher redshift. In the legend "corr n" means the redshift bin index of the two
observable differs with n, i.e. 1 is the adjacent bin. The top, middle and
bottom panels show respectively the fiducial, real space and no-wiggle
correlations.}
\label{many_auto_cross}
\xef

The cross-correlations filter away small scales also when the bins are
separated, although the distribution of radial distances ($\rpar$) changes.
Fig.16 include the cross-correlations between more bins. For larger redshift
bin separation the suppression of small scales in the cross-correlations become
stronger. On the other hand, for the BAO scale the first cross-correlations are
comparable to the auto-correlation. Here the gap between two redshift bins
introduce a lower limit on the galaxy pair separation. As the separation
between the bins increase, the distance filtered out gradually grows above the
BAO scale of 150 Mpc. For larger separation, as seen in the last
cross-correlations, the peak is also affected.

For angles above 3.5 deg., the last cross-correlation in Fig.
\ref{many_auto_cross} becomes negative. Unlike the auto-correlation which is
positive (for the relevant angles), the cross-correlation can also be negative.
In Fourier space (Cls), the negative cross-correlations can be understood from
the spherical Bessel function in Eq.\ref{cl_main} and \ref{psi}. For an infinitesimal thin
redshift bin the fluctuations $\delta_l(z,k)$ are proportional to $j_l(k
r(z))$, where $r(z)$ is the comoving distance to the redshift bin.  When
cross-correlating two redshifts the oscillations might be out of phase, which
generates a negative contribution. The integral over the redshift bins is (in
real space) a linear superpositition of two such Bessel functions. For thick
bins negative contributions average out and thin redshift bins increase the
probability of finding negative correlations.

\xbf
\xfigure{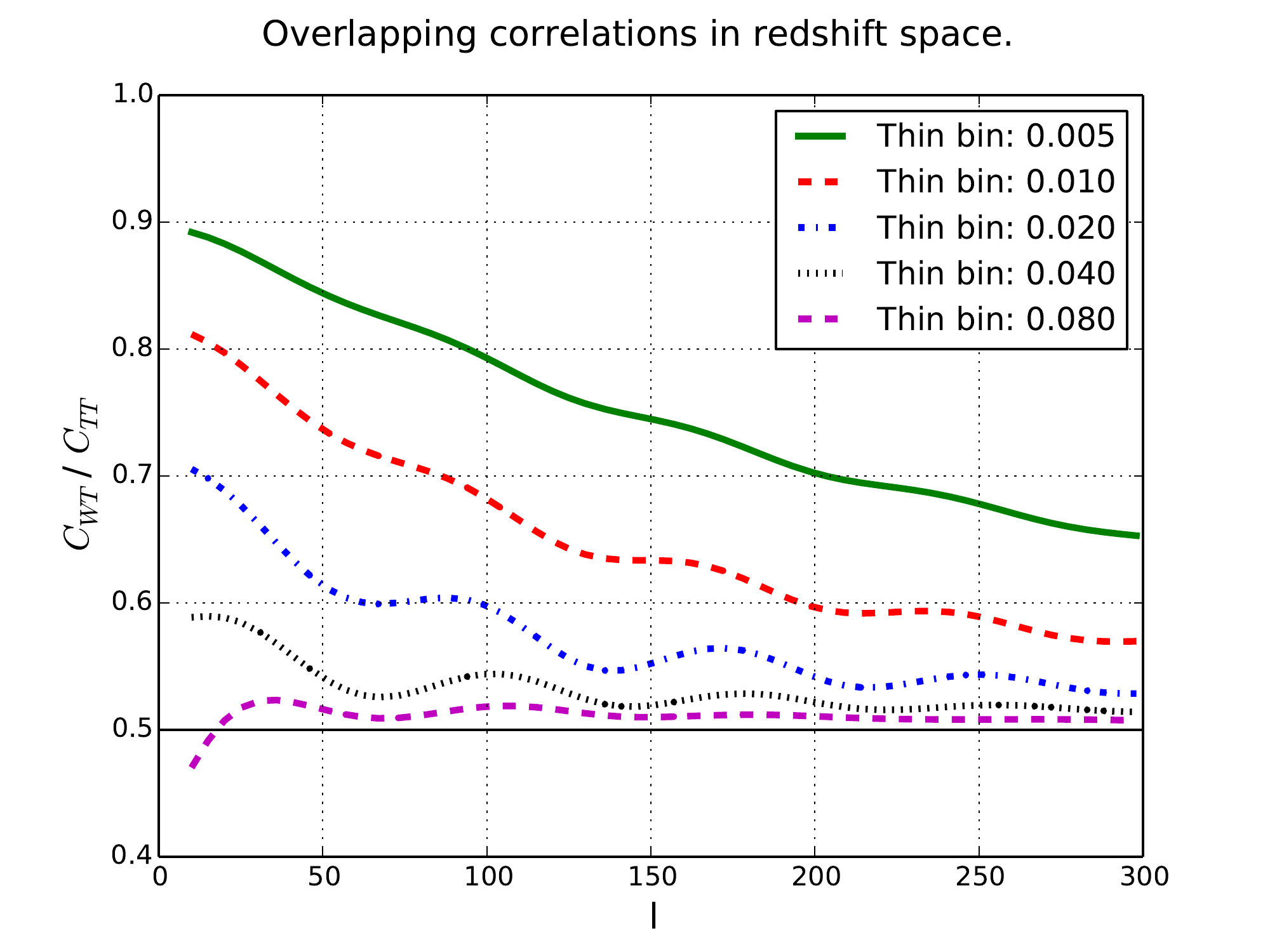}
\caption{Ratio $C_{WT}/C_{TT}$ where W and T are a wide and thin bin. Both bins
are centred around $z = 0.5$, with $\Delta Z_{Thick} = 2 \Delta Z_{Thin}$. The
lines corresponds to using $\dzbin/(1+z) = 0.005, 0.01, 0.02, 0.04, 0.08$ for the
thinnest redshift bin.}
\label{partial_overlap}
\xef

\subsubsection{Partial overlapping bins}
The correlations discussed so far has either been auto-correlations or
cross-correlations of non-overlapping redshift bins. One can for galaxy counts
use a multi-tracer strategy and split the galaxy sample into different
populations. For example including two galaxy populations with very different
bias reduce the sample variance. In the forecast we use the spectroscopic and
photometric surveys as two different populations. The spectroscopic binning is
7 times thinner (than the photometric sample) to capture the radial
information. When cross-correlating overlapping photometric and spectroscopic
surveys, it naturally leads to cross-correlations of redshift bins with
different width.

Fig. \ref{partial_overlap} show the cross-correlation of galaxy counts in two overlapping redshift
bins with $\Delta Z_{Wide} = 2 \Delta Z_{Thin}$. The ratio shown is $C_{WT} / C_{TT}$,
where T and W respectively denotes wide and thin bins. In the Limber
approximation, the auto-correlation is inverse proportional to the bin width.
If all the correlation of the overlapping cross-correlation is due to galaxy
pairs in the overlapping region, one expect $C_{WT} / C_{TT} = 0.5$ for the
Limber approximation in real space. For the relatively thick bins of $\dzbin =
0.04,0.05$, the ratio is close to the Limber ratio (0.5). Cross-correlation of
thinner bins increases the ratio. For the three thin redshift bins of 0.005,
0.01 and 0.02 the overlapping correlation is higher than what is expected from
counting galaxy pairs in the overlapping region. When using two overlapping
bins, the galaxies are not only correlating inside the overlapping redshift
region, therefore increasing the correlation.

\subsection{Errors and signal to noise.}
\label{subsec_err}
The cosmic variance errors when assuming Gaussian fluctuations are \mycite{dodelson}

\begin{align}
Cov(C_{ij}, C_{mn}) &= N^{-1}(l) \left(C_{im} C_{jn} + C_{in} C_{jm}\right) \\
Var(C_{ij}) &= N^{-1}(l) \left(C_{ii} C_{jj} + C_{ij}^2\right) \\
Var(C_{ii}) &= 2 N^{-1}(l) C_{ii}^2
\label{error_summary}
\end{align}

\noindent
where $N(l) = 2 f_{sky} (2l+1)$ is the number of modes and $f_{sky}$ is the sky
fraction covered by the survey. The first line give the general covariance
expression, while the second and third line respectively give the variance for
an auto and cross-correlation. Additionally the counts-counts correlation
($C_{g_i g_j}$) include a shot-noise from sampling a finite number of galaxies
and shape measurement errors affect the shear-shear correlations. Let
$\tilde{C}$ and $C$ respectively denote the correlations including or not the
measurement errors. The observed correlations are

\begin{align}
\tilde{C}_{g_i g_j} &= C_{g_i g_j} + \delta_{ij} \frac{1}{N_i} \\
\tilde{C}_{\gamma_i \gamma_j} &= C_{\gamma_i \gamma_j} + 
  \delta_{ij}\frac{\sigma^2_\gamma}{N_i}
\end{align}

\noindent
where $N_i$ is the observed galaxy number in bin $i$ per stereo-radian and
$\sigma_{\gamma}^2$ is the average shear measurement variance.

Fiducially to match the forecast, we assume a 0.4 gal/sq.arcmin. dense sample
magnitude limited to $i_{AB} < 22.5$ and a galaxy bias of $b(z) = 2 +
2(z-0.5)$. The correlation in this subsection use $z=0.5$, which means $b = 2$.
These values are selected to match the spectroscopic sample in the forecast.
Note that the errors shown in Fig.24 are dominated by cosmic variance. The
exact $n(z)$ details (see paper-II) is therefore less important. All
signal-to-noise (\sn) plots assume 1000 sq.deg. survey area.

\xbf
\xfigure{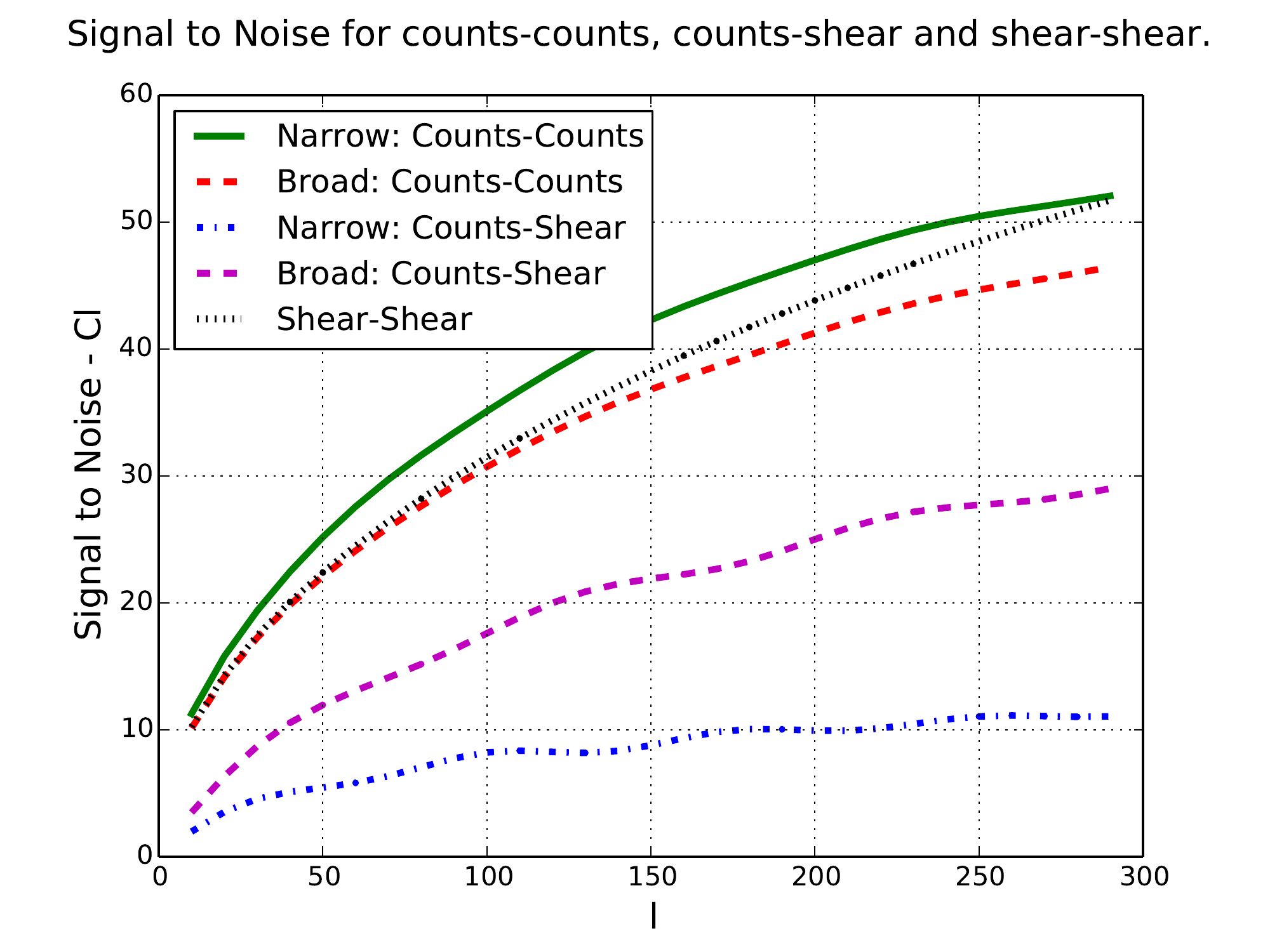}
\caption{Signal-to-Noise for different types of correlation. Two galaxy
populations are included, a foreground population for measuring galaxy counts
and a background population for Faint galaxies. In two lines corresponding to
Count-Counts and Counts-Shear, a thin bin of $\dzbin = 0.001$, $z=0.5$ is
used, while the other two uses a thick bin of $\dzbin = 0.01$, $z=0.5$.
The background bin at $z=1.1$ is $\dzbin = 0.15$ wide. Last the figure show
the Shear-Shear \sn  ratio for the background shear used for both the
Counts-Shear correlations.}
\label{sn_diff_corr}
\xef

Fig.\ref{sn_diff_corr} show the \sn for different counts-counts, counts-shear
and shear-shear correlations. An important point of this figure is to compare
how the redshift bin width affect the \sn. Therefore the counts-counts and
counts-shear correlations are both shown with a thin and a thick foreground bin
(see caption). For zero shot-noise, the \sn of the counts-counts
auto-correlations are

\begin{equation}
(\sn)[C_{g_i g_i}] \equiv \sqrt{N(l) / 2}
\end{equation}

\noindent
which is independent of the redshift bin width. The two count-counts auto
correlation lines differs in shot-noise covering different redshifts since
one set of bins are wider. Except this, the signal to noise for the two
auto-correlations does not depend on the redshift bin width. The counts-shear
\sn is directly dependent on the lens bin width. From the Limber approximation,
the counts-counts auto-correlation in the variance is inversely proportional to
the bin width ($C_{ii} \propto 1/\Delta_i$). On the other hand, the
counts-shear signal is independent of the foreground bin width when ignoring the
cosmological evolution in the lens bin. Combined these two expressions lead to

\begin{equation}
(\sn)[C_{g_i\gamma}] \propto \sqrt{\Delta_i}
\end{equation}

\noindent
where $\Delta_i$ is the redshift bin width. This means the \sn of a single
counts-shear cross-correlation decrease when using a thinner lens redshift bin.
In Fig. \ref{sn_diff_corr}, the two counts-shear \sn lines use $\dzbin = 0.01$
and $\dzbin = 0.1$ wide lens bins. A cross-correlation with 10 times thinner
redshift bins should result in around 3 times lower \sn. While each correlation
becomes noisier when decreasing the lens bin width, the reduced width allow for
more bins. The number of bins is proportional to $1/\Delta_i$. Therefore the
combined \sn for all counts-shear correlations scale with $\sqrt{\Delta_i}$. In
addition thinner lens bins has the advantage of less projection in redshift.

\xbf
\xfigure{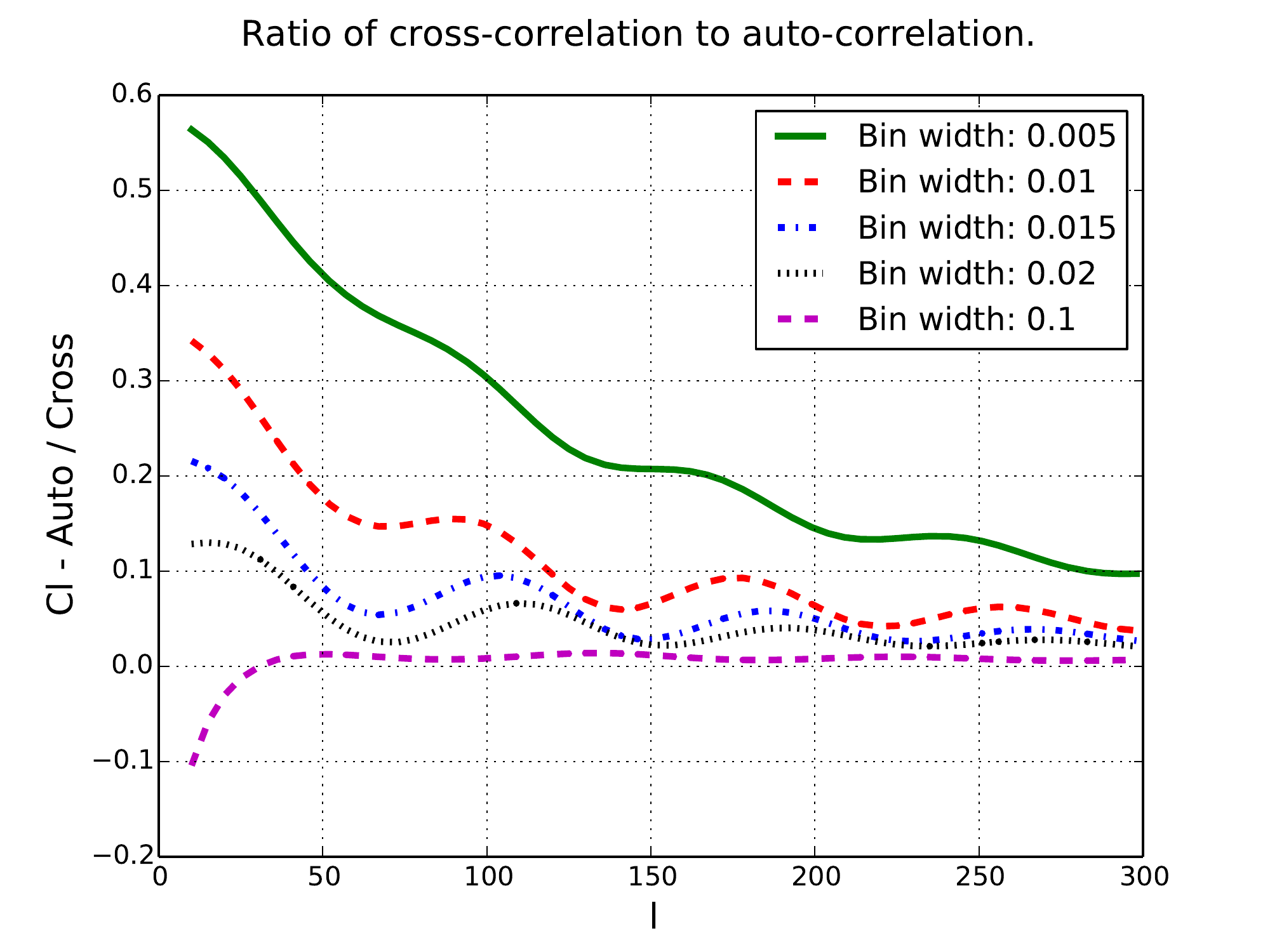}
\xfigure{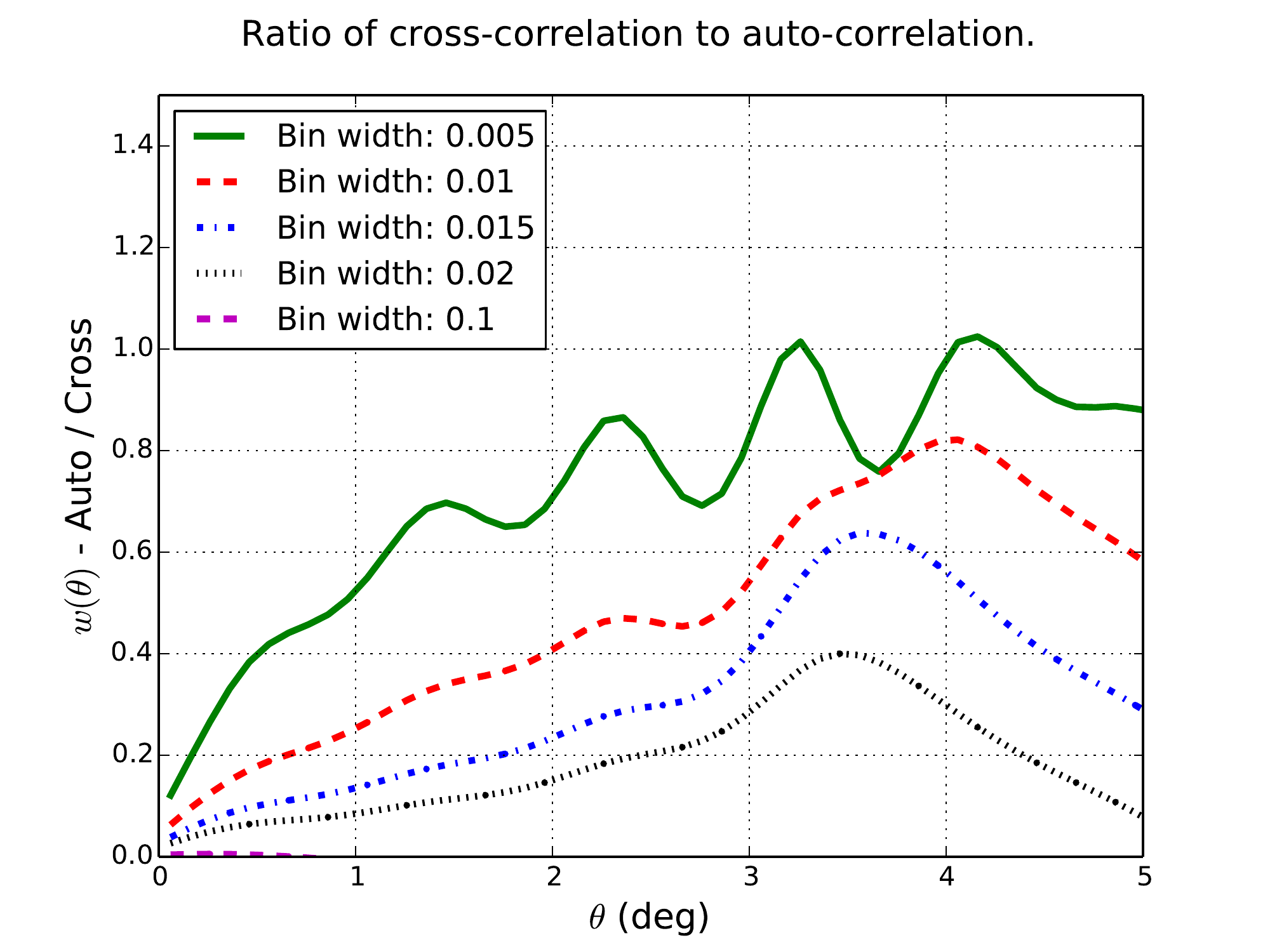}
\caption{Auto/cross-correlation ratio for galaxy counts. The auto-correlation
bin starts at $z=0.5$, with the four lines corresponding to $\dzbin =
0.005,0.01, 0.015, 0.02$.  The cross-correlation is between the
auto-correlation bin and the adjacent bin at higher redshift. Both redshift
bins are equally wide. In the top and bottom panel, the figure respectively
show the ratio in Fourier and configuration space.}
\label{ratio_cross_auto}
\xef

Subsection \ref{subsec_cross} studied the galaxy counts cross-correlation between
adjacent redshift bins. These correlations can, if the \sn is sufficient,
measure radial information. The \sn for the cross-correlations between
different redshift bins are directly related to the cross/auto correlations
ratio. This follow from

\begin{align}
\frac{(\sn)[C_{ij}]}{(\sn)[C_{ii}]} 
&= \sqrt{2} \frac{C_{ij}}{C_{ii}} \frac{C_{ii}}{\sqrt{C_{ii} C_{jj} + C_{ij}^2}} \\
&\approx \sqrt{2} \frac{C_{ij}}{C_{ii}} {\left( 1 + {(\frac{C_{ij}}{C_{ii}})}^2 \right) }^{-1/2} \\
&\approx \sqrt{2} \frac{C_{ij}}{C_{ii}}
\label{sn_cij_rule}
\end{align}

\noindent
where the second and third line respectively use $C_{ii} \approx C_{jj}$ and
$C_{ij} << C_{ii}$. Approximating the auto-correlations ($C_{ii} \approx
C_{jj}$) works for equally wide and thin bins. When $C_{ii}/C_{ij}=2.1, 6.9$
the last approximation is respectively accurate to 10\% and 1\%. The \sn ratio
can be understood from the cross-correlation variance being dominated by the
auto-correlation variance.

Fig.\ref{ratio_cross_auto} top panel show the ratio $C_{ij}/C_{ii}$ for various
bin widths. When increasing the bin width, the ratio decline quickly due to
$C_{ij}$ being sensitive to the redshift bin separation. If using $\dzbin =
0.005$ instead of $\dzbin = 0.01$, the $C_{ij}/C_{ii}$ ratio doubles. The
bottom panel show the $w_{ij}(\theta)/w_{ii}(\theta)$ ratios. Another
interesting aspect is looking at $w(\theta)$ the cross-correlations are having
a high signal at large angles. For example the cross-correlation with bin-width
of 0.01 is $40\%$ of the auto-correlation at 2 degrees and $80\%$ at 4 degrees.
This means the cross-correlations is gaining a higher signal to noise at larger
angles.

\xbf
\xfigure{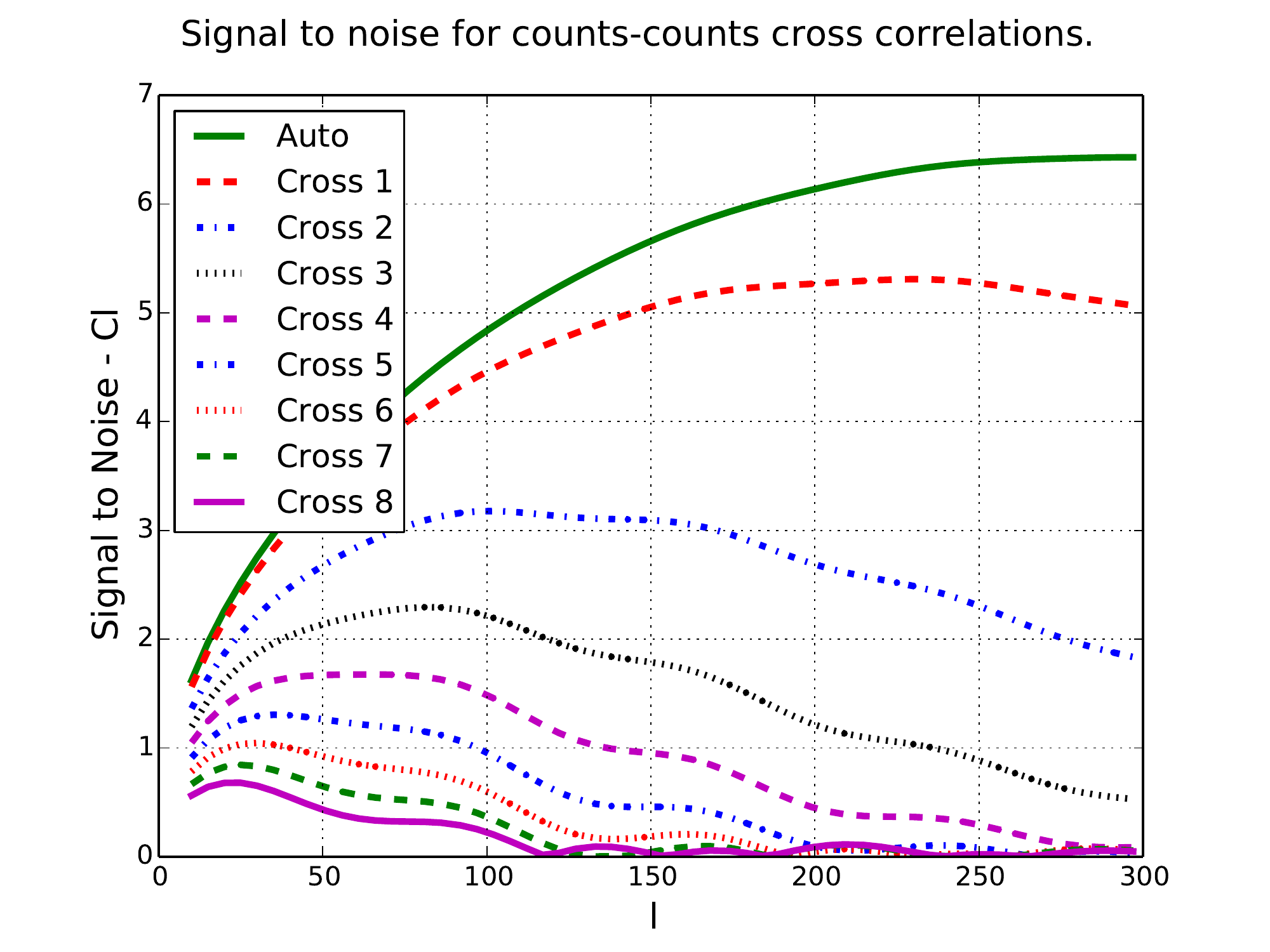}
\caption{Cross-correlations between very narrow redshift bins. All bins are
$\dzbin = 0.001(1+z)$ wide and the first redshift bin start at $ z =0.5$. The
auto-correlation is for the first redshift bin, while "Cross X" corresponds
to a cross-correlation of the first bin with a bin separated by X time $\Delta z$.}
\label{sn_cross_extreme}
\xef

To illustrate the effect of cross-correlations, Fig. \ref{sn_cross_extreme}
shows the signal-to-noise for extremely thin redshift bins. These redshift bins
are $\dzbin = 0.001(1+z)$ wide, which would correspond to 694 redshift bins for
$0.1 < z < 1.2$. From the figure, one see there is a sharp drop in \sn when
increasing the distance between the redshift bins. Also, the change is lower at
low l-values, which make the cross-correlations more important at large scales.
For the forecast in $\dzbin = 0.01 (1+z)$ wide bins, the main contribution
comes from the auto-correlation and cross-correlation with the adjacent
bin.

\xbf
\xfigure{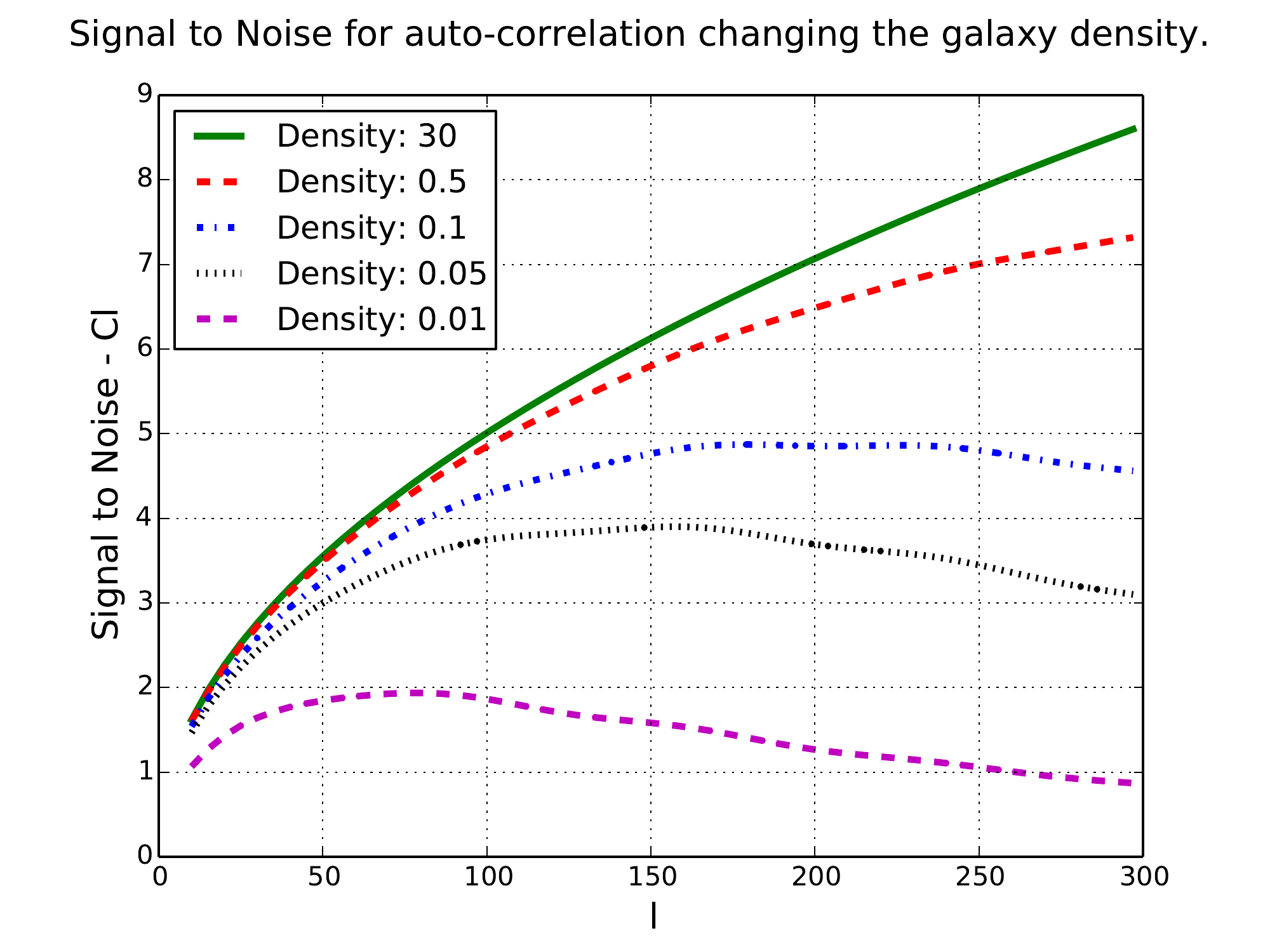}
\caption{Signal-to-noise for different galaxy densities. The auto-correlation
redshift bin is centred in $z = 0.5$ and is $\dzbin z = 0.01 (1+z)$ wide.  The
\sn is calculated for 0.01, 0.05, 0.1, 0.5 and 30 gal/sq.arcmin in the full
sample. In addition 50\% is removed to simulate various cuts.}
\label{sn_auto_dens}
\xef

Finally Fig. \ref{sn_auto_dens} shows the \sn for auto-correlations in the
redshift bin $z=0.5$, $\dzbin = 0.01(1+z)$ for different densities. The upper
line with 30 gal/sq.arcmin, which is lower than the expected LSST density of 40
gal/sq.arcmin \mycite{lsst2}, is approximately noiseless for the counts-counts
auto-correlation. A line with 0.5 gal/sq.arcmin is close to the spectroscopic
density used in the forecast (0.4 gal/sq.arcmin). For the l-values considered
for the forecast ($l \leq 300$), the dense spectroscopic sample has a \sn close
to the noiseless limit. These conclusions does however vary with lmax and also
with the redshift one study.

\section{Conclusion}
In this paper we have studied the modelling of galaxy clustering, RSD and WL
with angular cross-correlations. Lensing is often studied using 2D
correlations, while clustering and RSD is analyzed with the 3D power spectrum.
Combining WL, large scale structure and RSD is from a theoretical viewpoint
significantly easier using the angular correlation functions. Directly
constructing observable in angles (or multipoles) and redshifts avoids the
model assumptions that are needed in a 3D analysis when converting distances.
Moreover, the expression for the covariance is straight forward. Using only
angular correlations avoids double counting transverse information and can
naturally account for potential redshift uncertainties in the LSS analysis by
migration matrices \mycite{gazta}.

One practical concern is the efficiency of implementing a computer code
for predicting the angular correlation function. Analyzing a spectroscopic
sample using 2D-correlations requires a large number of redshift bins to capture
the radial information. In section \ref{sec_impl} we
introduced an algorithm specialised on calculating the cross-correlations
between many redshift bins and with multiple tracers. Instead of calculating
each correlation separately, all correlations are calculated at once. This allows for
reusing parts of the calculations and extensive use of matrix multiplication.
In particular, the integration between all correlation of redshift bins is
expressed using a matrix products. Being formulated in terms of array operations
and matrix multiplications allows for an efficient implementation, even in high
level languages (e.g. Python) with bindings to high performance linear algebra
implementations.

Section \ref{mod_sec_effects} began with studying the effect of BAO, RSD and
the Limber approximation for the auto and cross-correlations. For the
auto-correlation of $\dzbin = 0.1$ thick bins, the Limber approximation can
be sufficient for a small area survey. In narrow redshift bins ($\dzbin =
0.01$), which is needed for the forecast, the Limber approximation completely
breaks down. Redshift space distortions leads to 30\% larger amplitude for the
galaxy counts auto-correlations in broad bins.  For thin bins ($\dzbin =
0.01$), the RSD effect can result in 2.5-3 times higher auto-correlations at
low multipoles. In addition for thinner bins the effect of redshift space
distortions clearly shows a peak in angle. We showed that this second peak,
which we named the ghost BAO peak, results from the BAO peak being shifted in
redshift space.

The cross-correlation of nearby redshift bins unexpectedly has a larger BAO
contribution than the auto-correlations. When cross-correlating two redshift
bins, the bin separation affects the radial pair separation. In
cross-correlations the most probable radial galaxy separation is the distance
between the mean of the two redshift bins. Therefore cross-correlations include
pairs with higher radial separation, which suppress small-scale clustering and
lead to a larger BAO contribution. Galaxy pairs within a thin redshift bins are
mostly radially separated around the BAO scale, therefore measuring the angular
diameter distances. The cross-correlations between close redshift bins is
dominated by the radial BAO, which measures the comoving distance
\mycite{radialbao}. We have also shown in section \ref{subsec_err} that the
signal-to-noise (\sn) is larger in the cross-correlation at BAO scales.

We also studied the \sn for different correlations. The counts-counts
correlations has the highest signal-to-noise, while the counts-shear and
shear-shear correlation have a lower \sn ratio. In the count-shear correlation
the bin width of the galaxy counts in the lens bin directly affects the noise,
while the signal is only affected by projection effects. Each correlation of
narrow bins has a lower signal-to-noise. However the total \sn is higher since
the thinner bins result in more counts-shear cross-correlations. Finally we
looked at the sensitivity of galaxy density for the counts-counts
auto-correlations. For a dense galaxy sample (0.4 gal/sq.arcmin), used for the
spectroscopic sample in the forecast, the \sn for $z=0.5$ is close to the
noiseless limit.

Altogether our results show that this new approach to clustering analysis,
using angular cross-correlations in narrow redshift bins, is potentially
viable: it recovers all the 3D information \mycite{asorey, asorey2}, it can be
predicted with a fast algorithms and it contains new insights of physical
effects such as WL, RSD and BAO. In the following papers of this series we will
present different applications and results using the formalism presented here.

\section*{Acknowledgements}
We would like to thank Martin Crocce for help with the initial stages of this
long project. Funding for this project was partially provided by the Spanish
Ministerio de Ciencia e Innovacion (MICINN), project AYA2009-13936 and
AYA2012-39559, Consolider-Ingenio CSD2007- 00060, European Commission Marie
Curie Initial Training Network CosmoComp (PITN-GA-2009-238356) and research
project 2009- SGR-1398 from Generalitat de Catalunya. M.E. acknowledge support
from the European Research Council under FP7 grant number 27939. M.E. was also
supported by a FI grant from Generalitat de Catalunya.

\appendix
\section{Auto and cross correlations}
\label{effcompare}
\subsection{Redshift space distortions}
\newcommand{\runcomp}{ex151}

\xbf
\xfigure{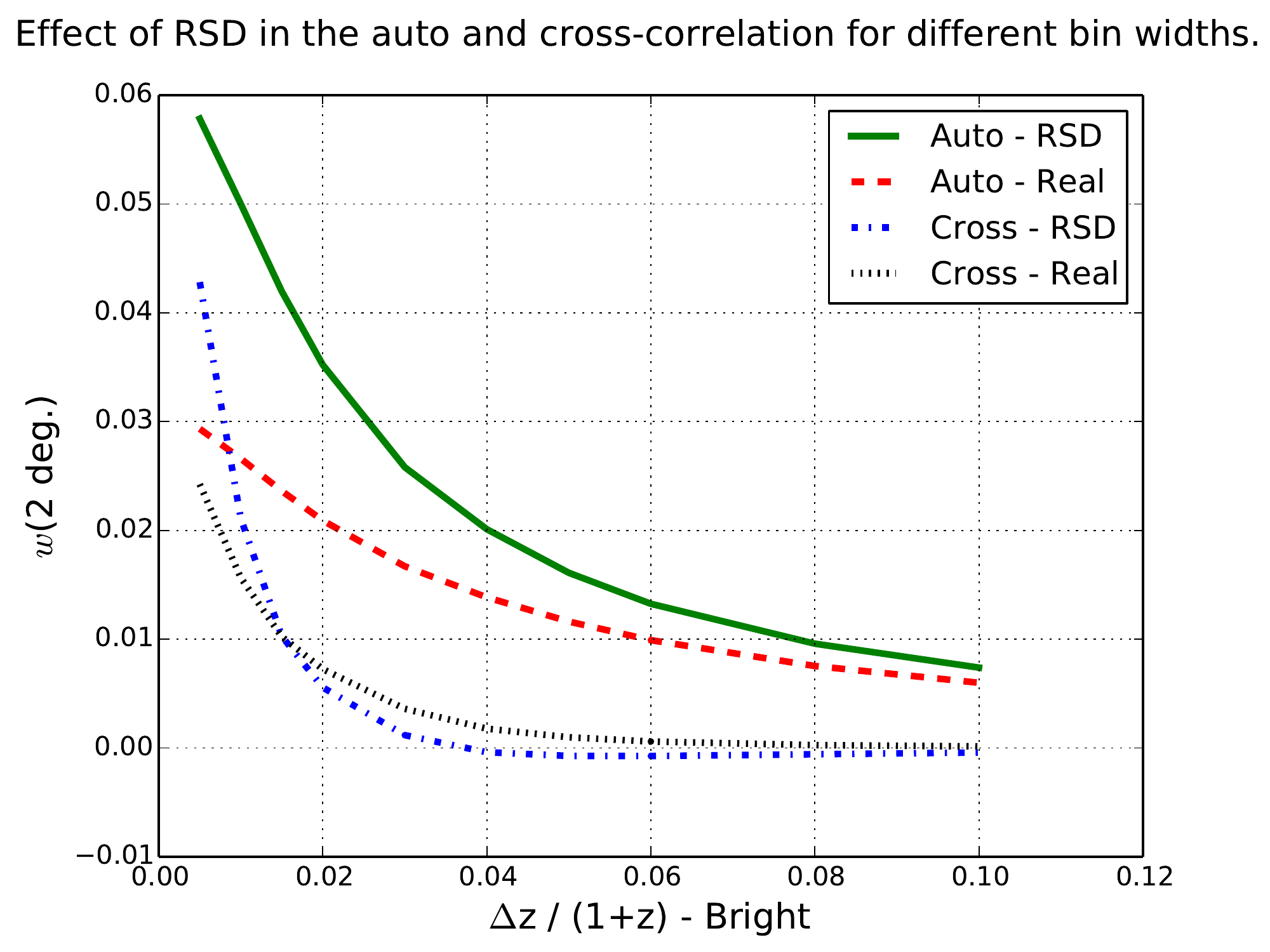}
\caption{Contribution of RSD for the auto- and cross-correlations with the
adjacent bin when varying the redshift bin width. Both redshift bins are
equally wide (at $z=0$), with the width given on the x-axis. The correlations
are shown at 2 degrees and the first redshift bin starts at $z = 0.5$.}
\label{comp_rsd_width}
\xef

The RSD dependence on the redshift bin width is illustrated in
Fig.\ref{comp_rsd_width}.  An additional redshift from peculiar line-of-sight
velocities can move galaxies between redshift bins, which cause the RSD signal.
In thin redshift bins more galaxies move between the bins, therefore thin bins
increase the RSD signal. For the thinnest bin ($\dzbin = 0.005$) the
auto-correlation has double the amplitude in redshift compared to real space.
When increasing the bin width, both the signal and fraction of RSD signal
decrease. For cross-correlations, the redshift space distortions can
contribute positive or negatively, depending on the bin width. In this
configuration, below $\dzbin \approx 0.015$ the RSD increase the
cross-correlation, which suppressing the signal for wider redshift bins. For
thick redshift bins ($\dzbin = 0.05 - 0.1$) the cross-correlations is negative
in redshift space.

\xbf
\xfigure{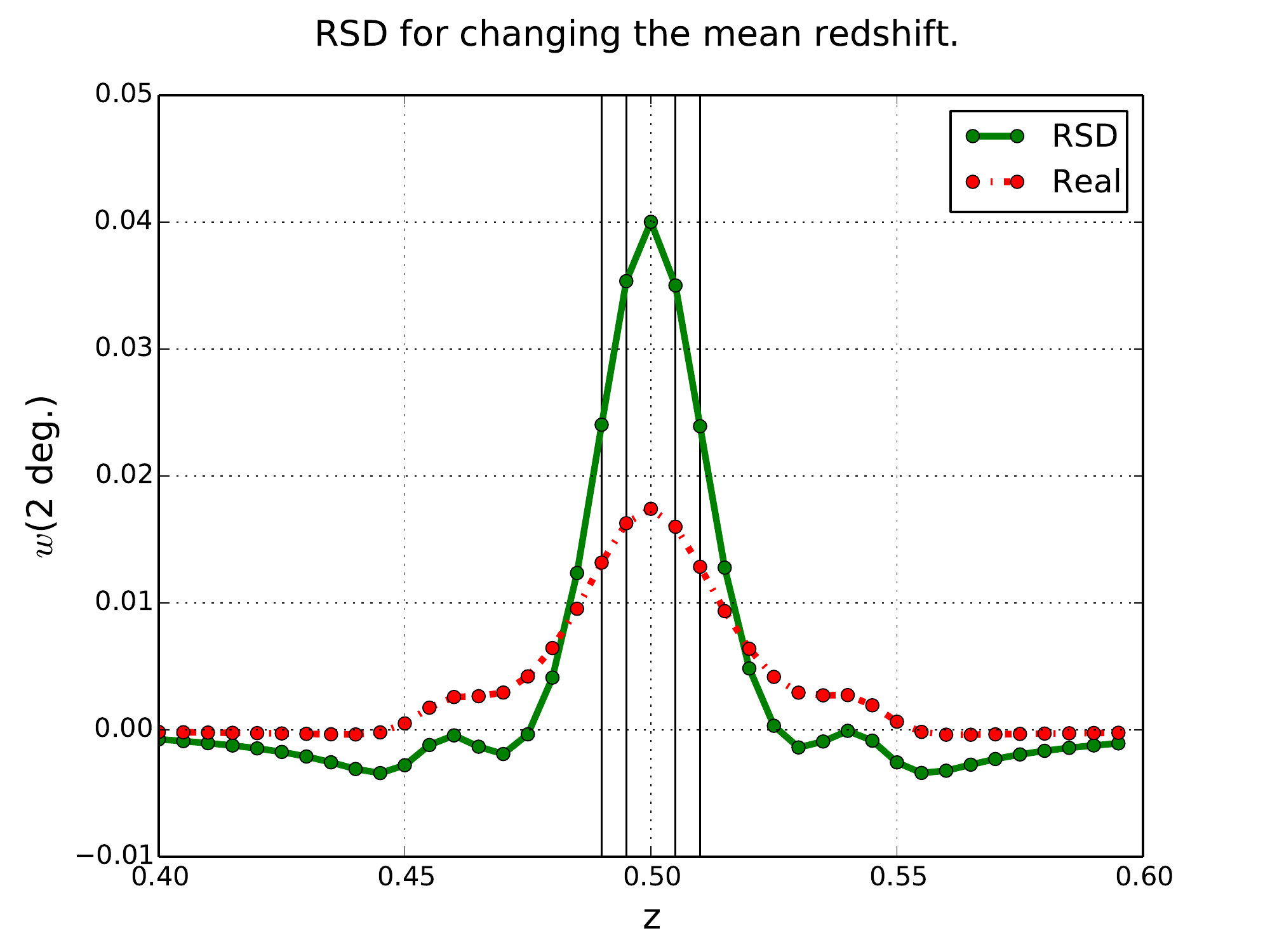}
\xfigure{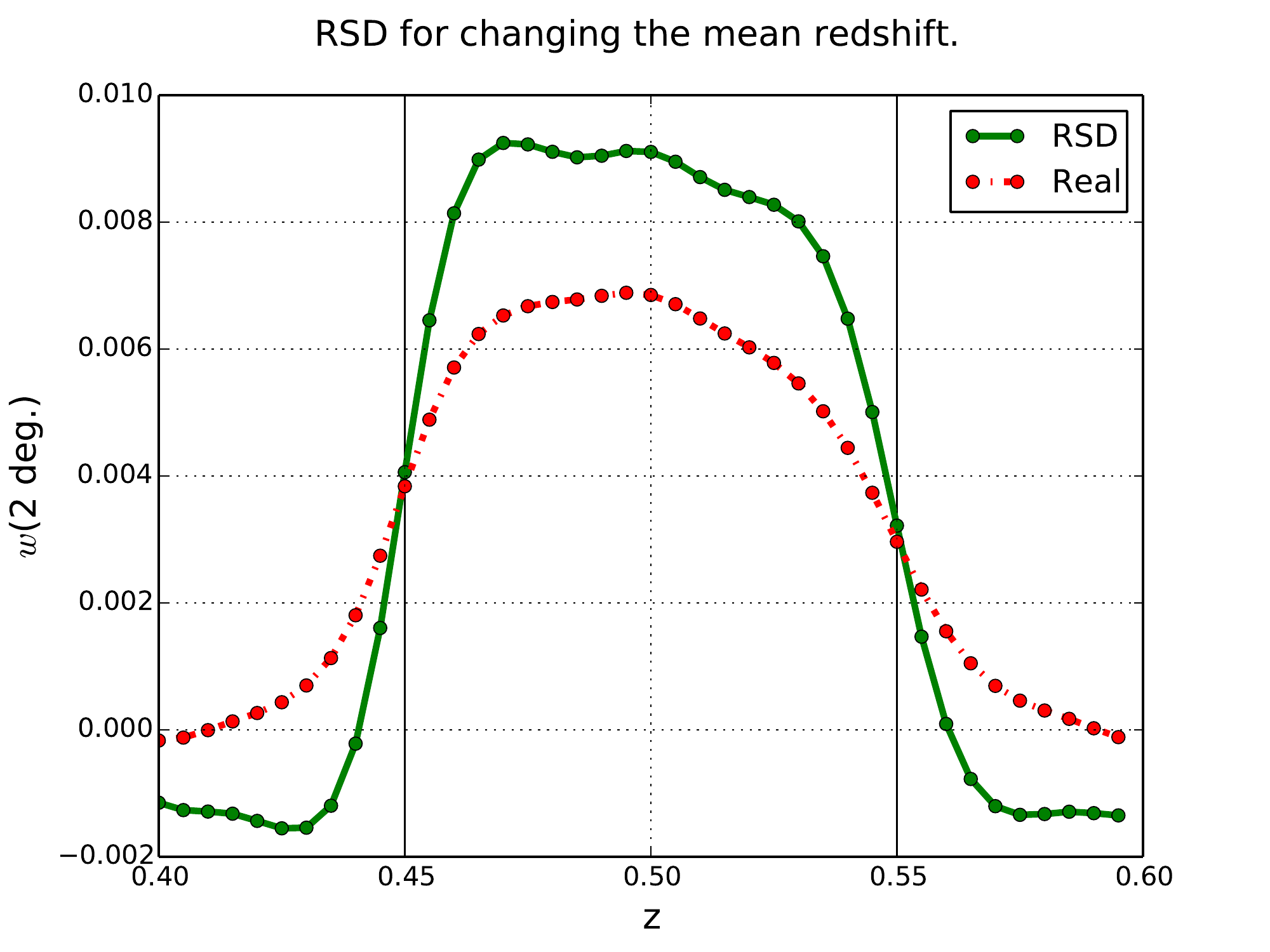}
\caption{The cross-correlation when changing one bin position. In both panels,
one redshift bin is fixed at $z=0.5$ and the second redshift mean vary as
indicated on the x-axis. The top panel correlate two thin ($\dzbin = 0.01$)
redshift bins. The two inner vertical lines ($z=0.495,0.505$) mark the fixed
redshift bin, while the two outer lines ($z=0.49,0.51$) mark when the bins have
no redshift overlap. In the bottom panel, the fixed bin is thick ($\dzbin =
0.1$), while the varying bin is thin ($\dzbin = 0.01$). In the fixed bin, the
bias $b(z) = 1.2 + 0.4 (z - 0.5)$ is used. Here the vertical lines
($z=0.45,0.55$) mark the fixed bin. The two lines show the correlation in
redshift and real space.}
\label{comp_rsd_mean}
\xef

Fig. \ref{comp_rsd_mean} shows a $w(\theta)$ cross-correlation amplitude when
shifting one redshift bin, while the other is centred around $z=0.5$. In the
top panel, both redshift bins are $\dzbin = 0.01$ wide and the inner vertical
lines mark the fixed bin. For two narrow and fully overlapping bins, the signal
double in redshift space.  When reducing the amount of overlap, both the
clustering in real space and redshift space decreases. The two outer vertical
lines marks having redshift bins side by side. In this configuration, the
correlations are still positive. For larger separation the redshift space
distortions suppress the signal, which was also seen in Fig.
\ref{comp_rsd_width}.

In the bottom panel the fixed bin is thick ($\dzbin = 0.1$), while the bin
changing position is still thin ($\dzbin = 0.01$). For fully overlapping bins
and close centers the signal is fairly flat. When the thin bin move closer to
the edge, but they still overlap, the signal falls off sharply. The decrease in
the cross-correlation amplitude comes from removing part of the non-overlapping
cross-correlations between the bins. When the bins move apart, the signal
clearly becomes negative in redshift space.

\subsection{Non-linear effects}
\xbf
\xfigure{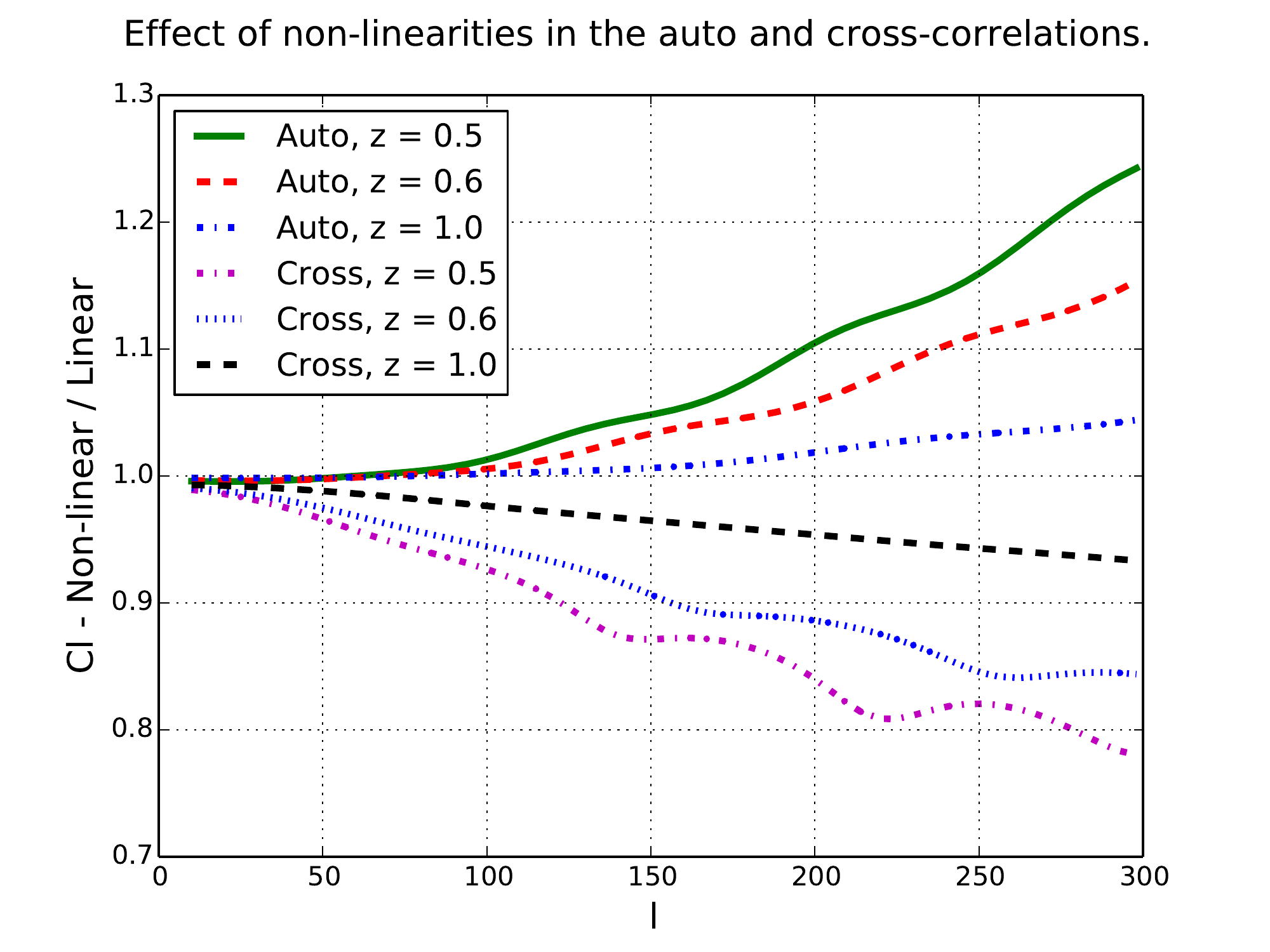}
\caption{The impact of non-linear effect in the auto and cross-correlations for different
redshifts. The figure show non-linear/linear correlation function ratio, with
non-linear including Halofit contributions to the linear EH power spectrum. All redshift
bins are $\dzbin = 0.01 (1+z)$ wide, the cross-correlation is with the adjacent bins
and the redshifts $z = 0.5,0.6,1.0$ give the start of the first redshift bin.}
\label{comp_nonlin}
\xef

Non-linear gravitational effects enhance the dark matter power spectrum on
small scales. The EH and CAMB \mycite{camb} power spectrum models only the
linear power spectrum. For modelling the non-linear effects, one can 
create fitting formulas based on n-body simulations or use pertubation theory.
The Halofit-II power spectrum model is based on a series of simulations to
model non-linear gravitational effects \mycite{halofit1, halofit2}. To include
the Halofit only require implementing the model and providing a linear power
spectrum, for which we use the Eisenstein-Hu.

Fig. \ref{comp_nonlin} shows the non-linear/linear Cl ratios.  The ratios are for
auto and adjacent cross-correlations in the redshift bins $z=0.5,0.6,1.0$ and with
$\dzbin = 0.01$. The non-linear effects in the power spectrum increase with the
comoving wavenumber ($k \equiv l/\chi(z)$). As expected, the effect increase
with l and lower redshifts has the highest non-linear contributions. The
oscillations seen are due to the BAO. In the cross-correlations, the non-linear
effects suppress the signal. Previously, we showed (Fig. \ref{delta_cross}) how
counts-counts cross-correlations in narrow bins has both a positive and
negative contribution from different scales. Since higher k-values contribute
negatively and the non-linear effect increase with scale, this leads to
non-linear effects suppressing the cross-correlations.

\section{Clenshaw-Curtis integration}
\label{cc_int}
\subsection{Overview}
The Clenshaw-Curtis integration algorithm expands the integrand in Chebyshev polynomials
and works well for oscillating integrals. Integrating $f$ over the interval
$[-1,1]$ can be approximated

\begin{equation}
\int_{-1}^1 f(x) dx \approx \sum_{n=0}^{N/2} \wght_n {f(\cos[n \pi/N]) + f(-\cos[n \pi /N])}
\label{cc_expand}
\end{equation}

\noindent
where $N$ is the number of integration points. The coefficients $\wght_n$ are given by

\begin{align}
d_i &= 
\begin{cases}
1 &i = 0 \\
\frac{1}{1-N^2} &i = N/2-1 \\
\frac{1}{1-{(2i)}^2} &\text{Otherwise} 
\end{cases} \\
D_{ij} &= \frac{2}{N} \cos{\left( \frac{2 i j \pi}{N} \right)} \\
\wght &= D^T d
\label{cc_weights}
\end{align}

\noindent
where the last equation uses matrix multiplication.The integration in
Eq.\ref{cc_expand} can be transformed to different integration limits.
For example when integrating over scales, one have

\begin{equation}
\int_{k_{min}}^{k_{max}} f(k) dk = k_w \int_{-1}^1 f(\bar{k} + k_w x) dx
\end{equation}

\noindent
where the integration variable is defined by $k = \bar{k} + k_w x$, $\bar{k}
\equiv \frac{1}{2} (k_{min} + k_{max})$ and $k_w \equiv \frac{1}{2} (k_{max} -
k_{min})$.

\subsection{Change of integral domain for the tomographic integration.}
\noindent
The Cl integrand is on the form $f(x) \equiv G_i(x) G_j(x)$. When trying to
integrate by multiplication, the expansion Eq.\ref{cc_expand} into two terms
creates an additional complication. Expanding the terms, one find

\begin{equation}
\begin{split}
\int_{-1}^1 G_i(x) G_j(x) dx = &\sum_{n=0}^{N/2} G_i(\cos(n \pi/N)) G_j(\cos(n \pi/N)) \\
  &+ G_i(\cos(-n \pi/N)) G_j\left(- \cos(n \pi /N)\right).
\end{split}
\end{equation}

\noindent
Through defining

\begin{align}
y^{+}_{in} &= G_i(\cos(n \pi/N)) \\
y^{+}_{jn} &= G_i(-\cos(n \pi/N))
\end{align}

\noindent
the integration can be written

\begin{equation}
\int_{-1}^1 G_i(x) G_j(x) dx = \sum_n w_n (y^{+}_{in} y^{+}_{jn} + y^{-}_{in} y^{-}_{jn}).
\end{equation}

\bibliographystyle{mn2e}
\bibliography{exbib.bib}{}
\end{document}